\documentclass[submission, PhysCore]{SciPost}
\usepackage[utf8]{inputenc}
\usepackage[T1]{fontenc}
\usepackage{graphicx}
\usepackage{longtable}
\usepackage{wrapfig}
\usepackage{rotating}
\usepackage[normalem]{ulem}
\usepackage{amsmath}
\usepackage{amssymb}
\usepackage{capt-of}
\usepackage{hyperref}
\usepackage{xcolor}
\usepackage{mathtools}
\usepackage{stackengine}
\usepackage{graphics}
\usepackage{subcaption}
\usepackage{float}
\usepackage{color,bm,cite}
\usepackage{url,hyperref,cite}
\hypersetup{colorlinks=true, linktoc=all, linkcolor=blue, citecolor=magenta}
\author{Albertus Johannes Jacobus Maria de Klerk}
\date{\today}
\title{}

\begin{document}

\definecolor{ggOrange}{cmyk}{0, 0.5, 1, 0}
\definecolor{ggRed}{cmyk}{0, 0.8, 1, 0}
\definecolor{Azure}{RGB}{0, 127, 255}

\definecolor{ggBlue}{RGB}{86, 180, 233}
\definecolor{ggGreen}{RGB}{0, 158, 115}

\makeatletter
\newsavebox{\@brx}
\newcommand{\llangle}[1][]{\savebox{\@brx}{\(\m@th{#1\langle}\)}%
  \mathopen{\copy\@brx\mkern2mu\kern-0.9\wd\@brx\usebox{\@brx}}}
\newcommand{\rrangle}[1][]{\savebox{\@brx}{\(\m@th{#1\rangle}\)}%
  \mathclose{\copy\@brx\mkern2mu\kern-0.9\wd\@brx\usebox{\@brx}}}
\makeatother

\begin{center}{
    \Large
    \textbf{Improved Hilbert space exploration algorithms for finite temperature calculations}
  }
\end{center}

\begin{center}
Albertus J. J. M. de Klerk,$^\ast$ and Jean-S\'ebastien
Caux$^\dagger$
\end{center}

\begin{center}
  Institute for Theoretical Physics, University of Amsterdam,\\
  Postbus 94485, 1090 GL Amsterdam, The Netherlands
\end{center}

\begin{center}
  $^\ast$a.j.j.m.deklerk@uva.nl\qquad
  $^\dagger$j.s.caux@uva.nl \\
  \vspace{3mm}\today
\end{center}

\section*{Abstract} {
\bf Computing correlation functions in strongly-interacting quantum systems is one
of the most important challenges of modern condensed matter theory, due to their
importance in the description of many physical observables. Simultaneously, this
challenge is one of the most difficult to address, due to the inapplicability of
traditional perturbative methods or the few-body limitations of numerical
approaches. For special cases, where the model is integrable, methods based on
the Bethe Ansatz have succeeded in computing the spectrum and given us
analytical expressions for the matrix elements of physically important
operators. However, leveraging these results to compute correlation functions
generally requires the numerical evaluation of summations over
eigenstates. To perform these summations efficiently, Hilbert space exploration
algorithms have been developed which has resulted most notably in the \textsc{ABACUS}
library \cite{caux_correlation_2009}. While this performs quite well for
correlations on ground states or low-entropy states, the case of high entropy
states (most importantly at finite temperatures or after a quantum quench) is
more difficult, and leaves room for improvement. In this work, we develop
a new Hilbert space exploration algorithm for the Lieb-Liniger model, specially
tailored to optimize the computational order on finite-entropy states for
correlations of density-related operators.
}

\vspace{5pt}
\noindent\rule{\textwidth}{1pt}
\tableofcontents\thispagestyle{fancy}
\noindent\rule{\textwidth}{1pt}

\section{Introduction}

The study of integrable models in one dimension has been a subject of interest
ever since Bethe's solution of the Heisenberg model \cite{bethe_zur_1931}, but
has gained broader attention in recent decades due to advances in both theory and
experiment.
On the theoretical side, Bethe Ansatz techniques have gone beyond the original
spectrum-limited results and given us efficient expressions for matrix element
of physically important operators
\cite{slavnov_calculation_1989,slavnov_nonequal-time_1990-2,korepin_quantum_1997,pozsgay_local_2011,piroli_exact_2015}.
On the experimental side, advances in the field of ultracold atoms
\cite{polkovnikov_colloquium_2011,langen_ultracold_2015} allowed for the
experimental realization of the Lieb-Liniger model and
demonstrated the implications of integrability on non-equilibrium dynamics and equilibration \cite{kinoshita_quantum_2006}.
Most importantly, integrable models have supplied a fertile ground to
investigate the general principles governing out-of-equilibrium physics and
helped improve our understanding of strongly correlated physics in one dimension
\cite{essler_quench_2016,calabrese_quantum_2016,cazalilla_quantum_2016,bernard_conformal_2016,caux_quench_2016-1,vidmar_generalized_2016,ilievski_quasilocal_2016,langen_prethermalization_2016,vasseur_nonequilibrium_2016,de_luca_equilibration_2016}.
The Lieb-Liniger model \cite{girardeau_relationship_1960-1,lieb_exact_1963,lieb_exact_1963-1,olshanii_atomic_1998},
perhaps the simplest non-trivial example of an integrable model in one
dimension, describes a gas of one-dimensional bosons interacting via a delta
function interaction potential and has received a lot of interest
\cite{caux_dynamical_2006,calabrese_dynamics_2007,kormos_interaction_2013,panfil_metastable_2013,panfil_finite-temperature_2014,kormos_analytic_2014,de_nardis_solution_2014,collura_stationary_2014,de_nardis_analytical_2014,nardis_relaxation_2015,de_nardis_probing_2017,atas_exact_2017,doyon_large-scale_2017-1,palmai_quasilocal_2018,doyon_soliton_2018,doyon_exact_2018,doyon_geometric_2018,de_nardis_edge_2018,bastianello_exact_2018,caux_hydrodynamics_2019-3,schemmer_generalized_2019-3,perfetto_quench_2019-2,koch_adiabatic_2021,mailoud_spectrum_2021,panfil_relevant_2021,de_rosi_hole-induced_2022,li_exact_2023,cheng_one-body_2023,cheng_exact_2022}.
It is this model that we consider throughout this paper, although we comment
on how the ideas in the paper can be extended to spin chains in Sec.
\ref{sec:conclusions}.

The Algebraic Bethe Ansatz \cite{korepin_quantum_1997} gives us efficient
expressions for important matrix elements, but even this combined with a
knowledge of the spectrum is not enough to analytically calculate important
correlation functions.
Though there exist some partial analytical results (e.g. few-spinon
contributions to zero-field correlations in spin-1/2 chains), in general one has
to resort to explicit numerical summations over intermediate states in order to
obtain quantitative results.
Studies into these algorithms led to a set of highly efficient scanning
algorithms encoded in \textsc{abacus} \cite{caux_correlation_2009}. 
However, no universally optimal approach for such algorithms is known, making it
interesting to investigate alternative approaches and see if we can improve upon
the state of the art.

In this work, we develop novel scanning algorithms, inspired by those
implemented in \textsc{abacus}, which we compare to the current version of
\textsc{abacus} by considering the dynamical structure factor.
We show that at zero temperature the performance of our algorithms is virtually
identical to those in the current version of \textsc{abacus}, whereas we note
that our algorithm performs more optimally for the finite temperature
calculation.
We also compare \textsc{abacus} to our algorithms for generating an optimal
basis for computing local observables following a quench in the interaction
strength \cite{robinson_computing_2021-1}.
In this case we notice a more dramatic increase in performance at finite
temperature, highlighting the advantages of our approach for problems where
states with multiple particle-hole excitations relative to the thermal state
play an important role.

We start by reviewing the Lieb-Liniger model in Sec. \ref{sec:the lieb-liniger model},
where we review the calculation of the DSF as well as the quench of interaction strength.
This shows the need for Hilbert space scanning algorithms whose general
principles we discuss in Sec. \ref{sec:scanning algorithms}.
We then start by writing down the simplest algorithm that satisfies the general
rules we require of all scanning algorithms in Sec. \ref{sec:a basic scanning algorithm}.
This initial algorithm does not target a specific momentum sector, but with a
simple addition to the rules of the algorithm it can be made to do so, as we show
in \ref{sec:imposing momentum conservation}.
In Sec. \ref{sec:imposing additional constraints} we show how this simple
algorithm can be improved upon by adding additional rules relating to the
preservation of the number of particle-hole pairs.
In Sec. \ref{sec:building the tree} we show how the algorithm can be improved
even more not by adding additional rules, but by cleverly prioritizing certain
parts of the calculation.
The resulting algorithm is then compared to the existing state of the art in
Sec. \ref{sec:comparison to the state of the art}, where we show
the advantages of our algorithm for finite temperature computations.
We conclude and discuss how our insights are relevant to scanning algorithms
for the spin chain in Sec. \ref{sec:conclusions}.

\section{The Lieb-Liniger model}
\label{sec:the lieb-liniger model}

The Lieb-Liniger model \cite{lieb_exact_1963-1,lieb_exact_1963}
describes
one-dimensional bosons with a delta function interaction potential, so its
Hamiltonian can, in first quantised form, be written as
\begin{equation}
  H = \sum_{i = 1}^N \left[ - \frac{\hbar^2}{2m} \frac{\partial^2}{\partial x_i^2} + 2c \sum_{i < j} \delta(x_i - x_j) \right].
\end{equation}
in the case where we have $N$ particles and $x_i$ is the position of the
$i^{\textrm{th}}$ particle.
Furthermore, $c$ represents the interaction strength and $m$ the mass.
We take $\hbar = 1 = 2m$ to define our units.
In the following we will often consider the second quantised form of this
Hamiltonian, which with $\hbar$ and $2m$ set to one reads
\begin{equation} 
  H = \int_0^L dx \left[ \partial_x \Psi^\dagger(x) \partial_x \Psi(x) 
    + c \Psi^\dagger(x) \Psi^\dagger(x) \Psi(x) \Psi(x) \right].
\end{equation} 
Here $\Psi^\dagger$ is the bosonic creation operator which satisfies the canonical
commutation relations, 
\begin{equation} 
  \left[ \Psi(x), \Psi^\dagger(y) \right] = \delta(x - y).
\end{equation} 
One of the interesting features of the model is that we can
solve its Schrödinger equation to obtain the complete spectrum and all the corresponding eigenstates,
making it a perfect candidate for studying features that are normally not
accessible in interacting quantum many-body systems, such as the time evolution
following a quantum quench.
Furthermore, it turns out that the resulting time-evolution is different from
those of generic quantum systems as found experimentally in cold atom
experiments \cite{kinoshita_quantum_2006}.

In the limit where the interaction strengh vanishes we are dealing with free
bosons so we can readily diagonalise the Hamiltonian using a Fourier transform.
In the opposite (Tonks-Girardeau) limit where the bosons are infinitely repulsive, we can map the
problem to that of non-interacting spinless fermions in one dimension
\cite{girardeau_relationship_1960}.
In the intermediate regime the eigenstates can be found by considering what is
known as the Bethe Ansatz, first used to solve the one-dimensional Heisenberg
spin chain \cite{bethe_zur_1931}.

The Bethe Ansatz approach to solving the Lieb-Liniger model begins by considering
the fundamental domain $D_N$, which for $N$ particles is defined as
\begin{equation}
  \label{eq:fundamental domain}
  D_N = \{ x \in \mathbb{R}^N ~ | ~ x_1 < x_2 < \dots < x_N \}.
\end{equation}
Note that the restriction of the problem to this domain does not constitute a
loss of generality as we can extend this solution to $\mathbb{R}^N$ by invoking
the symmetry requirements of the wavefunction when exchanging particles.

The Bethe Ansatz is then that the wavefunction is a superposition of all
permutations of plane waves with quasi-momenta $\lambda_j$ (often called rapidities) and
amplitudes $A_\sigma$ that are at this stage undetermined giving
\begin{equation}
  \label{eq:Bethe ansatz}
  \Psi_N(x) = \sum_{\sigma \in S_N} A_\sigma e^{\sum_{j=1}^N i \lambda_{\sigma(j)} x_j}.
\end{equation}
where $S_N$ is the group of permutations of the numbers $\{1, \dots, N \}$.
The coefficients $A_\sigma$ can then be determined by the boundary conditions
for the fundamental domain arising from the delta function interaction potential
for two-particle collisions.
This gives
\begin{equation}
  \label{eq:coefficients Bethe ansatz}
  A_\sigma = (-1)^{[\sigma]} \prod_{1 \leq l < j \leq N} (\lambda_{\sigma(j)} - \lambda_{\sigma(l)} + ic),
\end{equation}
where $(-1)^{[\sigma]}$ is the sign of the permutation.
Imposing periodic boundary conditions on the Bethe wave function gives rise to
what are called the Bethe equations, which determine the values of the
rapidities $\lambda_j$, which read
\begin{equation} 
  \label{eq:Bethe equations}
  e^{i \lambda_j L} = \prod_{l \neq j} \frac{\lambda_j - \lambda_l + i c}{\lambda_j - \lambda_l - ic}.
\end{equation}
Any set of rapidities $\{\lambda_i\}_{1 \leq i \leq N}$ satisfying Eq. \eqref{eq:Bethe equations}
thus gives rise to an eigenstate of the Lieb-Liniger model with energy $\sum_j \lambda_j^2$
and momentum $\sum_j \lambda_j$.

Instead of considering the Bethe equations directly, it is useful to take the logarithm
of Eq. \eqref{eq:Bethe equations} resulting in the logarithmic Bethe equations 
\begin{equation} 
  \label{eq:logarithmic Bethe equations}
  \lambda_j + \frac{2}{L} \sum_{k = 1}^N \textrm{atan} \left(\frac{\lambda_j -\lambda_k}{c}\right) - \frac{2\pi I_j}{L} = 0.
\end{equation}
where we introduced the quantum numbers $\{I_j \}_{1 \leq j \leq N}$ which are
integers when $N$ is odd and half-odd integers when $N$ is even.
The quantum numbers are more than a mathematical necessity introduced by the
logarithm, they turn out to be a convenient way of uniquely labelling the
eigenstates.
There is a one-to-one correspondence between the quantum numbers and the
rapidities which respects the ordering, meaning that if $I_j > I_k$ then
$\lambda_j > \lambda_k$
due to the monotonic nature of the second term on the left hand side of
Eq.\eqref{eq:logarithmic Bethe equations}.
Furthermore, the sum of the quantum numbers is proportional to the momentum of
the eigenstate via
\begin{equation}
  P = \sum_j \lambda_j = \frac{2 \pi}{L} \sum_j I_j.
  \label{eq:momentum from quantum numbers}
\end{equation}
Finally, no two quantum numbers can be equal since the wavefunction then
formally vanishes.

The solvability of the model is a result of the fact that all interactions
can be reduced to two-body interactions, a hallmark of integrability.
Another special property of the Lieb-Liniger model, also sometimes used to
characterize integrability, is that it has infinitely many non-trivial commuting
conserved charges whose eigenvalues are given by
\begin{equation}
  Q_n = \sum_{j=1}^N \lambda_j^n
\end{equation}
for any $n \in \mathbb{N}$.
Here $Q_1$ and $Q_2$ represent the momentum and energy respectively.

The ground state of the Lieb-Liniger model is the state whose quantum numbers
are as close to zero as possible.
Since the quantum numbers are not allowed to coincide and they can only take on
integer or half-odd integer values, this results in a configuration like a Fermi
sea. To be precise, the quantum numbers of the ground state are given by
\begin{equation}
  \left\{ - \frac{N + 1}{2}, - \frac{N + 1}{2} + 1, \dots , \frac{N - 1}{2} \right\}.
\end{equation}
We can therefore define a “Fermi momentum” in analogy with the Fermi sea, given
by $k_F = \frac{\pi}{L} (N - \tfrac{1}{2})$
so that $k_F$ is between the last occupied and first unoccupied mode.

Taking the thermodynamic limit of the logarithmic Bethe equations leads to the
thermodynamic Bethe Ansatz.
In this limit the states are described by the continuum version of the quantum
numbers called a root distribution.
Root distributions corresponding to finite temperature states can be determined
and discretized in order to obtain sets of quantum numbers at finite size that
best resemble these distributions.
These finite size approximations of the thermodynamic root distributions are
called representative states \cite{korepin_quantum_1997}.

Another important property of the Lieb-Liniger model is that efficient expressions
for matrix elements of operators such as the density operator
\begin{equation}
  \rho(x) = \Psi^\dagger(x) \Psi(x) 
\end{equation}
as well as the \(g_2\) operator
\begin{equation}
  g_2(x) = \Psi^\dagger(x) \Psi^\dagger(x) \Psi(x) \Psi(x)
\end{equation}
have been obtained using Algebraic Bethe Ansatz techniques
\cite{slavnov_nonequal-time_1990,pozsgay_local_2011,piroli_exact_2015}.
Knowledge of matrix elements of operators in combination with the solvability of
the model is what allows us to study the correlation functions as well as the
non-equilibrium time-evolution of observables in the Lieb-Liniger model.
The time evolution of local observables in integrable models is of particular
interest as, at long time after driving the system out of equilibrium,
expectation values do not thermalize and instead relax to values determined from
the Quench Action method
\cite{brockmann_quench_2014,brockmann_ne-xxz_2014,brockmann_overlaps_2014},
or (when applicable) a GGE
\cite{rigol_relaxation_2007,rigol_thermalization_2008,caux_constructing_2012}.
This has not only been understood theoretically, but has been directly observed
in experimental studies \cite{kinoshita_quantum_2006}.

A property of the matrix elements crucial to the algorithms in this article, 
is that on average, the off-diagonal matrix elements are largest when the bra
and ket states share the most quantum numbers.\footnote{Here it is good to note that how much the quantum they do not share are different is of secondary importance. For example, even when one of the quantum numbers $I_j$ and the corresponding rapidity $\lambda_j$ go off to $\pm \infty$, the term in the logarithmic Bethe equations, corresponding to the rapidity at infinity evaluates to $\pm \pi / L$ which effectively shifts $I_k$ by $\mp 1/2$. Thus even a massive change in one quantum number shifts the other quantum numbers only a little. }
To see why, note that Slavnov's formula for the overlap
$\langle \mu | \lambda \rangle$ between a Bethe state $|\lambda\rangle$ and an
arbitrary set of rapidities $| \mu \rangle$ has poles for coinciding rapidities
\cite{korepin_quantum_1997}.
As such, overlaps between Bethe states are maximal when the number of (close to)
coinciding rapidities is maximal.
The matrix elements of the density and $g_2$ operators are derived from this
formula for the overlaps by determining the action of the operator under
consideration on the bra or ket and using the overlap formula.
For example, acting with $\Psi(0)$ on a Bethe state results in a superposition
of states where one of the rapidities is removed and the others are shifted due
to the interactions.
Therefore, the off-diagonal matrix elements of the density operator are maximal
when the bra and ket differ by one quantum number.
As the number of differing quantum numbers increases, the matrix element becomes
smaller due to the smaller number of (close to) coinciding rapidities.
Similarly, the off-diagonal matrix elements of the $g_2$ operator are maximal
when the bra and ket differ by one or two quantum numbers.
The importance of the number of differences between the quantum numbers of the
bra and ket is also influenced by the interaction strength, with its importance
diminishing as the interaction strength decreases.

To see how the need for Hilbert space scanning algorithms arises, consider the
computation of the dynamical structure factor in the ground state of the
Lieb-Liniger model, given by
\begin{align}
  S(k, \omega)
    &= \int_0^L \textrm{d}x \int \textrm{d}t e^{-ikx + i \omega t} \langle \rho(x, t) \rho(0, 0) \rangle \\
    &= \frac{2\pi}{L} \sum_\alpha |\langle 0 | \rho_k | \alpha \rangle|^2 \delta(\omega - E_\alpha - E_0),
    \label{eq:DSF}
\end{align}
where $\rho_k$ is the Fourier transform of the density operator.
In order to numerically approximate Eq. \eqref{eq:DSF}, we need a set of
eigenstates $|\alpha \rangle$ for which we compute the energies $E_\alpha$,
the matrix elements $\langle 0 | \rho_k | \alpha \rangle$, and perform the summation.
As the Lieb-Liniger model possesses an infinite number of eigenstates, an
evaluation of the sum in Eq. (15) neccessitates a truncation and the accuracy of
the calculation depends on the number of eigenstates and their matrix elements
$\langle 0 | \rho_k | \alpha \rangle$.
The convergence is quantified by the $f$ sum rule
\cite{lifshitz_statistical_1980}, which states that
\begin{equation}
  \int \frac{\textrm{d} \omega}{2 \pi} \omega S(k, \omega) = \frac{N k^2}{L}.
  \label{eq:f-sum rule}
\end{equation}
Given Eq. \eqref{eq:f-sum rule}, we can convert the contributions to the
summation in Eq. \eqref{eq:DSF} into a weighing function for a given eigenstate
$|\alpha\rangle$ given by
\begin{equation}
  w_f(\alpha) = \frac{L}{N k^2} (E_\alpha - E_0) \left| \langle 0 | \rho_k | \alpha \rangle \right|^2
  \label{eq:f-sumrule weight}
\end{equation}
for $k \neq 0$ such that the summation over the weights of all eigenstates $|\alpha\rangle$ at
a fixed momentum value gives 1.
When considering the summation over the weights of Eq. \eqref{eq:f-sumrule weight},
there are two contributing factors that determine the importance of an
eigenstate, its energy and the matrix element. For the Lieb-Liniger model it
turns out that most matrix elements except for a tiny portion of states are
vanishingly small, dominating the effect the energy has. Therefore our focus
lies primarily on finding those states for which the matrix elements are large
in order to get a good saturation of the sumrule.
Hilbert space scanning algorithms are designed to preferentially generate
eigenstates for which for example $w(|\alpha \rangle) = |\langle 0 | \rho_k | \alpha \rangle|^2$ or
$w(|\alpha \rangle) = (E_\alpha - E_0) |\langle 0 | \rho_k | \alpha \rangle|^2$ are maximal
\cite{caux_dynamical_2006,caux_correlation_2009,panfil_finite-temperature_2014}.

Another problem where the need for generating appropriate eigenstates arises, is
when choosing a basis for truncated spectrum methods
\cite{yurov_truncated_1990,yurov_truncated-fermionic-space_1991,konik_numerical_2007}.
Consider a quench of the Lieb-Liniger model where we change the interaction
strength at $t = 0$ from $c_i$ to $c_f$.
Truncated spectrum methods can be used to compute the time evolution of the
initial state $|\Psi_0 \rangle$ in terms of a set of eigenstates of the
Lieb-Liniger model at interaction strength $c_f$, i.e.
\begin{equation}
  |\Psi_0 (t) \rangle = \sum_\alpha b_\alpha e^{-iE_\alpha t} |\alpha \rangle 
  \label{eq:TSA}
\end{equation}
where $b_\alpha = \langle \alpha | \Psi_0 \rangle$ are the coefficients being
approximated by the truncated spectrum methods.
However, the accuracy of the expansion in Eq. \eqref{eq:TSA} depends on the
choice of basis states $|\alpha \rangle$.
By choosing a weighing function that approximates $\langle \alpha | \Psi_0 \rangle$
we can leverage our scanning algorithms to generate a close to ideal basis for this
quench \cite{robinson_computing_2021}.

\section{Scanning Algorithms}
\label{sec:scanning algorithms}

The algorithms for exploring Hilbert space that we present in this article can
all be generally described as generating a single-rooted tree where each node
represents an eigenstate.
The algorithms differ in the rules that determine which new eigenstates are
generated from a given node, resulting in trees with different topologies even
when they are generated from the same initial eigenstate, which we will herein
call the seed state.
In some cases, the seed state for an algorithm will be directly related to the
observable we are trying to compute (for example, for the calculation of the
dynamical structure factor, Eq. \eqref{eq:DSF}, the seed state will be the ground
state) while for other problems it may not be (this is the case for the quench
problem, which we discuss later).
This approach using tree-building algorithms, was shown to be very successful
for the computation of correlation functions in integrable models
\cite{caux_correlation_2009}.
The purpose of this article is to introduce new algorithms for scanning and
comparing their properties for different problems.

There are two properties we require of all of our Hilbert space exploration
algorithms:
\begin{enumerate}
\item \emph{Uniqueness:} No eigenstate should come up more than once when
generating a tree of eigenstates.
\item \emph{Completeness}: All eigenstates in a pre-defined sector of Hilbert
space must occur in the tree if we give the algorithm infinite computation time.
\end{enumerate}
Property 1 ensures that we can use the eigenstates generated in a tree to
perform a summation such as the ones in Eq. \eqref{eq:DSF} and Eq.
\eqref{eq:TSA} without having to separately keep track of which eigenstates we
already included.
Property 2 comes from the fact that we want to be able to approximate summations
such as the one in Eq. \eqref{eq:DSF} arbitrarily well given infinite
computational resources.
If some states were not generated by the algorithm, this would not be
possible.

In the absence of a UV cut-off, any momentum sector of the Lieb-Liniger model is
infinite dimensional even in a finite volume system.
Therefore we can never truly generate the full corresponding tree with finite
computational resources.
As such, it is not only relevant what the final tree would look like, but also
how it is built, i.e. what it looks like after some finite time.
In order to ensure that we spend our time wisely, we can pause the generation of
new eigenstates from certain branches of the tree.
The goal here is to pause the generation of eigenstates in the algorithm until
they are the most important ungenerated eigenstates that are remaining.
The result is that after some finite time we end up with a truncated tree where
most of the nodes on the outside of the tree could be used to generate additional new eigenstates.

\begin{figure}

\centering

\begin{subfigure}[b]{0.3\textwidth}
  \centering
  \includegraphics{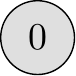}
  \caption{Zeroth step}
  \label{fig:Erroneous truncation subfig 0}
\end{subfigure}
\hfill
\begin{subfigure}[b]{0.3\textwidth}
  \centering
  \includegraphics{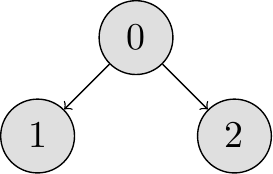}
  \caption{First step}
  \label{fig:Erroneous truncation subfig 1}
\end{subfigure}
\hfill
\begin{subfigure}[b]{0.3\textwidth}
  \centering
  \includegraphics{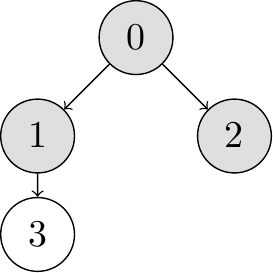}
  \caption{Second step}
  \label{fig:Erroneous truncation subfig 2}
\end{subfigure}
\\ \vspace{1cm}

\centering

\begin{subfigure}[b]{0.3\textwidth}
  \centering
  \includegraphics{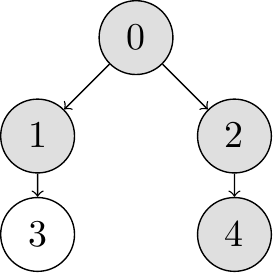}
  \caption{Third step}
  \label{fig:Erroneous truncation subfig 4}
\end{subfigure}
\hfill
\begin{subfigure}[b]{0.3\textwidth}
  \centering
  \includegraphics{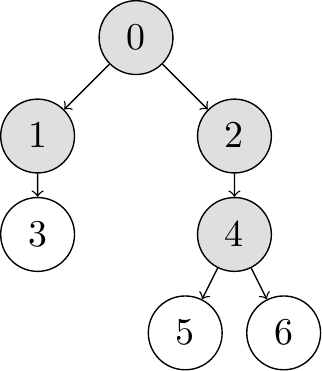}
  \caption{Fourth step}
  \label{fig:Erroneous truncation subfig 5}
\end{subfigure}
\hfill
\begin{subfigure}[b]{0.3\textwidth}
  \centering
  \includegraphics{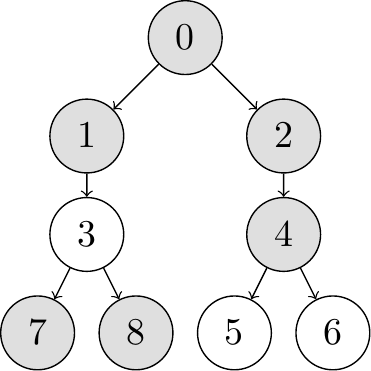}
  \caption{Fifth step}
  \label{fig:Erroneous truncation subfig 6}
\end{subfigure}

\caption{
Illustration of the state of the tree after the first six steps of a non-optimal
algorithm. The white circles represent eigenstates whose weight is smaller
than some threshold value, whereas the grey circles represent eigenstates whose
weight is larger than the same threshhold. An optimal algorithm would therefore
not generate grey circles after white circles, which we see occuring in the
fifth step of the algorithm.
}
\label{fig:Erroneous truncation}
\end{figure}

The way in which we determine which branches to pause at a given time and the
interplay between this and the rules for growing the tree as determined by our
algorithm can have a strong influence on the way the tree is grown and therefore
the quality of our algorithm.
To see this, consider an algorithm that satisfies the completeness and
overcounting criteria.
Now suppose we run this algorithm for some time which gives rise to a growing
tree as depicted in Fig. \ref{fig:Erroneous truncation}.
The white nodes represent eigenstates whose weight is below some threshold value
whereas the weight of the grey nodes is bigger, allowing us to visually
distinguish between high-weight and low-weight nodes.
The algorithm showcased here is not ideal since the high-weight nodes 7 and 8
are generated after the low-weight nodes 5 and 6.
When scaled up this means we can, at finite runtime, miss important states
because they are effectively locked behind low-weight states.
In an ideal algorithm the descendents of a node would therefore always have a
lower weight than the parent node.

The extent to which the descendents of nodes have a weight lower than their
parents depends on an interplay between the rules of the algorithm and the
weighing function considered. 
Therefore there may be some algorithms which are more compatible with certain
weighing functions and others that are more compatible with others.
For the Lieb-Liniger model, which we consider in this article, we can identify a
criterion that is common to the weighing functions we want to consider.
This criterion is the number of particle-hole excitations as it is closely related
to the matrix element value of the operators we want to consider as explained in Sec.
\ref{sec:the lieb-liniger model}.
This allows us to create close to optimal algorithms for this system.

\section{A basic scanning algorithm}
\label{sec:a basic scanning algorithm}

In order to introduce a way of visualising the quantum numbers, consider a set
of quantum numbers $ \{I_i\}_{i \leq 5} $ which we assume to be ordered, an assumption we retain for the remainder of this article. 
Since no two quantum numbers can be the same, we can think of them as five
particles on a one-dimensional lattice, which can be visualized as follows:
\begin{equation*}
  \cdots \stackunder{$\circ$}{-4 } \stackunder{$\bullet$}{-3 } \stackunder{$\circ$}{-2 } \stackunder{$\bullet$}{-1 } \stackunder{$\bullet$}{ 0 } \stackunder{$\bullet$}{ 1 } \stackunder{$\circ$}{ 2 } \stackunder{$\bullet$}{ 3 } \stackunder{$\circ$}{ 4 } \cdots
\end{equation*}
The numbers below the circles represent the number corresponding to the position
on the line, so this state represents $\{-3, -1, 0, 1, 3\}$.
In this article we visualize the quantum numbers in order to illustrate the
principles of our algorithms. For this purpose we are not interested in the
absolute values of the quantum numbers, but rather only the differences between
the quantum numbers allowing us to drop the numbers below the circles going
forward.

One of the simplest ways of generating a new set of quantum numbers from a given
set is by changing one of the quantum numbers by $\pm 1$, the minimal possible
value provided that this does not render two quantum numbers equal. 
Such a change can be identified with a particle hopping to the left or right on the lattice,
where a particle hopping to the right looks like
\begin{equation*}
  \begin{split}
  \stackunder{$\cdots \; {\circ} \; {\bullet} \; {\circ} \; {\bullet} \; {\bullet} \; {\bullet} \; {\circ} \; {\bullet} \; {\circ} \; \cdots$}{$\downarrow$} \\
  \stackunder{$\cdots \; {\circ} \; {\circ} \; {\textcolor{ggBlue}{\bullet}} \; {\bullet} \; {\bullet} \; {\bullet} \; {\circ} \; {\bullet} \; {\circ} \; \cdots$}{}
  \end{split}
\end{equation*}
and a particle hopping to the left looks like
\begin{equation*}
\begin{split}
 \stackunder{$\cdots \; {\circ} \; {\bullet} \; {\circ} \; {\bullet} \; {\bullet} \; {\bullet} \; {\circ} \; {\bullet} \; {\circ} \; \cdots$}{$\downarrow$} \\
 \stackunder{$\cdots \; {\textcolor{ggBlue}{\bullet}} \; {\circ} \; {\circ} \; {\bullet} \; {\bullet} \; {\bullet} \; {\circ} \; {\bullet} \; {\circ} \; \cdots$}{}
\end{split}
\end{equation*}
In both cases the first line represents the quantum numbers of the initial state
and the line following that represents the quantum numbers of its descendent where
one coloured particle has hopped to a neighbouring lattice site.

Such particle hops can be used to formulate rules for generating descendents and
building a tree, but we need to impose rules to avoid overcounting.
We start with an algorithm where the descendents of a node are those where a
single particle has hopped one position to the right or left provided that this
does not result in a collision of particles.
In the following we identify the issues with this algorithm and propose
solutions, the result of which will be our first real scanning algorithm which
does not overcount states.

The first issue we consider is that when a particle first moves to the right and
then back it will result in the same state we started with.
Without any restrictions we can generate subtrees that look like
\begin{equation*}
\begin{split}
 \stackunder{$\cdots \; {\circ} \; {\bullet} \; {\circ} \; {\bullet} \; {\bullet} \; {\bullet} \; {\circ} \; {\bullet} \; {\circ} \; \cdots$}{$\downarrow$} \\
 \stackunder{$\cdots \; {\circ} \; {\bullet} \; {\circ} \; {\bullet} \; {\bullet} \; {\circ} \; {\bullet} \; {\bullet} \; {\circ} \; \cdots$}{$\downarrow$} \\
 \stackunder{$\cdots \; {\circ} \; {\bullet} \; {\circ} \; {\bullet} \; {\bullet} \; {\bullet} \; {\circ} \; {\bullet} \; {\circ} \; \cdots$}{} \\
\end{split}
\end{equation*}
The first rule of our scanning algorithms is therefore that once a particle has
moved to either the left or the right, it can only continue moving in that
direction.
We refer to particles that have moved to the right as rightmovers whereas we
refer to particles that have moved to the left as leftmovers.
In our visualizations we colour the leftmovers blue and the rightmovers orange.

The second issue is due to there being no preferred ordering of moving particles
to the left or right, as with the current rules the following subtree could be
generated
\begin{eqnarray}
& \stackunder{$\cdots$}{$\swarrow$} \; {\circ} \; {\circ} \; {\bullet} \; {\bullet} \; {\bullet} \; {\circ} \; {\circ} \; \stackunder{$\cdots$}{$\searrow$} & \nonumber \\ 
{\circ} \; {\textcolor{ggBlue}{\bullet}} \; {\circ} \; \stackunder{${\bullet}$}{$\downarrow$} \; {\bullet} \; {\circ} \; {\circ} & &
{\circ} \; {\circ} \; {\bullet} \; \stackunder{${\bullet}$}{$\downarrow$} \; {\circ} \; {\textcolor{ggOrange}{\bullet}} \; {\circ} \nonumber \\
{\circ} \; {\textcolor{ggBlue}{\bullet}} \; {\circ} \; {\bullet} \; {\circ} \; {\textcolor{ggOrange}{\bullet}} \; {\circ} & &
{\circ} \; {\textcolor{ggBlue}{\bullet}} \; {\circ} \; {\bullet} \; {\circ} \; {\textcolor{ggOrange}{\bullet}} \; {\circ} \nonumber
\end{eqnarray}
Here we see that the bottom two states are the same even though they are not the
same node in the tree and no particle has changed direction.
In order to avoid overcounting we impose the rule that if a state has a
rightmover then its descendents can not have additional leftmovers. In the
subtree we just considered this means that the state on the bottom right would
not have been generated as there is already a rightmover (the right most
particle in the above figure).

The final issue that arises is similar to the previous one, only now it involves
only moves to the left or right.
To understand the issue, note that with our current rules the following subtree
could be generated
\begin{eqnarray}
& \stackunder{$\cdots$}{$\swarrow$} \; {\circ} \; {\bullet} \; {\bullet} \; {\circ} \; {\bullet} \; {\bullet} \; {\circ} \; \stackunder{$\cdots$}{$\searrow$} & \nonumber \\ 
{\textcolor{ggBlue}{\bullet}} \; {\circ} \; {\bullet} \; \stackunder{${\circ}$}{$\downarrow$} \; {\bullet} \; {\bullet} \; {\circ} & &
{\circ} \; {\bullet} \; {\bullet} \; \stackunder{${\textcolor{ggBlue}{\bullet}}$}{$\downarrow$} \; {\circ} \; {\bullet} \; {\circ} \nonumber \\
{\textcolor{ggBlue}{\bullet}} \; {\circ} \; {\bullet} \; {\textcolor{ggBlue}{\bullet}} \; {\circ} \; {\bullet} \; {\circ} & &
{\textcolor{ggBlue}{\bullet}} \; {\circ} \; {\bullet} \; {\textcolor{ggBlue}{\bullet}} \; {\circ} \; {\bullet} \; {\circ} \nonumber 
\end{eqnarray}
Again, we see that the two lowest configurations of integers are identical, despite being on different branches of the tree. 
This issue can be avoided by only allowing particles to hop to the left if they
are the rightmost leftmover or to its right.
This would rule out the configuration on the bottom right, as it is generated by
creating a leftmover to the left of the existing leftmover.
We can run into the same problem when considering rightmovers giving rise to
\begin{eqnarray}
& \stackunder{$\cdots$}{$\swarrow$} \; {\circ} \; {\bullet} \; {\bullet} \; {\circ} \; {\bullet} \; {\bullet} \; {\circ} \; \stackunder{$\cdots$}{$\searrow$} & \nonumber \\ 
{\circ} \; {\bullet} \; {\circ} \; \stackunder{${\textcolor{ggOrange}{\bullet}}$}{$\downarrow$} \; {\bullet} \; {\bullet} \; {\circ} & &
{\circ} \; {\bullet} \; {\bullet} \; \stackunder{${\circ}$}{$\downarrow$} \; {\bullet} \; {\circ} \; {\textcolor{ggOrange}{\bullet}} \nonumber \\
{\circ} \; {\bullet} \; {\circ} \; {\textcolor{ggOrange}{\bullet}} \; {\bullet} \; {\circ} \; {\textcolor{ggOrange}{\bullet}} & &
{\circ} \; {\bullet} \; {\circ} \; {\textcolor{ggOrange}{\bullet}} \; {\bullet} \; {\circ} \; {\textcolor{ggOrange}{\bullet}} \nonumber
\end{eqnarray}
which can be avoided by only allowing particles to hop to the right if
they are the leftmost rightmover or to its left.

The rules we have imposed thus far constitute the first scanning algorithm
that is complete and does not count the same state twice as we will show shortly.
For now, let us summarise the rules of the algorithm.

Let $\{I_i\}_{i \leq N}$ be the set of quantum numbers at a node.
We denote the leftmoving and rightmoving quantum numbers by $\{I^L_j\}_{j \leq
N_L}$ and $\{I^R_j\}_{j \leq N_R}$ respectively.
\begin{itemize}
  \item \textbf{Move quantum numbers to the right:} Generate a descendent for every
        $I_l \in \{I_i\}_{i \leq N}$ for which $I_l \notin \{I_j^L\}_{j \leq
        N}$, $I_l \leq I_0^R$, and $I_l+1 \neq I_{l+1}$ if $l \neq N$ with quantum
        numbers given by $\{I_i + \delta_{l,i} \}_{i \leq N}$.
  \item \textbf{Move quantum numbers to the left:} If $\{I_k^R\} = \emptyset$
        generate a descendent for every $I_l \in \{I_i\}_{i \leq N}$ such that
        $I_l \geq I_{N_L}^L$, and $I_{l-1} \neq I_{l} - 1$ if $l \neq 0$ with
        quantum numbers given by $\{I_i - \delta_{l,i} \}_{i \leq N}$.
\end{itemize}
The maximal number of descendents is therefore $2 N$.
What remains is the proof that this algorithm generates a tree containing every
set of allowed quantum numbers exactly once regardless of the seed state chosen.

Consider an arbitrary seed state $\{I_i^{SS}\}_{i \leq N}$ and an arbitrary
target state $\{ \tilde{I}_i \}_{i \leq N}$.
In order to show completeness and the absence of overcounting it is sufficient
to show that the target state occurs in the generated tree precisely once.
This is equivalent to there being one way of applying the rules of the algorithm
to get from the seed state to the target state.
The quantum numbers of the node we consider at an intermediate step of the
algorithm will be denoted by $\{I_i \}_{i \leq N}$.

Since our algorithm does not allow particles to hop to the left in the presence
of rightmovers, we first have to move all particles for which $\tilde{I}_l <
I_l^{SS}$.
We first consider the leftmost particle of this type with index $l$ and claim
that we can move it all the way to its target position without collisions.
To show this, assume the contrary, which implies that $l \neq 1$ and
$I^{SS}_{l-1} \geq \tilde{I}_l$.
However, since $l$ was the index of the leftmost particle that had to move to
the left, we know that $\tilde{I}_{l-1} \geq I_{l-1}^{SS}$.
Combining these statements gives $\tilde{I}_{l-1} \geq I_{l-1}^{SS} \geq
\tilde{I}_l$ which is a contradiction.
Repeating this argument for all quantum numbers of the seed state that have to
be decreased to reach the target state starting with the smallest one, we can
move all such quantum numbers to the right position in a unique way.

Having fixed the quantum numbers that have to be decreased, we are left with
quantum numbers for which $\tilde{I}_l \geq I_l^{SS}$.
This time we claim that we can start from the rightmost particle for which this
inequality holds and the rules of our algorithm allow it to be put in its place
without collisions.
To prove this, we again assume the contrary which implies that $l \neq N$ and
that $I_{l+1} \leq \tilde{I}_l$.
However, since $I_l$ was the rightmost quantum number that had to be increased,
and since we already fixed the leftmovers we have that $I_{l+1} = \tilde{I}_{l+1}$.
Together this gives $\tilde{I}_{l+1} = I_{l+1} \leq \tilde{I}_l$ giving the
contradiction we require.
This finishes the proof of completeness as well as showing that there is no
overcounting.

Throughout the remainder of this article, we refer to the scanning algorithm
developed in this section as stepwise scanning (SWS).

\section{Imposing momentum conservation}
\label{sec:imposing momentum conservation}

In the previous section we introduced a scanning algorithm that would generate
every state in Hilbert space exactly once, given infinite computational resources.
However, often we are interested in a particular momentum sector, so for such
problems it is not particularly well-suited.
After all, it would mean that we are only interested in a tiny subset of the states
that we generate.
In this section we introduce a variant of the previous algorithm which restricts
itself to a given momentum sector.

To turn stepwise scanning into an algorithm that generates descendents whose
momentum is equal to that of their parent, we combine the rules we have for
moving particles to the left and right.
The momentum of a state is proportional to the sum of the quantum numbers,
Eq. \eqref{eq:momentum from quantum numbers}, so moving one particle one step to
the right and another one step to the left ensures that momentum remains
preserved.
Furthermore, this approach eliminates the second problem we encountered in the
previous section due to which we imposed the rule that no particle can hop left
in the presence of rightmovers.
The resulting algorithm goes as follows.

Let $\{I_i\}_{i \leq N}$ be the set of quantum numbers at a node.
We denote the leftmoving and rightmoving quantum numbers by $\{I^L_j\}_{j \leq
N_L}$ and $\{I^R_j\}_{j \leq N_R}$ respectively.
\begin{itemize}
  \item \textbf{Generate rightmovers:} Generate an intermediate descendent $C_r$
        for every $I_l \in \{I_i \}_{i \leq N}$ such that $I_l \notin
        \{I_j^L\}_{j \leq N_L}$, $I_l \leq I^R_0$, and $I_l + 1 \neq
        I_{l+1}$ if $l \neq N$.
  \item \textbf{Generate leftmovers:} For every intermediate descendent $C_r$
        generate a descendent for every $I_l \in \{I_i^{C_r} \}_{i \leq N}$ such
        that $I_l \notin \{I_j^{C_r,R}\}_{j \leq N_{C_r,R}}$, $I_l \geq I^L_{N_{C_r,L}}$, and
        $I_l - 1 \neq I_{l-1}$ if $ l \neq 0$.
\end{itemize}
The scanning routine described here, which we call momentum preserving stepwise
scanning (SWS-MP), generates at most $N^2$ descendents in contrast to the at most
$2N$ descendents in regular stepwise scanning.
The difference arises due to the fact that we imposed momentum conservation,
which led us to essentially apply first the first step of regular stepwise
scanning, generating at most $N$ intermediate descendents, and then applying the
second step of the stepwise scanning algorithm to these intermediate descendents.

To show that this algorithm is also complete in the sense that it can generate
any state whose momentum is equal to that of the seed state, consider a random
seed state $\{I^{SS}_i\}_{i \leq N}$ and a random target state $\{\tilde{I}_{i
\leq N}\}$ with the same momentum.
There are indices $k \in \{k_1, \dots, k_{N_L}\}$ such that $\tilde{I}_k \leq
I_k^{SS}$ as well as indices $l \in \{l_1, \dots, l_{N_R}\}$ such that
$\tilde{I}_l \geq I_l^{SS}$ which represent the indices of what will become the
leftmovers and rightmovers respectively.
Note that the size of these sets of indices can be different, as momentum
preservation only requires the number of hops to the right and left to be
preserved.
In order to reach the target state from the seed state we again have to start by
moving the leftmost leftmover, i.e. $I_{k_1}$, and the rightmost rightmover, i.e.
$I_{N_R}$.
The proof that these particles can hop to their target positions without
collisions is exactly the same as the proof of completeness for regular stepwise
scanning, so for that we refer the reader to the previous section.
The same argument holds for the rightmovers being able to move to their target
positions.
Since there is still a unique order of moves by which we can reach the target
state given the seed state, we also have no overcounting.

To understand the problems that arise when we try to apply this algorithm to
situations where the momentum of the states we want to scan for is not equal to
the natural candidate for the seed state, consider the example of the dynamical
structure factor, see Eq. \eqref{eq:DSF}.
Here the reference state we are interested in is the ground state $|0\rangle$
and we want to use a scanning algorithm to find the states $|\alpha\rangle$ at
some fixed value of momentum most important to the summation.
However, since the states in the intermediate summation do not belong to the
same momentum sector as the ground state we cannot use a momentum preserving
stepwise scanning algorithm in order to find them.
In principle we can choose some other state from the targeted momentum sector
and use it as a seed state to the algorithm, but this raises the question of
which state to choose.
Even though in the limit where we have infinite computational resources this
does not matter, it does affect what the tree looks like after a finite amount
of time.
After all, a bad choice can lead to the situation where important contributions
are only generated after a long time because they are far down in the tree or
are "hidden" as descendents of unimportant states.
In fact, it is unclear if there even exists a seed state that would not lead to
a Bethe tree with undesirable properties in this case.

In order to avoid having to choose a seed state for the target momentum sector,
we choose to tweak the rules of the scanning algorithm such that we can use
a given reference state (in the case of the dynamical structure factor, the
ground state) despite its momentum not being equal to that of the momentum
sector we are interested in.
After all, we want to generate the states with few particle-hole pairs first as
these are the states we generally expect to have the largest matrix elements for
local operators. Taking the reference state to be the seed state ensures that
these few particle-hole states are generated because the rules of our algorithm
generate descendents with at most two additional particle-hole pairs.
In order to generate states in the targeted momentum sector from a given reference
state we add the rule that if the momentum of a node is smaller than the target
momentum, its descendents are generated by following the first step of the
algorithm, whereas if it is larger the rules of the second step are used.
Note that applying only one of the two steps of momentum preserving stepwise
scanning generates descendents whose momentum is changed by the minimal amount
compared to their parent state.
As such, there can be no overshooting of the target momentum sector.
Furthermore, every branch created from the seed state reaches the target momentum
sector in $\tfrac{k_{target}L}{2 \pi}$ steps or dies off before then.

In order to assess the quality of this algorithm, we consider again the
dynamical structure factor introduced in Eq. \eqref{eq:DSF}.
Starting from the ground state, we can generate a tree of eigenstates where we use
the weighing function that selects for states which contribute most strongly
for the saturation of the $f$ sum rule as defined in Eq. \eqref{eq:f-sum rule}.
For now, we generate descendents node by node starting each time with one of the
highest weight nodes as we elaborate more in section \ref{sec:building the tree}.
The results for this calculation for three different values of the interaction
strength are shown in Fig.
    \ref{subfig:sumrule saturation T=0, c=1, kf rho, B_full}, 
    \subref{subfig:sumrule saturation T=0, c=4, kf rho, B_full},
    \subref{subfig:sumrule saturation T=0, c=16, kf rho, B_full}
We see that in all three cases we reach close to optimal convergence with very
few states meaning the most important states are generated first.

\begin{figure}
  \subcaptionbox{$c=1$, $T=0$ \label{subfig:sumrule saturation T=0, c=1, kf rho, B_full}} {
    \centering
    \includegraphics[width=0.3\textwidth]{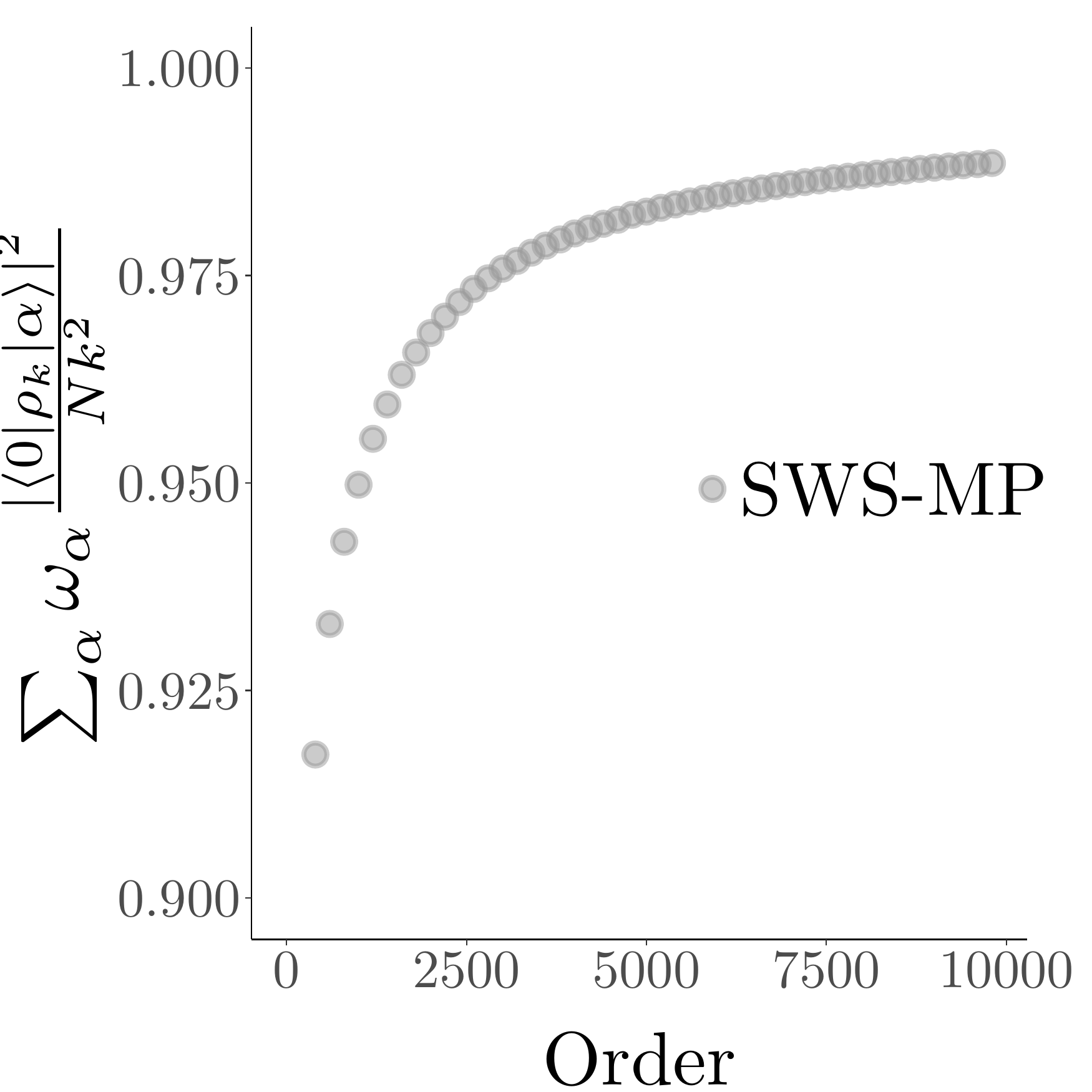}
  }
  \subcaptionbox{$c=4$, $T=0$ \label{subfig:sumrule saturation T=0, c=4, kf rho, B_full}} {
    \centering
    \includegraphics[width=0.3\textwidth]{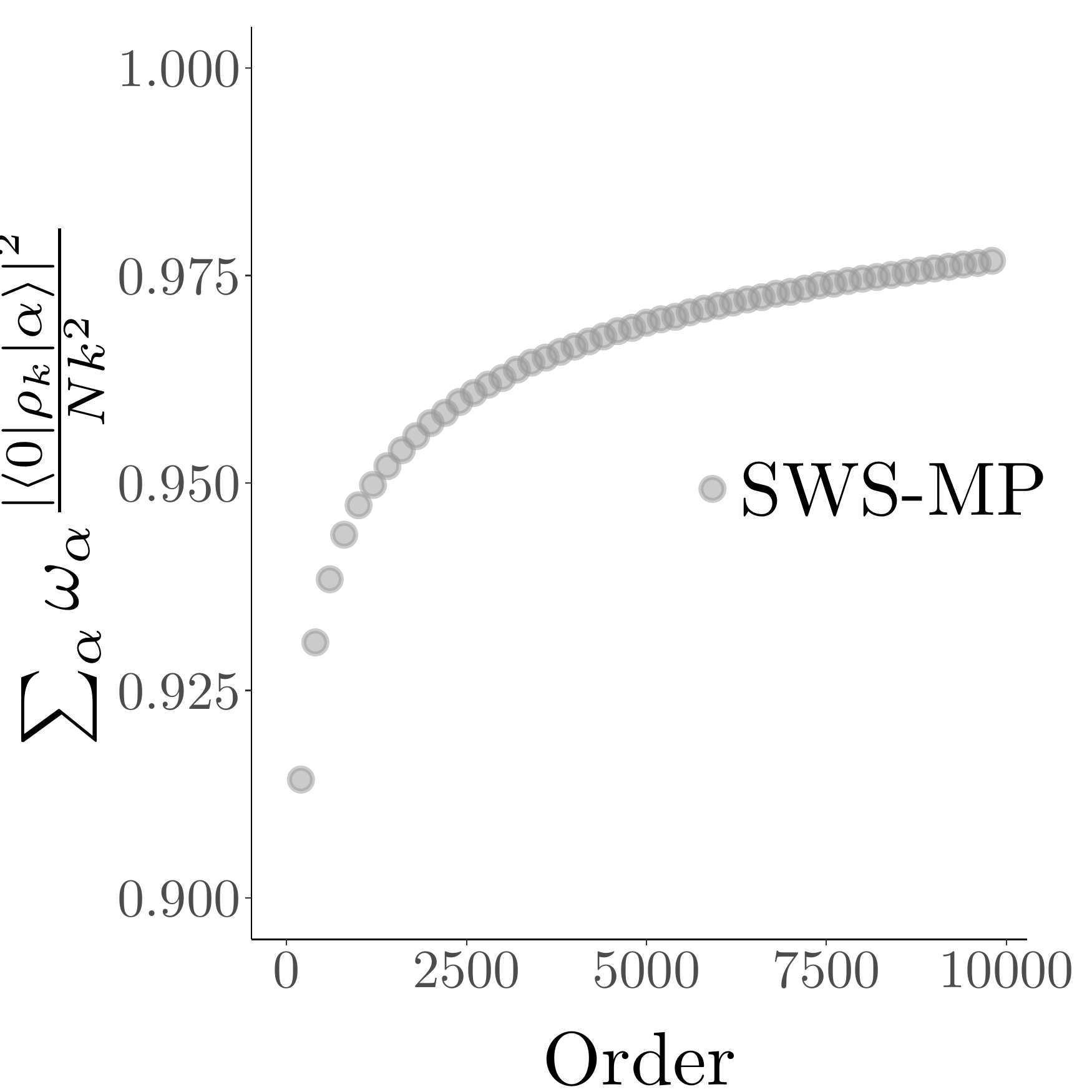}
  }
  \subcaptionbox{$c=16$, $T=0$ \label{subfig:sumrule saturation T=0, c=16, kf rho, B_full}} {
    \centering
    \includegraphics[width=0.3\textwidth]{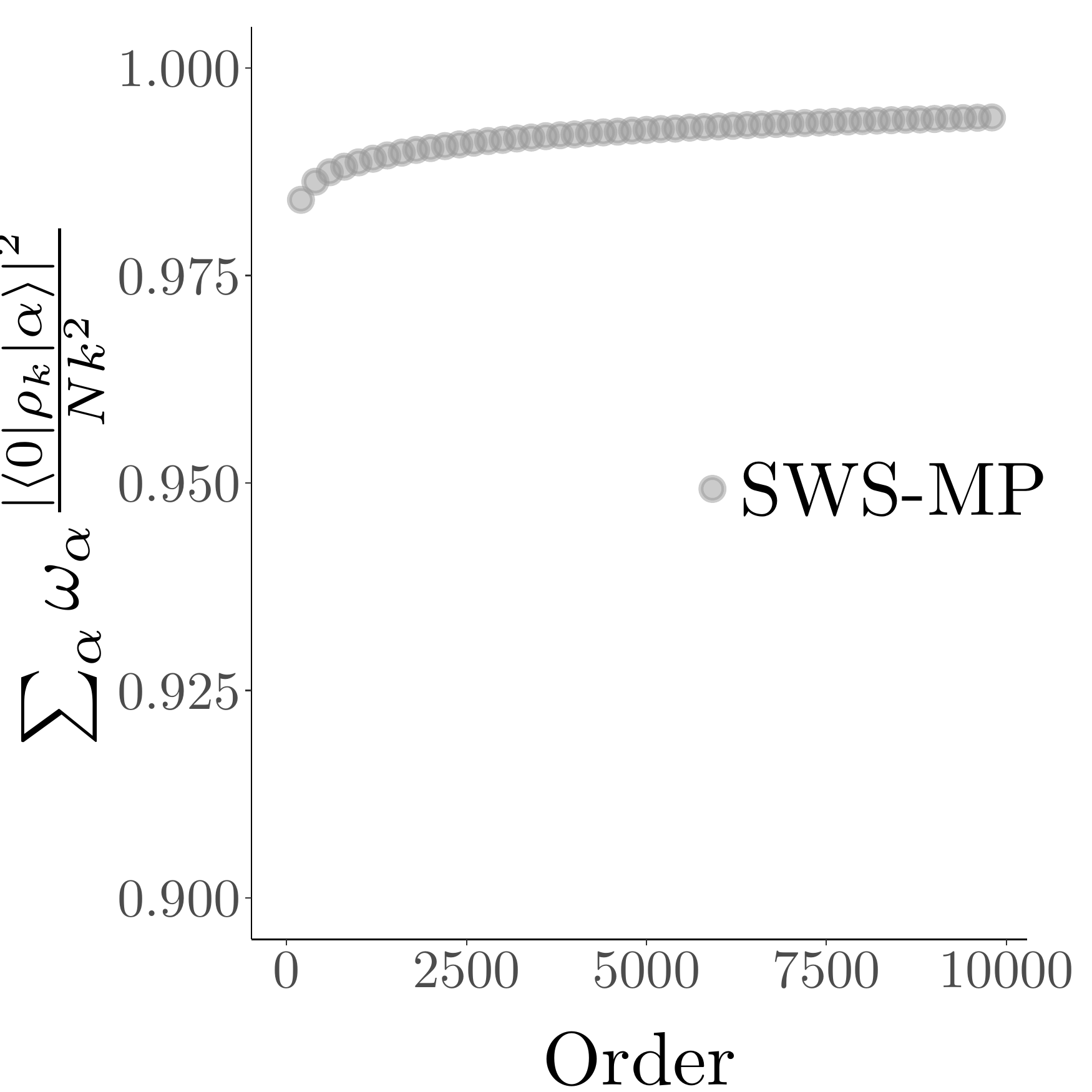}
  }
  \\
  \subcaptionbox{$c=1$, $T=1$ \label{subfig:sumrule saturation T=1, c=1, kf rho, B_full}} {
    \centering
    \includegraphics[width=0.3\textwidth]{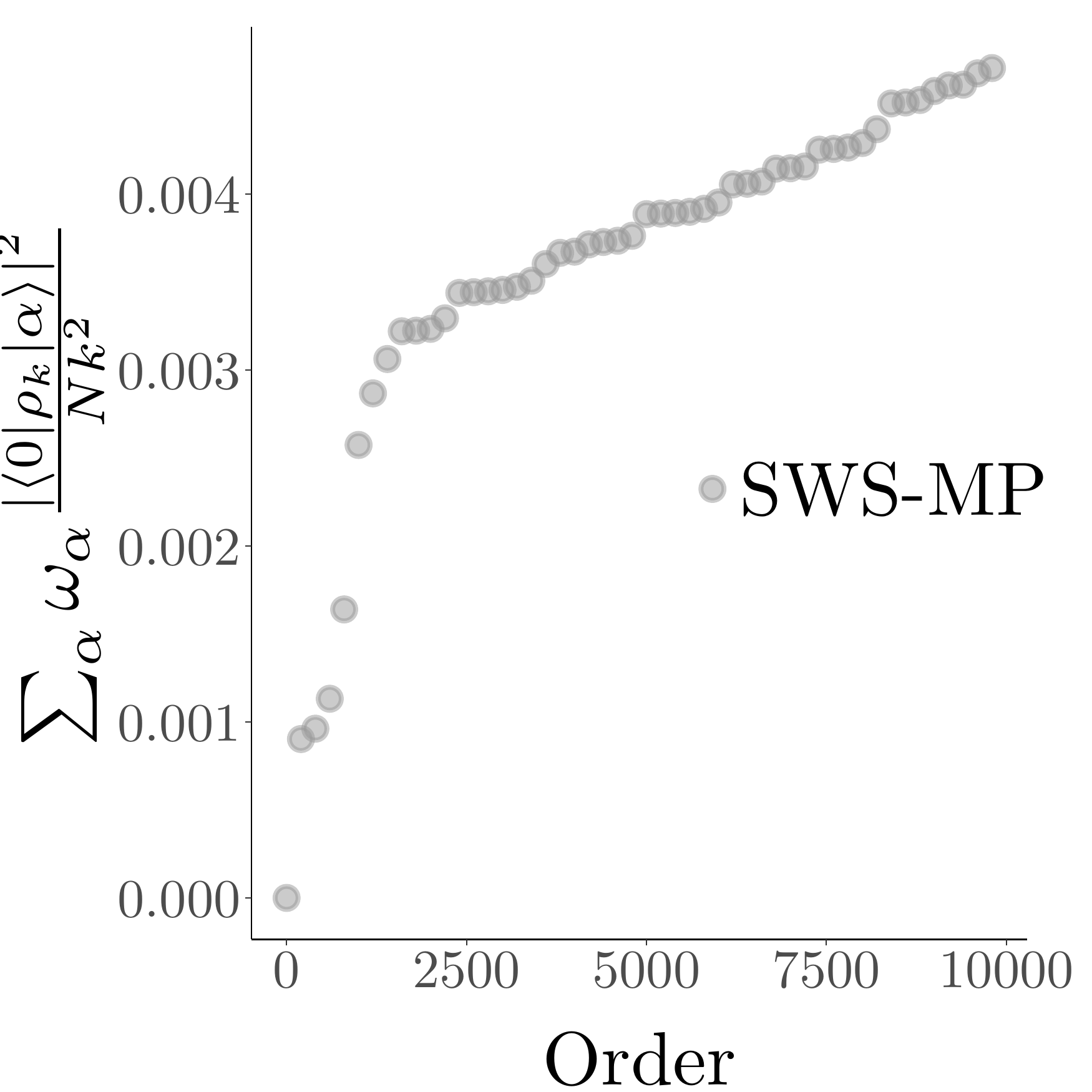}
  }
  \subcaptionbox{$c=4$, $T=1$ \label{subfig:sumrule saturation T=1, c=4, kf rho, B_full}} {
    \centering
    \includegraphics[width=0.3\textwidth]{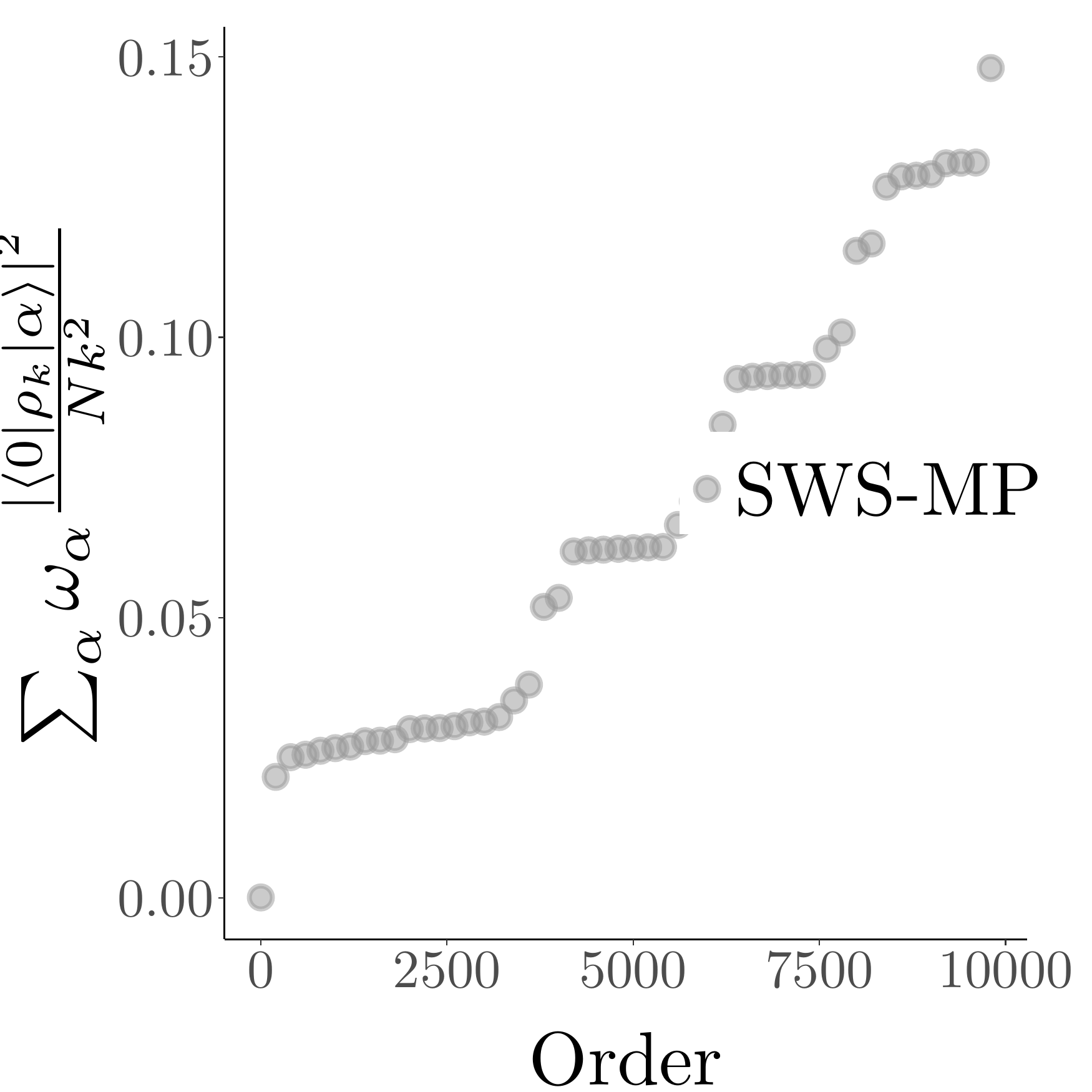}
  }
  \subcaptionbox{$c=16$, $T=1$ \label{subfig:sumrule saturation T=1, c=16, kf rho, B_full}} {
    \centering
    \includegraphics[width=0.3\textwidth]{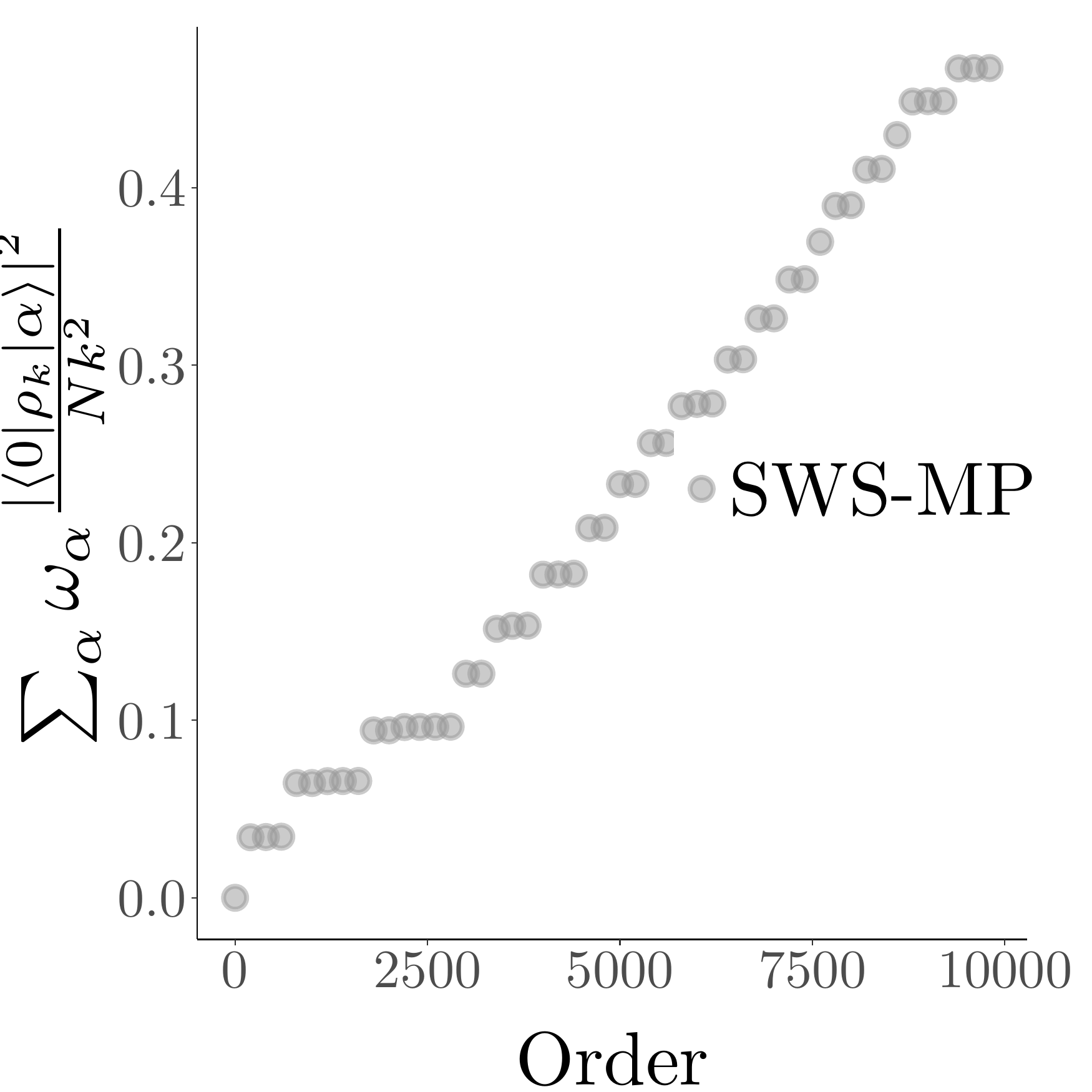}
  }
  \caption{
  Saturation of the $f$ sumrule with the number of states included in the
  summation. Starting from the ground state for $(a)$-$(c)$ and the
  representative thermal state at $T=1$ for $(d)$-$(f)$, we generated 10,000
  states using momentum preserving stepwise scanning for a target momentum of
  $k=\pi$, and $N = 128 = L$. We plot the sum rule saturation after every 200
  states for $c=1$ in $(a)$ and $(d)$, for $c=4$ in $(b)$ and $(e)$, and for
  $c=16$ in $(c)$ and $(f)$. Convergence is near perfect after very few states
  in the zero temperature case, whereas in the finite temperature case
  convergence is poor for the number of states considered. Furthermore, the
  interaction strength is seen to be an important variable in the finite
  temperature case, with smaller interaction strengths corresponding to poorer
  convergence.
  }
  \label{fig:c=16,kf,rho,N=128,B_full} 
\end{figure}

At finite temperature, we can do the same calculation provided that we replace
the ground state with the representative state of the thermal state we wish to
consider \cite{korepin_quantum_1997}.
The calculation of the dynamical structure factor at finite temperature is
inherently more difficult, but it also turns out that the simple scanning
algorithm we have developed is not optimal for the finite temperature case.
To understand why, consider Fig. 
    \ref{subfig:sumrule saturation T=1, c=1, kf rho, B_full},
    \subref{subfig:sumrule saturation T=1, c=4, kf rho, B_full},
    \subref{subfig:sumrule saturation T=1, c=16, kf rho, B_full}
where we again consider the \(f\) sumrule convergence.
Besides the significantly poorer rate of convergence, we observe clear plateaus
where convergence stagnates.
This indicates that the algorithm is not succesfully generating the states that
contribute to the finite-temperature correlation function most strongly first.
In the next section we propose changes to our algorithm that lead to better
convergence.

\section{Imposing additional constraints}
\label{sec:imposing additional constraints}

Thus far we have proposed algorithms where we generate descendents by
moving quantum numbers by the minimal amount at each step of the algorithm,
assuming this minimal change would result in the most important states first
being generated first.
However, it turns out that there is a property of the quantum numbers defining
an eigenstate which is an important indicator of its importance for most of the
weighing functions we are interested in that we have ignored thus far.
This property is the number of particle-hole pairs of a state with respect to
a given state of importance to the calculation at hand, herein the reference state.
In this section we explain how these particle-hole pairs are defined and we
present algorithms that use knowledge of this property to produce more efficient
algorithms.

In order to define the concept of particles and holes in this context, consider
a seed state $\{I_i^{SS}\}_{i \leq N}$, visualised by
\begin{equation*}
\begin{split}
 \circ \circ \bullet \bullet \bullet \bullet \bullet \circ \circ.
\end{split}
\end{equation*}
Then for another state $\{I_i \}_{i \leq N}$, the quantum numbers $I_l$ which do
not occur in $\{I_i^{SS} \}_{i \leq N}$ are called particles whereas the lattice
positions which are now empty as a result are called holes.
For example, for the following three states the particles and holes are labelled
with a $p$ and $h$ respectively.
\begin{equation*}
\begin{split}
 {\circ} \; {\underset{p}{\textcolor{ggBlue}{\bullet}}} \; {\underset{h}{\circ}} \; {\bullet} \; {\bullet} \; {\bullet} \; {\bullet} \; {\circ} \; {\circ} \\
 {\circ} \; {\underset{p}{\textcolor{ggBlue}{\bullet}}} \; {\underset{h}{\circ}} \; {\bullet} \; {\bullet} \; {\bullet} \; {\underset{h}{\circ}} \; {\underset{p}{\textcolor{ggOrange}{\bullet}}} \; {\circ} \\
 {\circ} \; {\circ} \; {\bullet} \; {\bullet} \; {\underset{h}{\circ}} \; {\bullet} \; {\bullet} \; {\circ} \; {\underset{p}{\textcolor{ggOrange}{\bullet}}}
\end{split}
\end{equation*}
Note that we always have an equal number of particles and holes enabling us to
talk about particle-hole pairs.

What makes the number of particle-hole pairs an important property to consider
is that generally the fewer particle-hole pairs a state has, the bigger the
off-diagonal matrix element between it and the reference state for local operators.
This means that the states with few particle-hole pairs are generally those
with the largest weights, meaning we should generate them first.
For the regular and momentum preserving stepwise scanning algorithms, however, a
descendent can have fewer particle-hole pairs than its parent.
For example, consider a seed state given by 
\begin{equation*}
\begin{split}
 {\circ} \; {\bullet} \; {\circ} \; {\bullet} \; {\circ} \; {\bullet} \; {\circ} \; {\circ} \; {\circ} \\
\end{split}
\end{equation*}
then it generates the following subtree where a particle and hole annihilate one another
\begin{equation*}
\begin{split}
 {\underset{p}{\textcolor{ggBlue}{\bullet}}} \; {\underset{h}{\circ}} \; {\circ} \; {\bullet} \; \stackunder{${\circ}$}{$\downarrow$} \; {\underset{h}{\circ}} \; {\underset{p}{\textcolor{ggOrange}{\bullet}}} \; {\circ} \; {\circ} \\
 {\underset{p}{\textcolor{ggBlue}{\bullet}}} \; {\underset{h}{\circ}} \; {\underset{p}{\textcolor{ggBlue}{\bullet}}} \; {\underset{h}{\circ}} \; \stackunder{${\circ}$}{$\downarrow$} \; {\underset{h}{\circ}} \; {\circ} \; {\underset{p}{\textcolor{ggOrange}{\bullet}}} \; {\circ} \\
 {\underset{p}{\textcolor{ggBlue}{\bullet}}} \; {\textcolor{ggBlue}{\bullet}} \; {\circ} \; {\underset{h}{\circ}} \; {\circ} \; {\underset{h}{\circ}} \; {\circ} \; {\circ} \; {\underset{p}{\textcolor{ggOrange}{\bullet}}}
\end{split}
\end{equation*}
Taking on board that matrix elements of local operators depend upon the number
of particle-hole excitations in a structured way, we can modify the previously
proposed algorithms to more efficiently explore the Hilbert space.

One way to proceed is to simply forbid the annihilation of particle-hole pairs
within a modified algorithm.
However, this approach comes at a cost since adding this restriction to the
rules of the regular and momentum preserving stepwise scanning algorithms breaks
completeness.
In order to regain this necessary feature, we allow quantum numbers that have
not moved yet to hop into the position of a hole if it lies between it and one
of the neighbouring quantum numbers.
For example, this allows the following subtree to be generated:
\begin{equation*}
\begin{split}
 {\circ} \; {\bullet} \; {\circ} \; {\bullet} \; \stackunder{${\bullet}$}{$\downarrow$} \; {\bullet} \; {\circ} \; {\bullet} \; {\circ} \\
 {\underset{p}{\textcolor{ggBlue}{\bullet}}} \; {\underset{h}{\circ}} \; {\circ} \; {\bullet} \; \stackunder{${\bullet}$}{$\downarrow$} \; {\bullet} \; {\circ} \; {\bullet} \; {\circ} \\
 {\underset{p}{\textcolor{ggBlue}{\bullet}}} \; {\textcolor{ggBlue}{\bullet}} \; {\circ} \; {\underset{h}{\circ}} \; {\bullet} \; {\bullet} \; {\circ} \; {\bullet} \; {\circ} \\
\end{split}
\end{equation*}
Note that this means that this allows for quantum number jumps of more than one
position.
Applying these changes to stepwise scanning scanning leads to the following
rules, which constitute the leapwise scanning algorithm (LWS).

Let $\{I_i\}_{i \leq N}$ be the set of quantum numbers at a node.
We again denote the leftmoving and rightmoving quantum numbers by $\{I^L_j\}_{j \leq
N_L}$ and $\{I^R_j\}_{j \leq N_R}$ respectively.
Furthermore, we label the positions of the holes as $\{I_j^h \}_{j \leq N_h}$
and the particles by $\{I^p_j \}_{j \leq N_p}$.
\begin{itemize}
  \item \textbf{Generate higher momentum descendents:} Generate a descendent for every
        $I_l \in \{I_i\}_{i \leq N}$ for which $I_l \notin \{I_j^L\}_{j \leq
        N_L}$, $I_l \leq I_0^R$, and either
        \begin{itemize}
          \item[$\circ$] $I_l+1 \neq I_{l+1}$ if $l \neq N$ with quantum numbers
            given by $\{I_i + \delta_{l,i} \}_{i \leq N}$, or
          \item[$\circ$] there exists a $k$ such that $I_l < I_k^h < I_{l+1}$ 
            with quantum numbers given by $\{I_i + \delta_{i,l}(I_k^h - I_i)
            \}_{i \leq N}$
        \end{itemize}
  \item \textbf{Generate lower momentum descendents:} If $\{I_k^R\} = \emptyset$
        generate a descendent for every $I_l \in \{I_i\}_{i \leq N}$ such that
        \begin{itemize}
          \item[$\circ$] $I_l \geq I_{N_L}^L$, $(I_{l} - 1) \notin \{I_j^h\}_{j \leq
            N_h}$, and $I_{l-1} \neq I_{l} - 1$ if $l \neq 0$ with quantum
            numbers given by $\{I_i - \delta_{l,i} \}_{i \leq N}$
          \item[$\circ$] there exists a $k$ such that $I_{l-1} < I_k^h < I_l$ with
            quantum numbers given by $\{I_i + \delta_{i, l}(I_k^h - I_i) \}$
        \end{itemize}
\end{itemize}
The proof that this algorithm is complete and does not overcount carries over
directly from the proof for the stepwise scanning algorithm.

Like stepwise scanning, leapwise scanning is not suitable for targeting a fixed
momentum sector.
One step towards a solution of this problem is to, like before, combine the
first and second step of the leapwise scanning algorithm.
However, in the current case the resulting algorithm does not preserve momentum
since the momentum increasing move from the first step and the momentum
decreasing momentum from the second step may not cancel.
The solution to this issue is the same as the solution to the problem we had
with the momentum preserving stepwise scanning algorithm when we wanted to
consider a target momentum sector whose momentum was different from that of the
reference state.
We generate the momentum increasing descendents for states whose momentum is
smaller than the target momentum and momentum decreasing descendents for states
whose momentum is larger.
For states at the right momentum we combine the steps as we do in momentum
preserving stepwise scanning.
This algorithm, which we refer to as momentum preserving leapwise scanning
(LWS-MP), can be summarised as follows.

\begin{figure}
  \subcaptionbox{$c=1$, $T=0$ \label{subfig:sumrule saturation T=0, c=1, kf rho, D_full}} {
    \centering
    \includegraphics[width=0.3\textwidth]{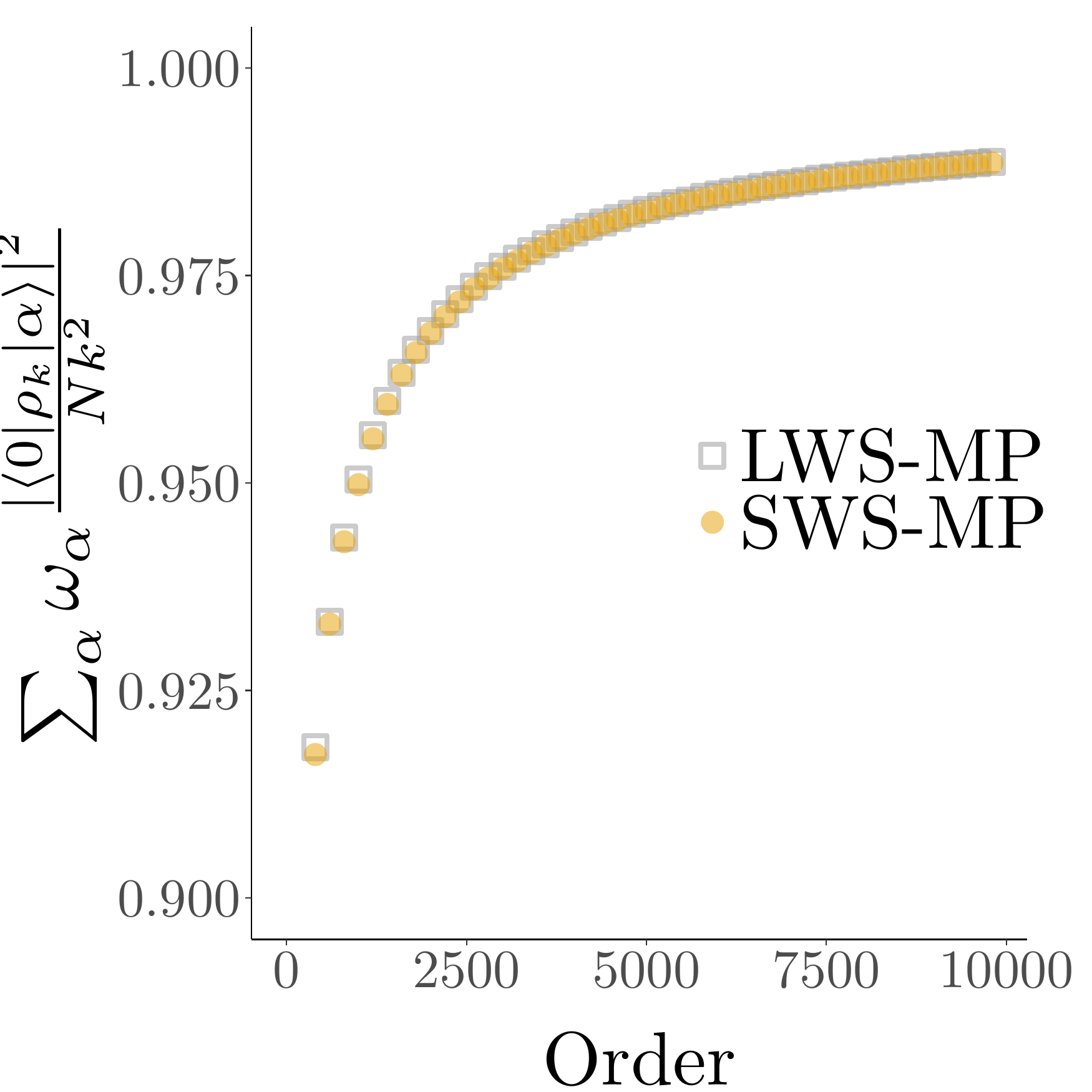}
  }
  \subcaptionbox{$c=4$, $T=0$ \label{subfig:sumrule saturation T=0, c=4, kf rho, D_full}} {
    \centering
    \includegraphics[width=0.3\textwidth]{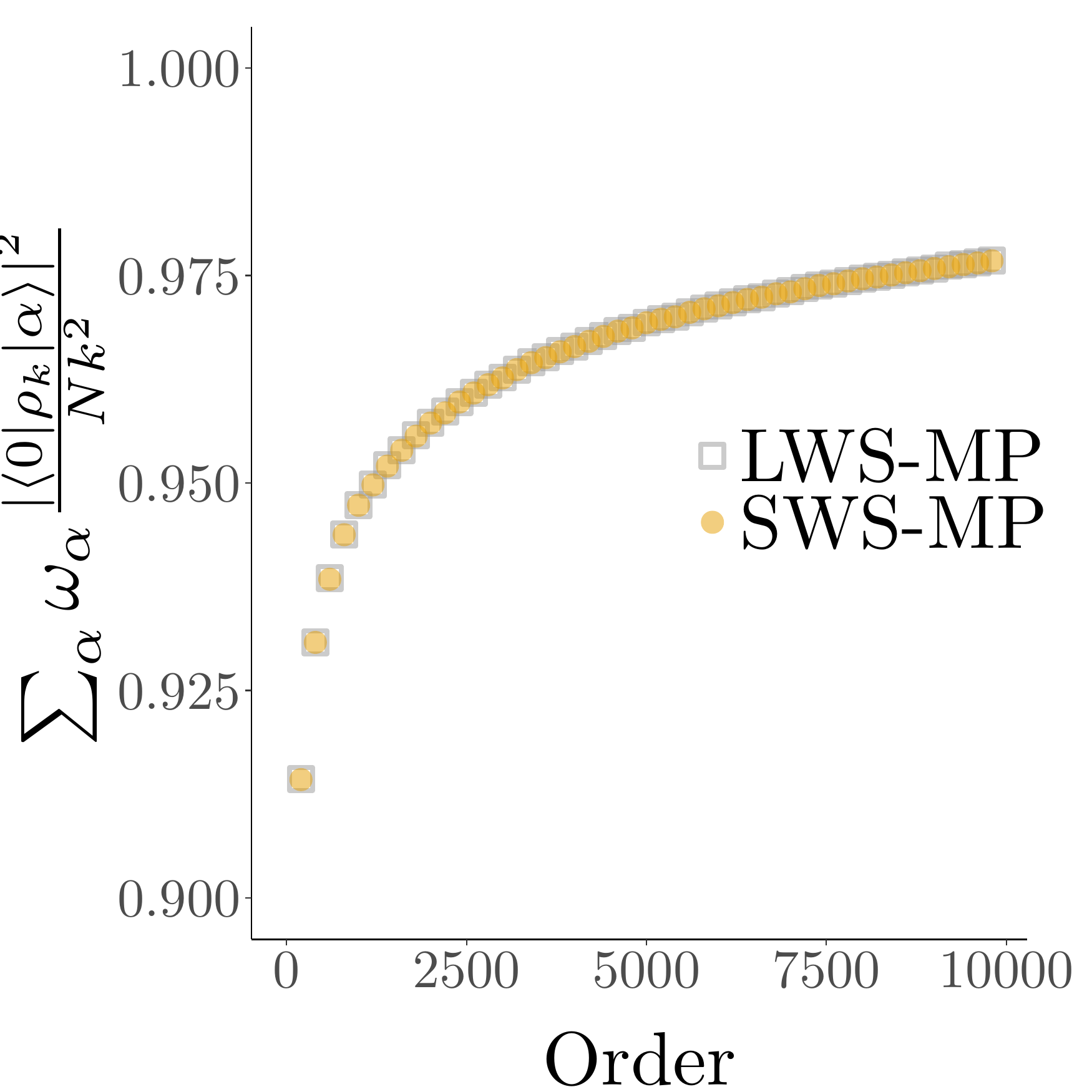}
  }
  \subcaptionbox{$c=16$, $T=0$ \label{subfig:sumrule saturation T=0, c=16, kf rho, D_full}} {
    \centering
    \includegraphics[width=0.3\textwidth]{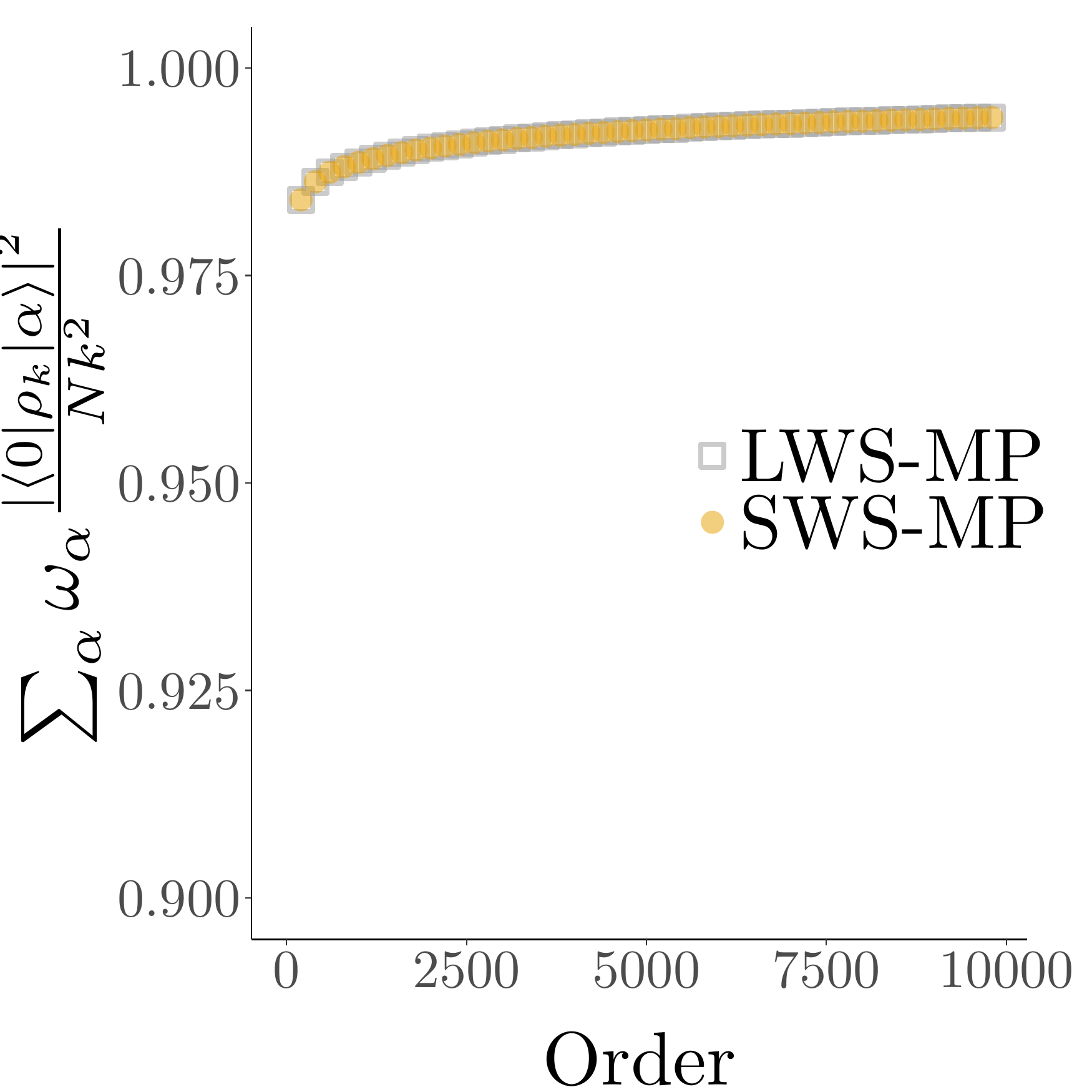}
  }
  \\
  \subcaptionbox{$c=1$, $T=1$ \label{subfig:sumrule saturation T=1, c=1, kf rho, D_full}} {
    \centering
    \includegraphics[width=0.3\textwidth]{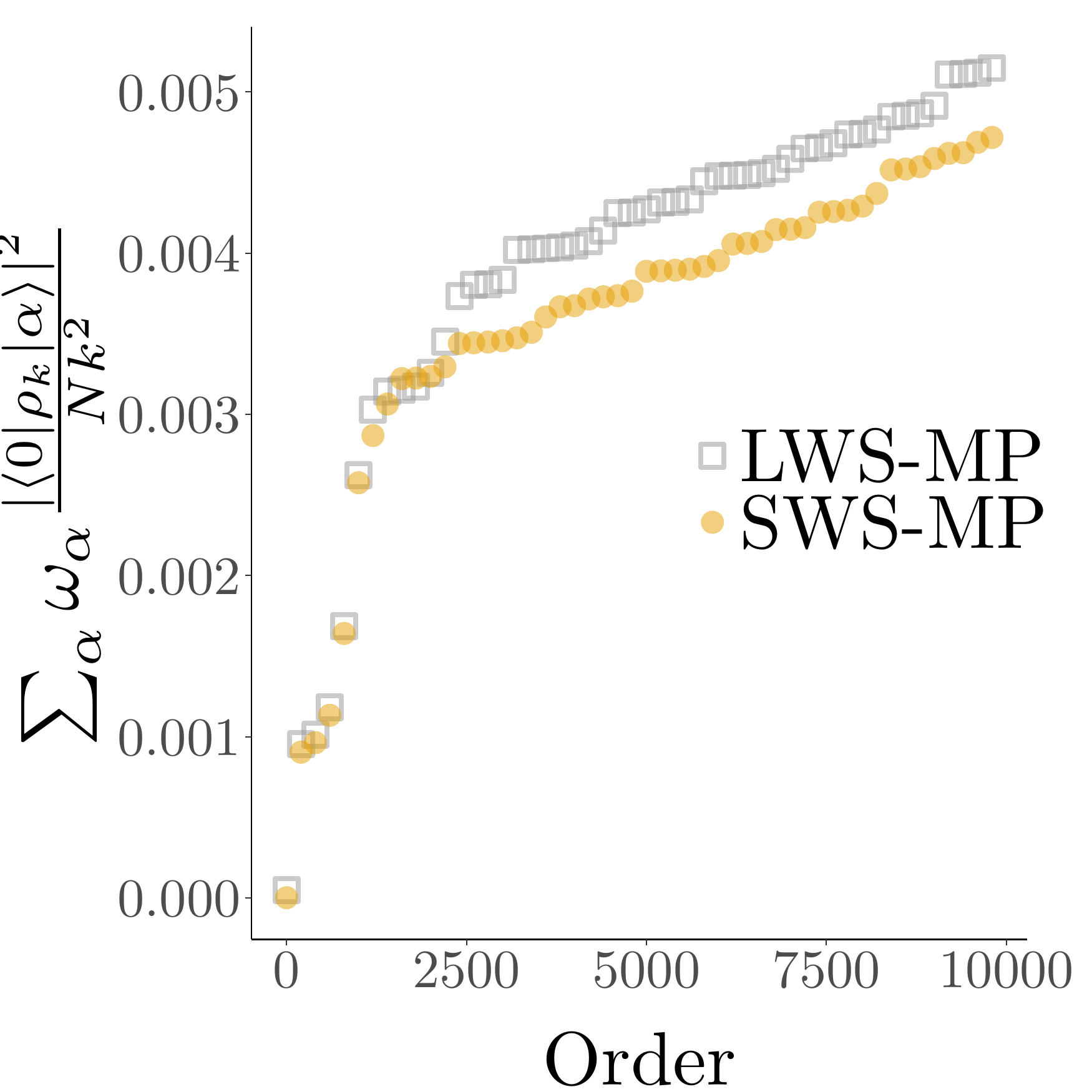}
  }
  \subcaptionbox{$c=4$, $T=1$ \label{subfig:sumrule saturation T=1, c=4, kf rho, D_full}} {
    \centering
    \includegraphics[width=0.3\textwidth]{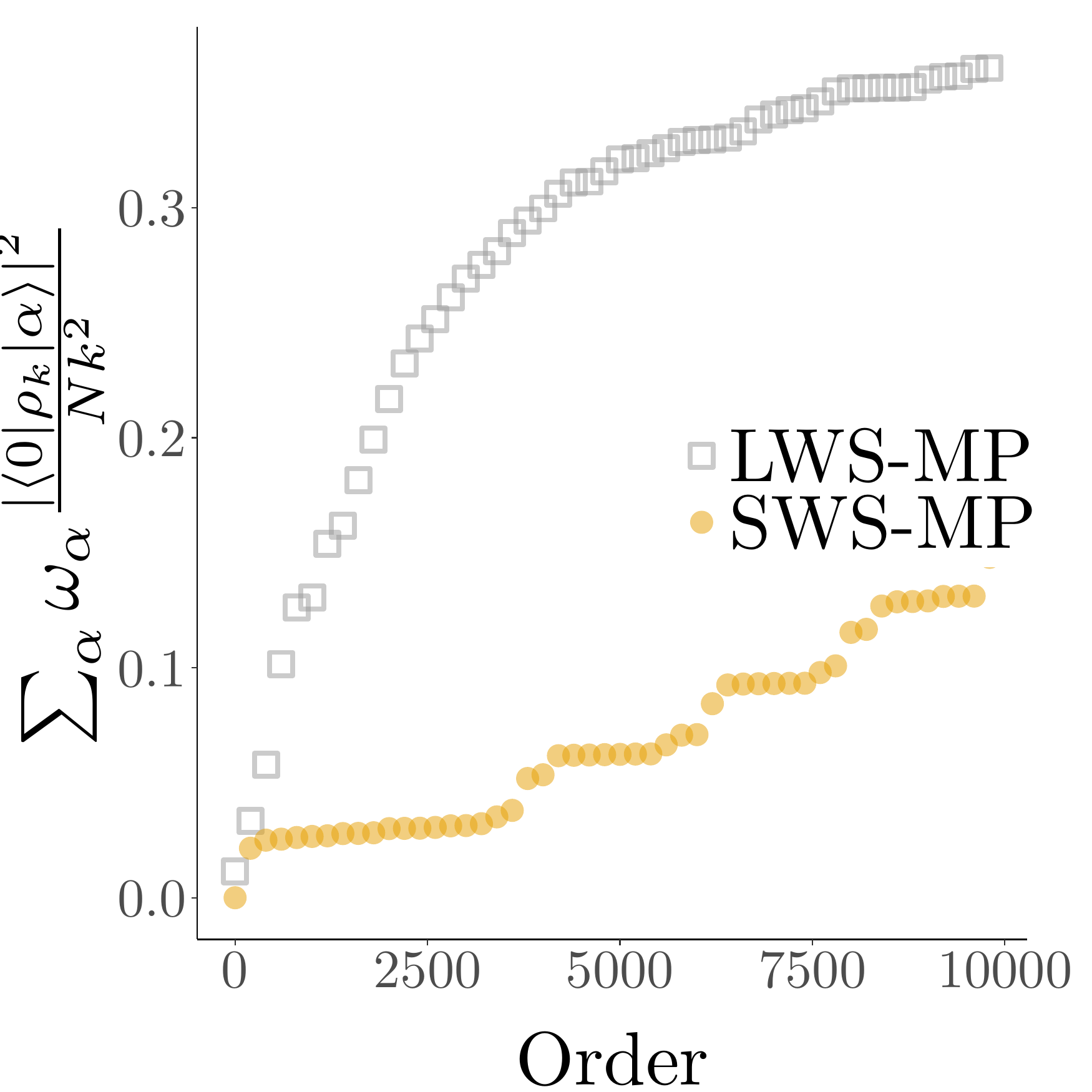}
  }
  \subcaptionbox{$c=16$, $T=1$ \label{subfig:sumrule saturation T=1, c=16, kf rho, D_full}} {
    \centering
    \includegraphics[width=0.3\textwidth]{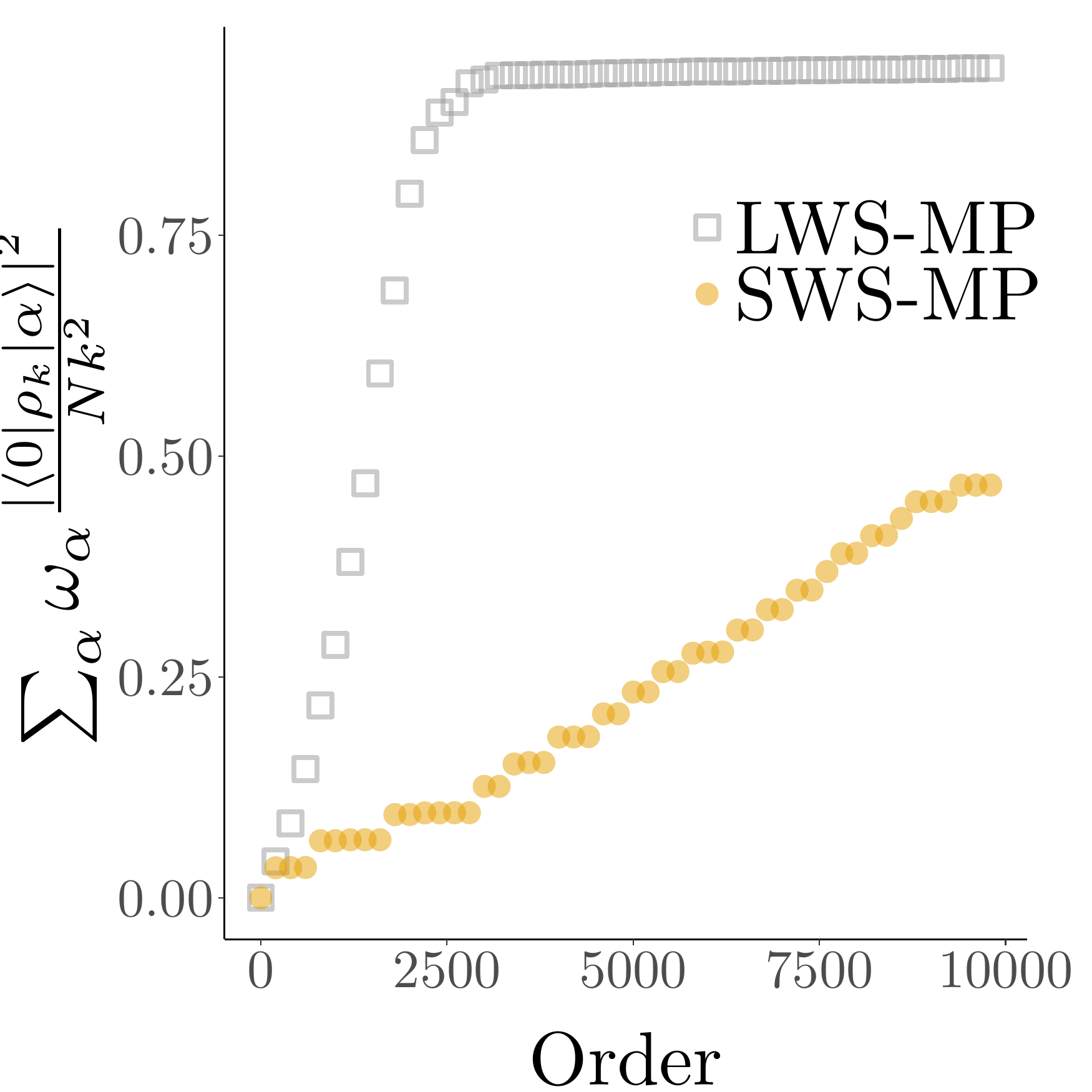}
  }
  \caption{
  Comparison of the saturation of the $f$ sumrule with the number of states
  included in the summation between momentum preserving stepwise (SWS-MP) and momentum
  preserving leapwise (LWS-MP) scanning. Starting from the ground state for $(a)$-$(c)$
  and the representative thermal state at $T=1$ for $(d)$-$(f)$, we generate
  10,000 states for a target momentum of $k=\pi$, and $N = 128 = L$. We plot the
  sum rule saturation after every 200 states for $c=1$ in $(a)$ and $(d)$, for
  $c=4$ in $(b)$ and $(e)$, and for $c=16$ in $(c)$ and $(f)$. Convergence is
  identical for both momentum preserving stepwise and leapwise scanning zero
  temperature, but not for the finite temperature case where momentum preserving
  leapwise scanning performs better. The plateaus at large and intermediate
  interaction strengths are barely noticeable for momentum preserving leapwise
  scanning whereas they remain pronounced in the case where $c=1$.
  \label{fig:c=16,kf,rho,N=128,B_full,D_full}
  } 
\end{figure}

Let $\{I_i\}_{i \leq N}$ be the set of quantum numbers at a node.
We denote the leftmoving and rightmoving quantum numbers by $\{I^L_j\}_{j \leq
N_L}$ and $\{I^R_j\}_{j \leq N_R}$ respectively.
Furthermore, we label the positions of the holes as $\{I_j^h \}_{j \leq N_h}$
and the particles by $\{I^p_j \}_{j \leq N_p}$.
\begin{itemize}
  \item \textbf{Generate rightmovers:} Generate an intermediate descendent $C_r$
        for every $I_l \in \{I_i \}_{i \leq N}$ such that $I_l \notin
        \{I_j^L\}_{j \leq N}$, $I_l \leq I_0^R$ and either
        \begin{itemize}
          \item[$\circ$] $I_l+1 \notin \{I_j^h\}_{j \leq N_h}$, and $I_l+1 \neq
            I_{l+1}$ if $l \neq N$ with quantum numbers given by $\{I_i +
            \delta_{l,i} \}_{i \leq N}$, or 
          \item[$\circ$] there exists a $k$ such that $I_l < I_k^h < I_{l+1}$ with
            quantum numbers given by $\{I_i + \delta_{i,l}(I_k^h - I_i) \}_{i
            \leq N}$
        \end{itemize}
  \item \textbf{Generate leftmovers:} For every intermediate descendent $C_r$
        generate a descendent for every $I_l \in \{I_i^{C_r} \}_{i \leq N}$ such
        that $I_l \notin \{I_j^R\}_{j \leq N_R}$, $I_l \geq I_{N_L}^L$, and
        either
        \begin{itemize}
          \item[$\circ$] $(I_{l} - 1) \notin \{I_j^h\}_{j \leq N_h}$, and $I_{l-1} \neq
            I_{l} - 1$ if $l \neq 0$ with quantum numbers given by $\{I_i -
            \delta_{l,i} \}_{i \leq N}$, or
          \item[$\circ$] there exists a $k$ such that $I_{l-1} < I_k^h < I_l$ with
            quantum numbers given by $\{I_i + \delta_{i, l}(I_k^h - I_i) \}$
        \end{itemize}
\end{itemize}
If the momentum of the node is smaller than the target momentum, we only do the
first step and the intermediate descendents become the descendents.
On the other hand, if the momentum of the node is smaller than the target momentum
we apply the second step to the node.
The proof that the resulting algorithm is complete and does not overcount is the
same as the proof for the momentum preserving stepwise scanning algorithm.

At zero temperature momentum preserving leapwise scanning gives practically the same results as momentum preserving stepwise scanning as can be seen in Fig.
  \ref{subfig:sumrule saturation T=0, c=1, kf rho, D_full},
  \subref{subfig:sumrule saturation T=0, c=4, kf rho, D_full},
  \subref{subfig:sumrule saturation T=0, c=16, kf rho, D_full}.
To understand this, note that at $T=0$ we start from the ground state which
  consists of a single block of neighbouring quantum numbers without holes. The
  only initial excitations that are then allowed by our rules are ones that move
  the outermost quantum numbers out leaving behind holes. This in turn gives
  space for more particle-hole pairs to be created, but no jumps over vacant
  positions that are not holes will occur during momentum preserving leapwise
  scanning. As a result, the descendents generated by momentum preserving
  stepwise and leapwise scanning as well as the resulting trees are virtually
  the same.

At finite temperature momentum preserving leapwise scanning outperforms momentum
preserving stepwise scanning as can be seen in part Fig.
  \ref{subfig:sumrule saturation T=1, c=1, kf rho, D_full},
  \subref{subfig:sumrule saturation T=1, c=4, kf rho, D_full},
  \subref{subfig:sumrule saturation T=1, c=16, kf rho, D_full}.
In this case, we start from a seed state where there is no longer a zero
  temperature Fermi sea, but instead there are vacant positions between the
  quantum numbers, i.e. the quantum numbers of the seed state are no longer of
  the form $\{a, a+1, a+2, \dots \}$ but rather something like $\{a, a+3, a+4, a+7, \dots\}$ for example.
  Therefore, when descendents are generated where quantum numbers neighbouring
  vacant positions are moved, they leave behind holes surrounded by vacant
  positions. For momentum preserving leapwise scanning, states with the same
number of particle-hole excitations can be created in such a scenario by letting
another quantum number jump over the vacant positions to this newly created hole
position, whereas in the case of momentum preserving stepwise scanning a bunch
of intermediate states would have to be created. Not having to generate these
intermediate states is what leads to the increase in efficiency.

\section{Beyond the topology of the tree}
\label{sec:building the tree}

The algorithms we have introduced thus far determine the topology of the tree
corresponding to a given seed state and target momentum sector, but they do not
determine the order in which the nodes of the tree are generated.
For the numerics displayed in the previous sections we adopted the rule that, by
default, we pause all branches and after each step of generating new descendents
we find the highest weight state all of whose descendents we then generate.
However, in Fig. 
  \ref{subfig:sumrule saturation T=1, c=1, kf rho, D_full},
  \subref{subfig:sumrule saturation T=1, c=4, kf rho, D_full},
  \subref{subfig:sumrule saturation T=1, c=16, kf rho, D_full}, we
saw that this approach produces plateaus which indicate that less important
states are generated before their more important counterparts.
The main reason for this is that at a given node there are descendents with
different levels of importance (as, for example, they have different numbers of
particle-hole excitations).
By generating all descendents of a node at the same time, states with more
particle-hole pairs can be generated before states with fewer particle-hole
pairs further down the tree.
In this section we show how splitting the descendents of the momentum preserving
leapwise scanning algorithm into three groups and treating them separately
results in a more efficient algorithm.

The division of the descendents of a given node into three groups is done based
on the number of additional particle-hole pairs they have compared to their parent,
which is either zero, one, or two for the algorithms in this paper.\footnote{This is due to the fact that the algorithms we consider change at most two quantum numbers per step of the algorithm. The approach outlined in this section can also be straightforwardly generalized to algorithms that can generate descendents with more than two additional particle-hole pairs by including additional groups.}
Given a node of the tree, we expect the states in the group of descendents with
the same number of particle-hole pairs as the parent node to be of similar
importance as the parent.
Since we always consider the paused node with the highest weight, its
descendents with the same number of particle-hole pairs are the ones we expect
to be the most important unexplored eigenstates.
Therefore we always start by generating these descendents if they have not yet
been generated. In constrast to regular momentum preserving leapwise scanning,
we do not also generate the descendents with additional particle-hole pairs at
the same time.

\begin{figure}
  \subcaptionbox{$c=1$, $T=0$ \label{subfig:sumrule saturation T=0, c=1, kf rho, D_full, D_partial}} {
    \centering
    \includegraphics[width=0.3\textwidth]{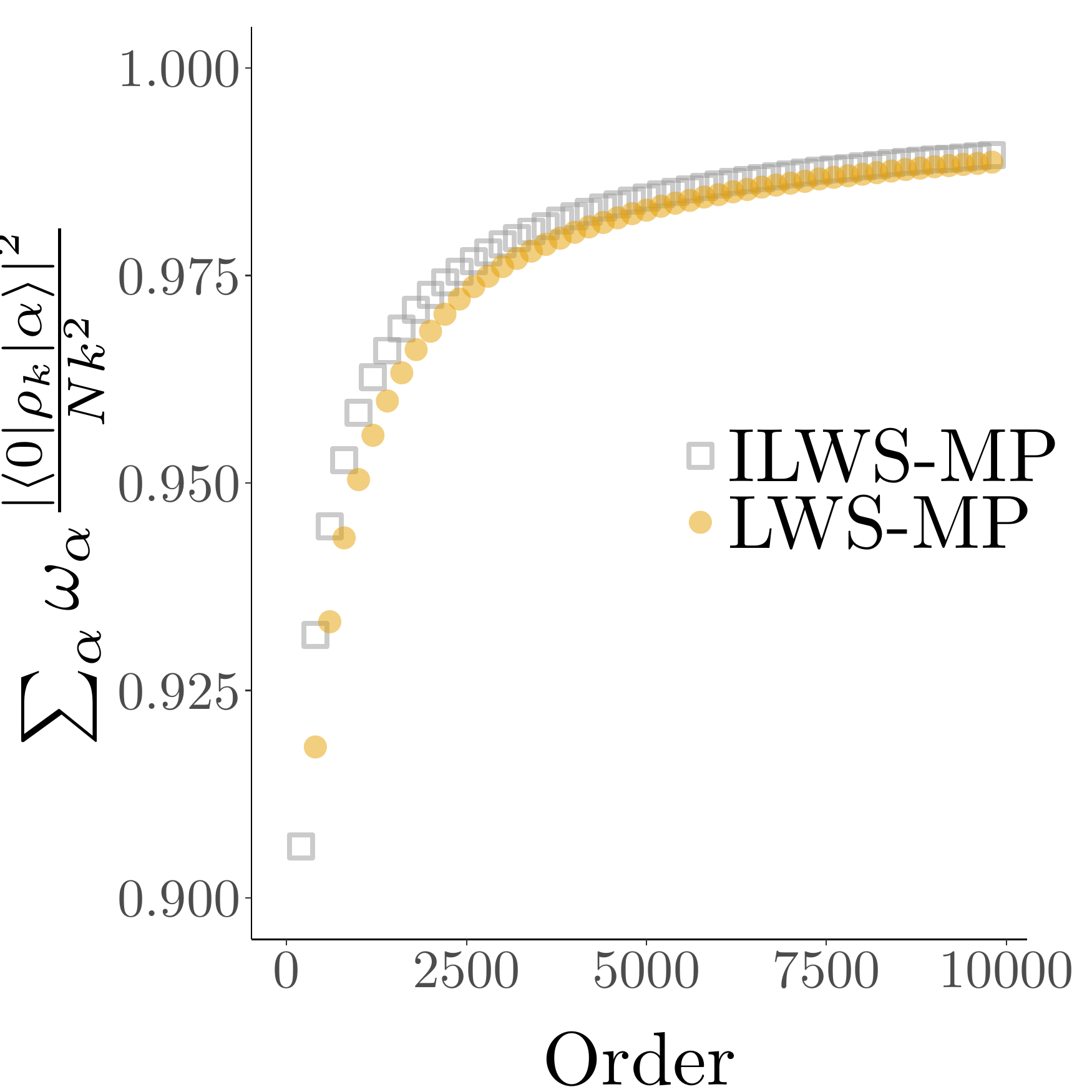}
  }
  \subcaptionbox{$c=4$, $T=0$ \label{subfig:sumrule saturation T=0, c=4, kf rho, D_full, D_partial}} {
    \centering
    \includegraphics[width=0.3\textwidth]{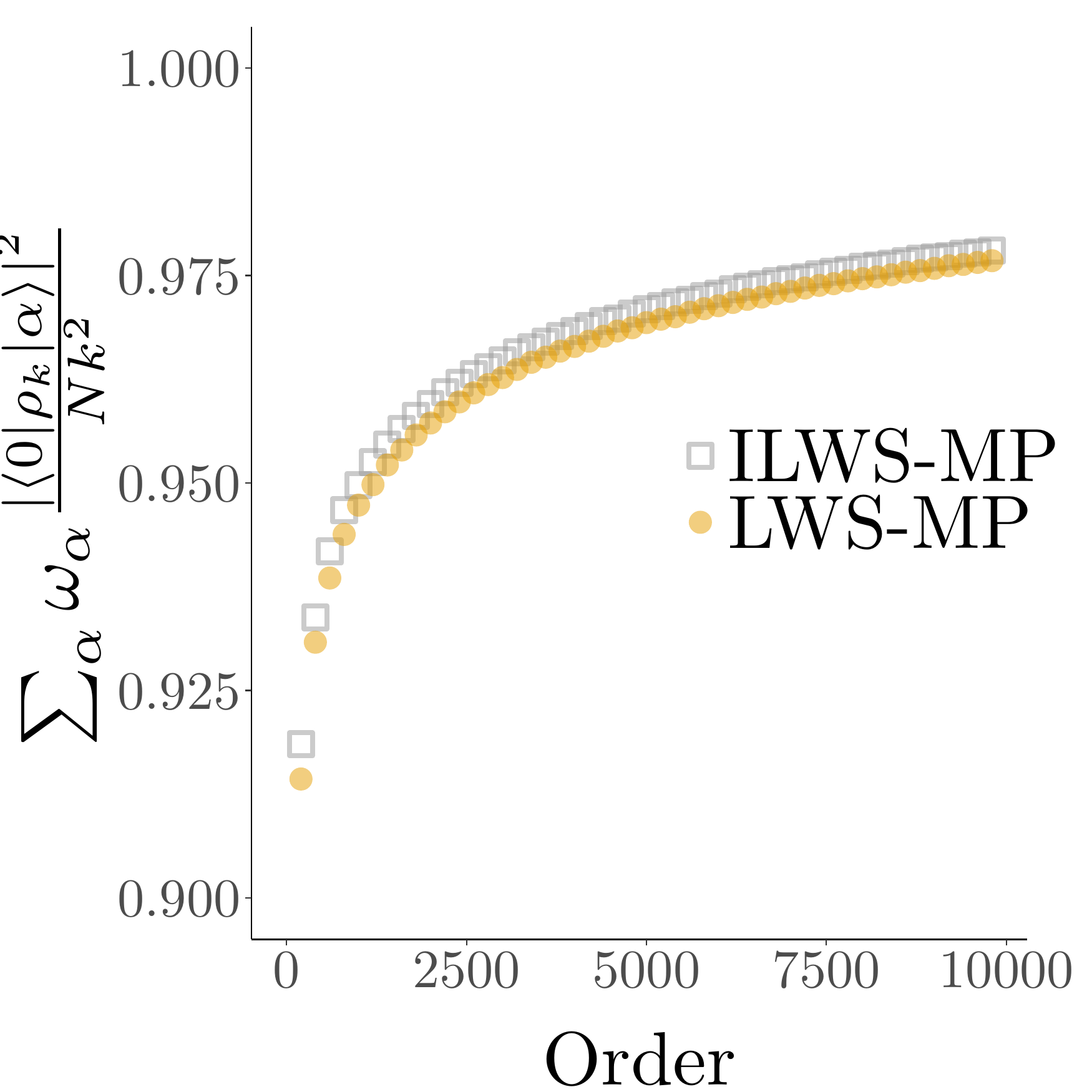}
  }
  \subcaptionbox{$c=16$, $T=0$ \label{subfig:sumrule saturation T=0, c=16, kf rho, D_full, D_partial}} {
    \centering
    \includegraphics[width=0.3\textwidth]{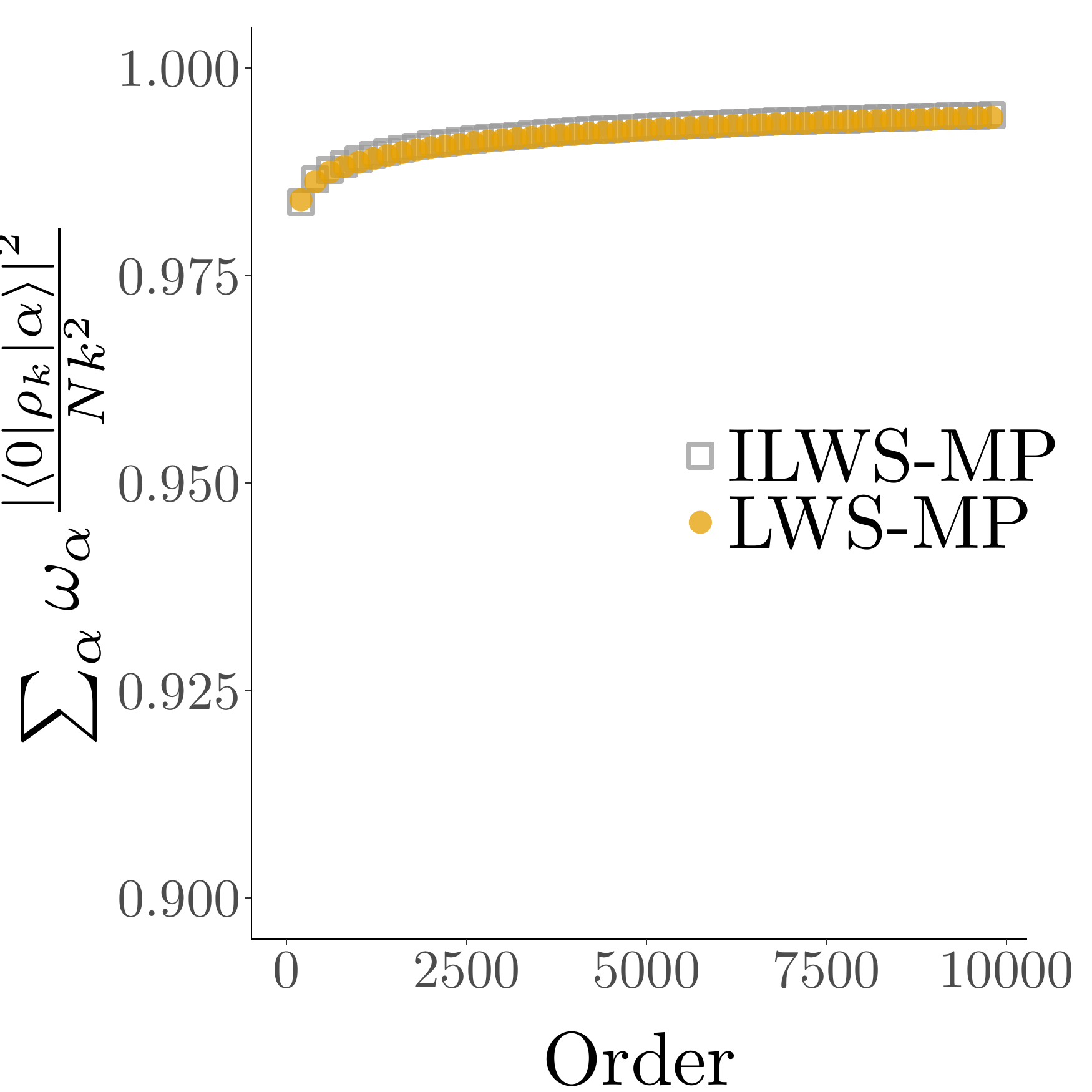}
  }
  \\
  \subcaptionbox{$c=1$, $T=1$ \label{subfig:sumrule saturation T=1, c=1, kf rho, D_full, D_partial}} {
    \centering
    \includegraphics[width=0.3\textwidth]{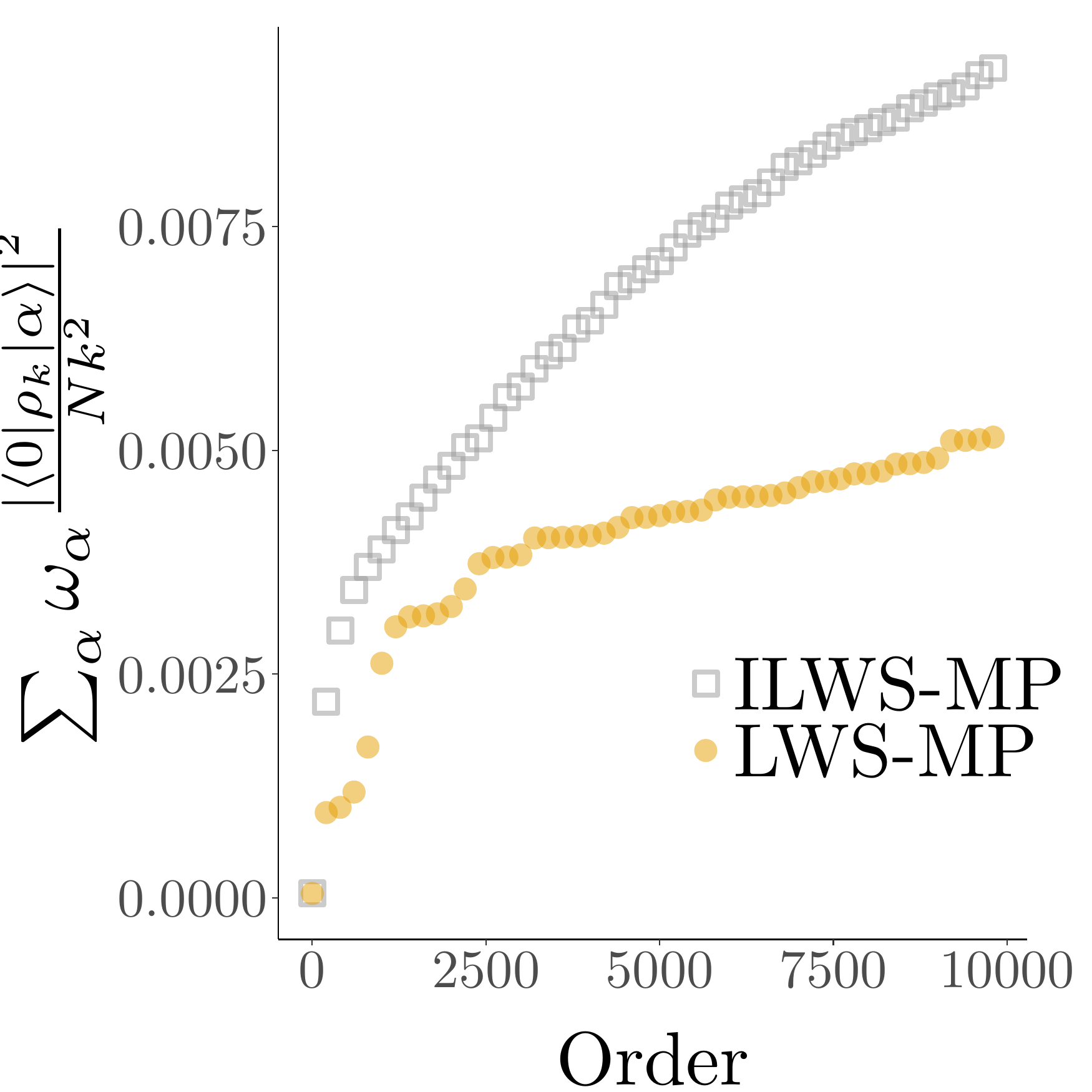}
  }
  \subcaptionbox{$c=4$, $T=1$ \label{subfig:sumrule saturation T=1, c=4, kf rho, D_full, D_partial}} {
    \centering
    \includegraphics[width=0.3\textwidth]{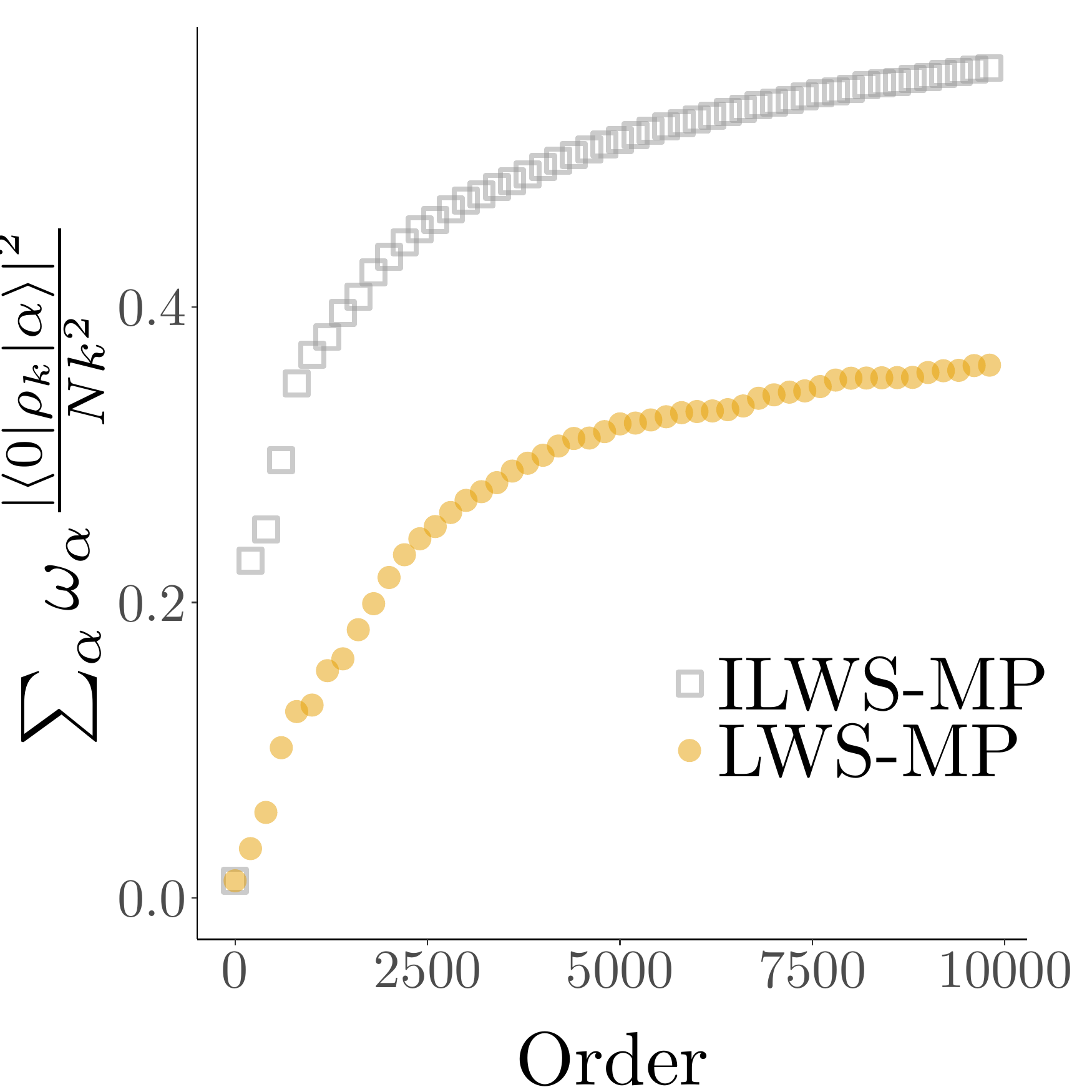}
  }
  \subcaptionbox{$c=16$, $T=1$ \label{subfig:sumrule saturation T=1, c=16, kf rho, D_full, D_partial}} {
    \centering
    \includegraphics[width=0.3\textwidth]{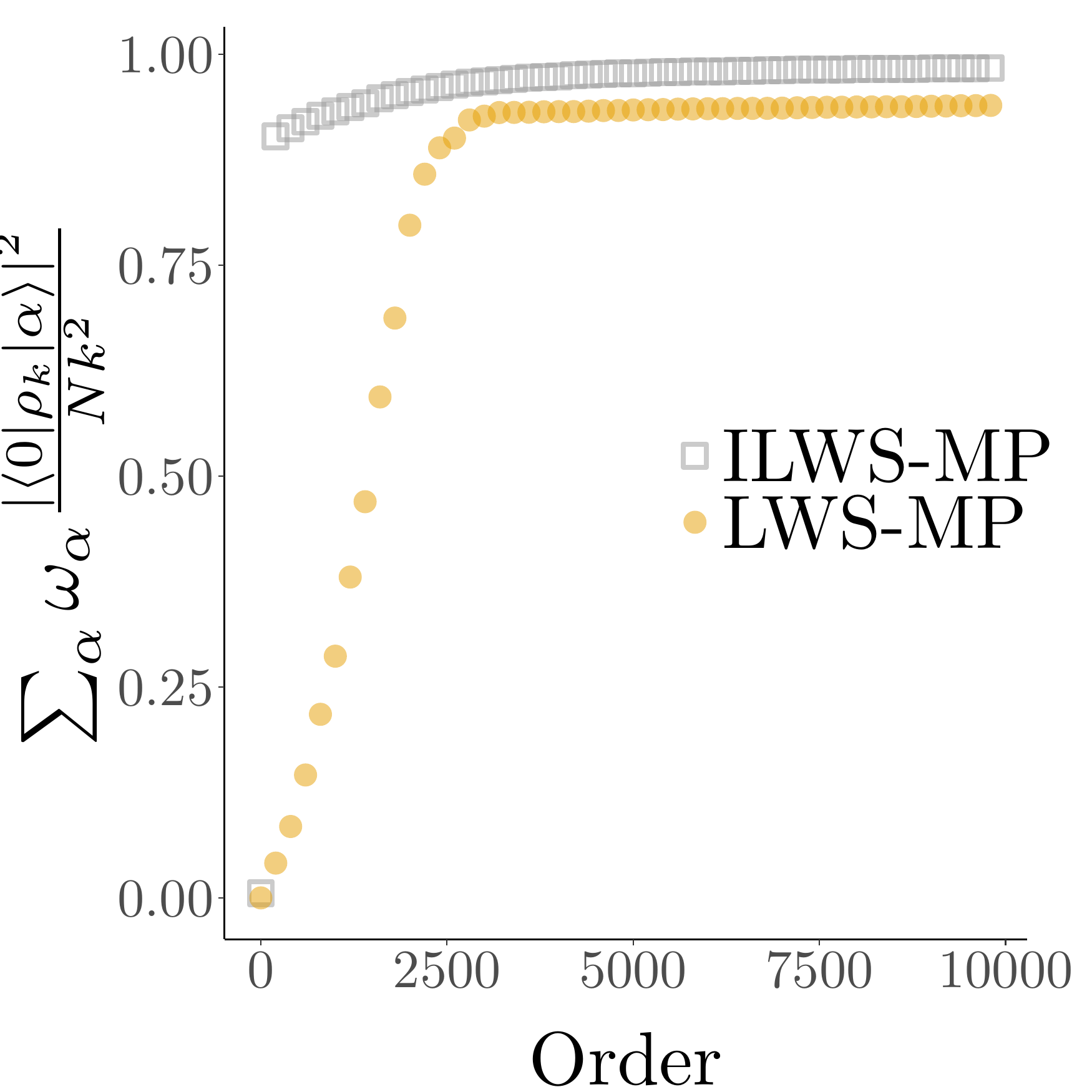}
  }
  \caption{
    Comparison of the saturation of the $f$ sumrule with the number of states
    included in the summation between momentum preserving leapwise scanning (LWS-MP)
    and improved momentum preserving leapwise scanning (ILWS-MP).
    Starting from the ground state for $(a)$-$(c)$ and the representative
    thermal state at $T=1$ for $(d)$-$(f)$, we generate 100,000 states for a
    target momentum of $k=\pi$, and $N = 128 = L$. We plot the sum rule
    saturation after every 200 states for $c=1$ in $(a)$ and $(d)$, for $c=4$ in
    $(b)$ and $(e)$, and for $c=16$ in $(c)$ and $(f)$. Convergence at zero
    temperature sees a tiny improvement for improved momentum preserving
    leapwise scanning to regular momentum preserving leapwise scanning and
    remains near perfect for very few states. At finite temperature, we see a
    dramatic increase in performance in both overall convergence as well as the
    number of states required to achieve this convergence for improved momentum
    preserving leapwise scanning, eliminating all plateaus. Still, absolute
    convergence remains a challenge at lower values of the interaction strength
    for the number of states considered regardless of the algorithm used.
  }
  \label{fig:c=16,kf,rho,N=128,D_partial,D_full} 
\end{figure}

Once we have generated the descendents with the same number of particle-hole
pairs for a given node, we can generally not yet forget about this node as we
could previously, since we still have to generate some of its descendents with
additional particle-hole pairs. Therefore, we store it in a different list which
tracks the states for which the descendents with no additional particle-hole
pairs have been generated. Furthermore, we give it a weight by choosing a random
descendent with an additional particle-hole pair and computing its weight, which
is then used to weigh the parent node in this secondary list. Having populated
both the initial list of paused nodes and this secondary list, we can choose to
generate new descendents by either considering a state from the original list
and generating more states with the same number of particle-hole pairs as their
parent, or by considering a state from the secondary list, and generating the
descendents of the node with one additional particle-hole pair. Which choice we
make depends on where the state with the highest weight is, as it could be in
the first list where its weight is that of the eigenstate, whereas in the second
list it would be the weight of a randomly chosen descendent with an additional
particle-hole pair.

In some cases, the full list of descendents would also have included states
with two additional particle-hole pairs with respect to their parent node.
In this case we follow a similar procedure and move the node from the secondary
list to a third where we keep nodes whose descendents with zero and one
additional particle-hole pair have been generated and associate to it another
representative weight obtained by computing the weight of one of the states with
two additional particle-hole
pairs. We then repeat the procedure outlined above with three lists.
The resulting procedure is what we refer to as improved momentum preserving
leapwise scanning (ILWS-MP).

The advantage of this approach is that, at the cost of computing at most two
additional matrix elements per node, we can ensure that we are generating the
states most important to the calculation under consideration.
Despite this representing an additional computational effort, it can still
represent a net gain as it can allow us to generate far fewer states for a given
accuracy of the calculation under consideration.
Whether the additional cost of computing these matrix elements is worth it can
depend, for example, on the seed state under consideration.
To illustrate this, consider the zero temperature dynamical structure factor
calculation whose results are illustrated in Fig.
  \ref{subfig:sumrule saturation T=0, c=1, kf rho, D_full, D_partial},
  \subref{subfig:sumrule saturation T=0, c=4, kf rho, D_full, D_partial},
  \subref{subfig:sumrule saturation T=0, c=16, kf rho, D_full, D_partial}.
Here we see that the performance the of regular and improved momentum
preserving leapwise scanning is virtually identical.
On the other hand, when considering the finite temperature equivalent as
illustrated in Fig. 
  \ref{subfig:sumrule saturation T=1, c=1, kf rho, D_full, D_partial},
  \subref{subfig:sumrule saturation T=1, c=4, kf rho, D_full, D_partial},
  \subref{subfig:sumrule saturation T=1, c=16, kf rho, D_full, D_partial},
we see a big difference in performance.
The algorithm where we allow some descendents to be generated at a later point
in time outperforms the algorithm where all descendents are generated.
Not only does it generate the same states quicker, but it also appears to
achieve higher sum rule saturations in part because the algorithm where
all descendents are generated gets stuck generating unimportant states.

\section{Comparison to the state of the art}
\label{sec:comparison to the state of the art}

In this section, we compare the improved momentum preserving leapwise scanning
algorithm to the most recent version of the state of the art software for the
computation of correlation functions in the Lieb-Liniger model called
\textsc{abacus} \cite{caux_correlation_2009}.
In Sec. \ref{subsec:the dynamical structure factor} we consider the dynamical
structure factor at zero and finite temperature.
Since \textsc{abacus} was developed in part to compute the dynamical structure
factor, it makes it the ideal candidate for a fair comparison.
In Sec. \ref{subsec:generating a basis for the interaction quench} we consider
the generation of an optimal basis for the interaction quench.
In both cases, we use commit \textsc{08c85cf590} of \textsc{abacus} as available
at https://jscaux.org/git/jscaux/ABACUS.
This version of \textsc{abacus} generates a tree topologically equivalent to the one
generated by stepwise scanning, generating all its descendents by moving quantum
numbers by at most one position.
The difference between \textsc{abacus} and our momentum preserving stepwise scanning is
the way the tree is built.
Most crucially, \textsc{abacus} deals with the need to generate states with few
particle-hole pairs by forcing the generation of branches of the tree where a
particle and hole recombine.
Therefore we can view it as stepwise scanning with forced recombinations, which
we refer to with its abbreviation (SWS-FR) throughout this section.
As we shall see, this is not the optimal way to deal with finite temperature
states.


\subsection{The dynamical structure factor}
\label{subsec:the dynamical structure factor}

\begin{figure}
  \subcaptionbox{$c=1$, $T=0$ \label{subfig:sumrule saturation T=0, c=1, kf rho, ABACUS, D_partial}} {
    \centering
    \includegraphics[width=0.3\textwidth]{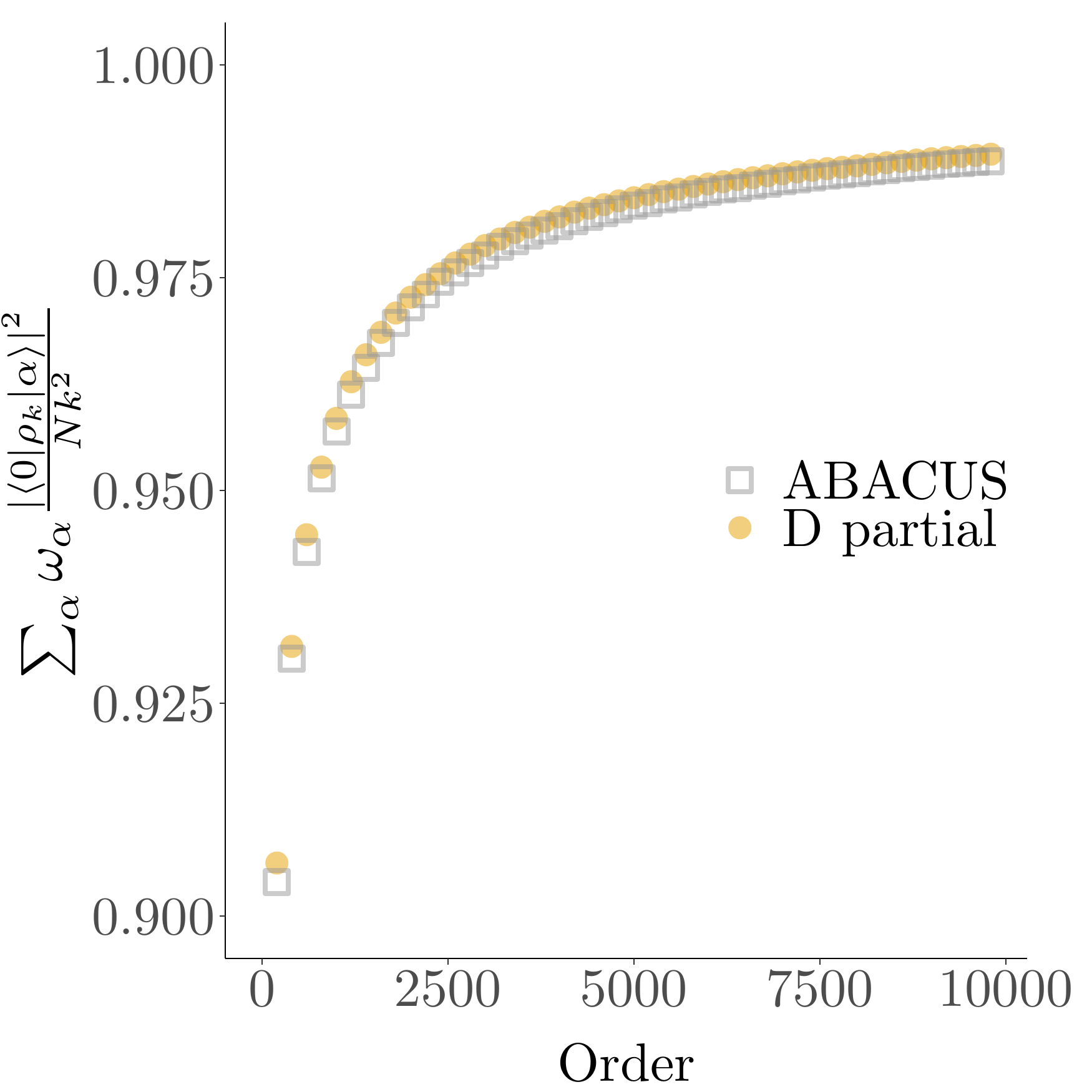}
  }
  \subcaptionbox{$c=4$, $T=0$ \label{subfig:sumrule saturation T=0, c=4, kf rho, ABACUS, D_partial}} {
    \centering
    \includegraphics[width=0.3\textwidth]{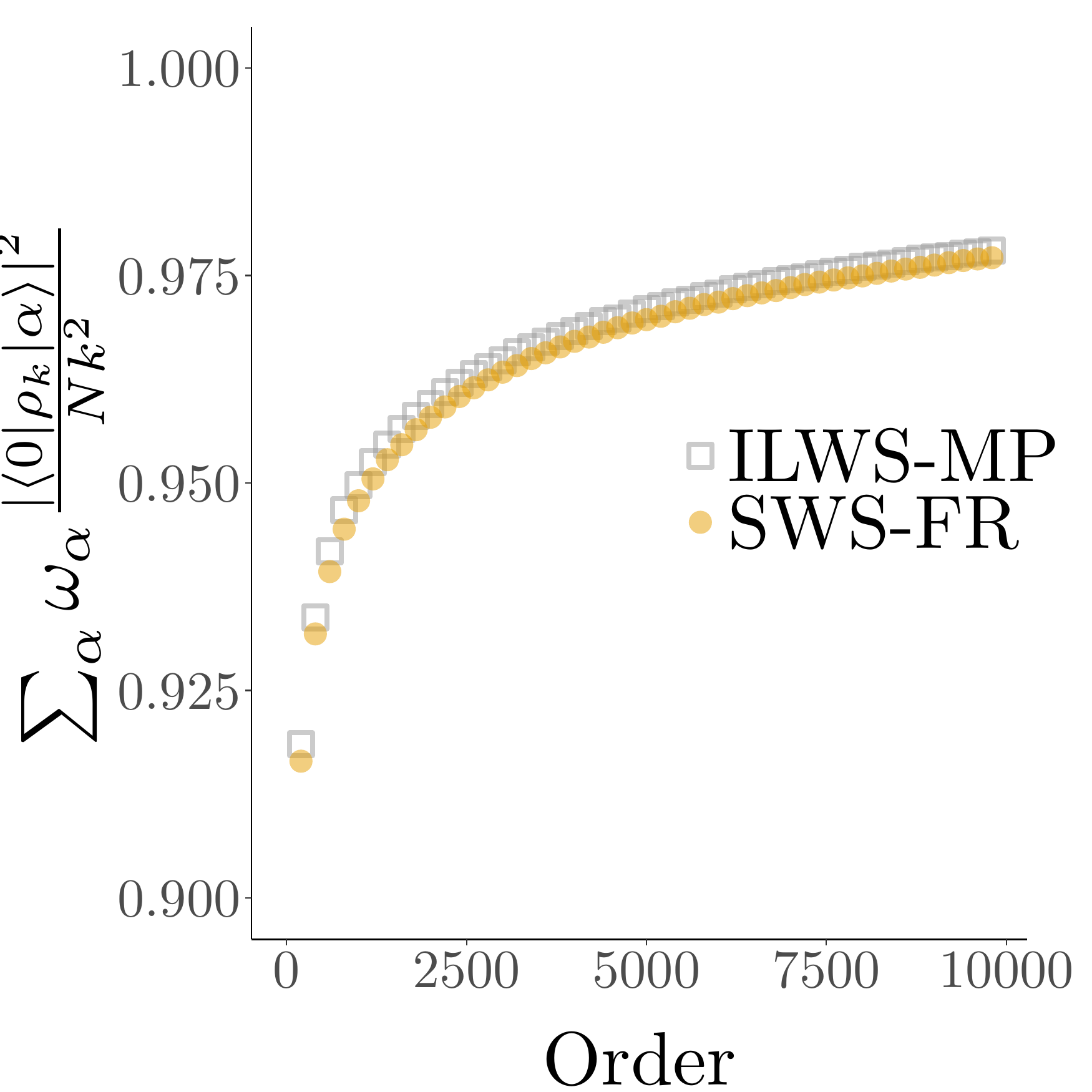}
  }
  \subcaptionbox{$c=16$, $T=0$ \label{subfig:sumrule saturation T=0, c=16, kf rho, ABACUS, D_partial}} {
    \centering
    \includegraphics[width=0.3\textwidth]{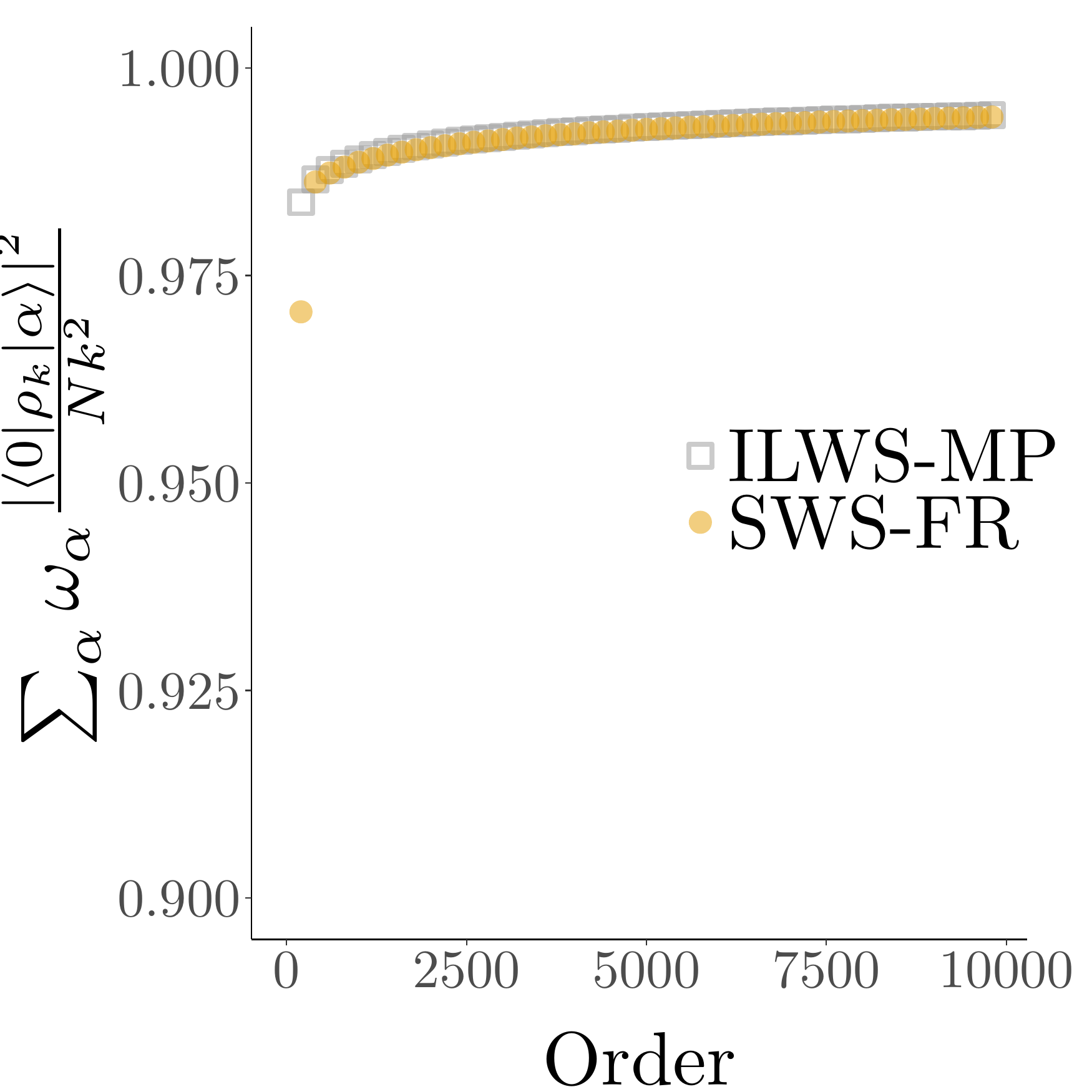}
  }
  \\
  \subcaptionbox{$c=1$, $T=1$ \label{subfig:sumrule saturation T=1, c=1, kf rho, ABACUS, D_partial}} {
    \centering
    \includegraphics[width=0.3\textwidth]{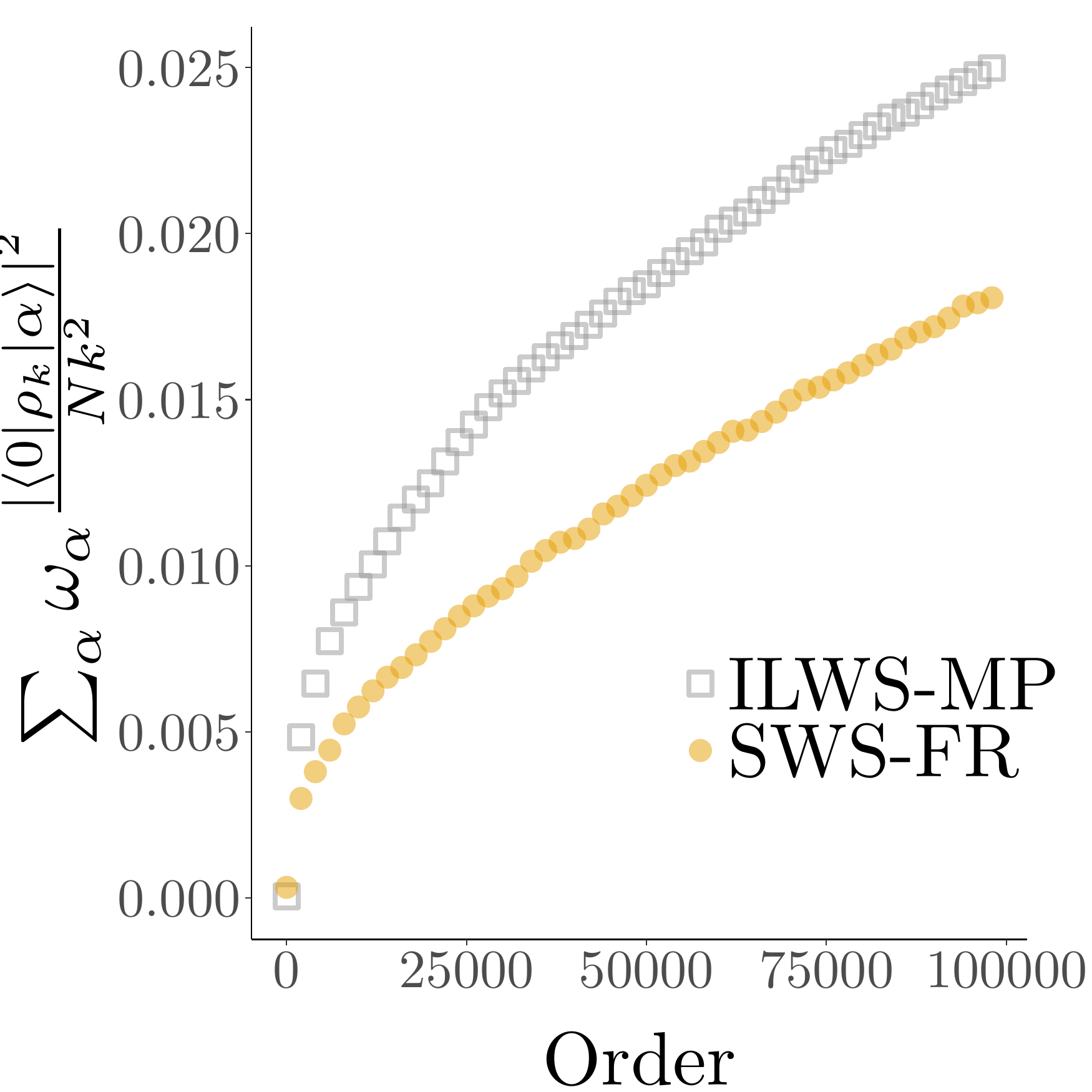}
  }
  \subcaptionbox{$c=4$, $T=1$ \label{subfig:sumrule saturation T=1, c=4, kf rho, ABACUS, D_partial}} {
    \centering
    \includegraphics[width=0.3\textwidth]{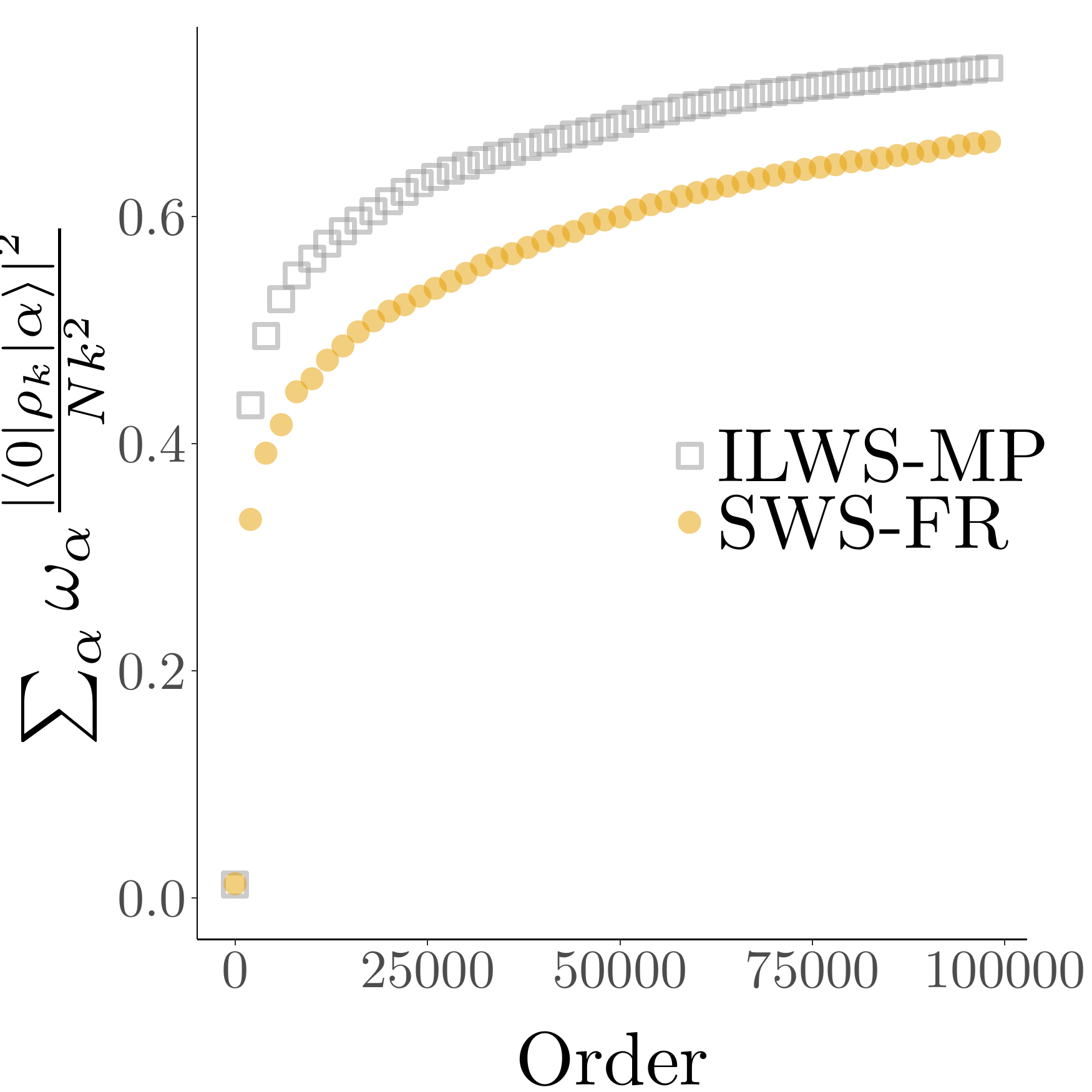}
  }
  \subcaptionbox{$c=16$, $T=1$ \label{subfig:sumrule saturation T=1, c=16, kf rho, ABACUS, D_partial}} {
    \centering
    \includegraphics[width=0.3\textwidth]{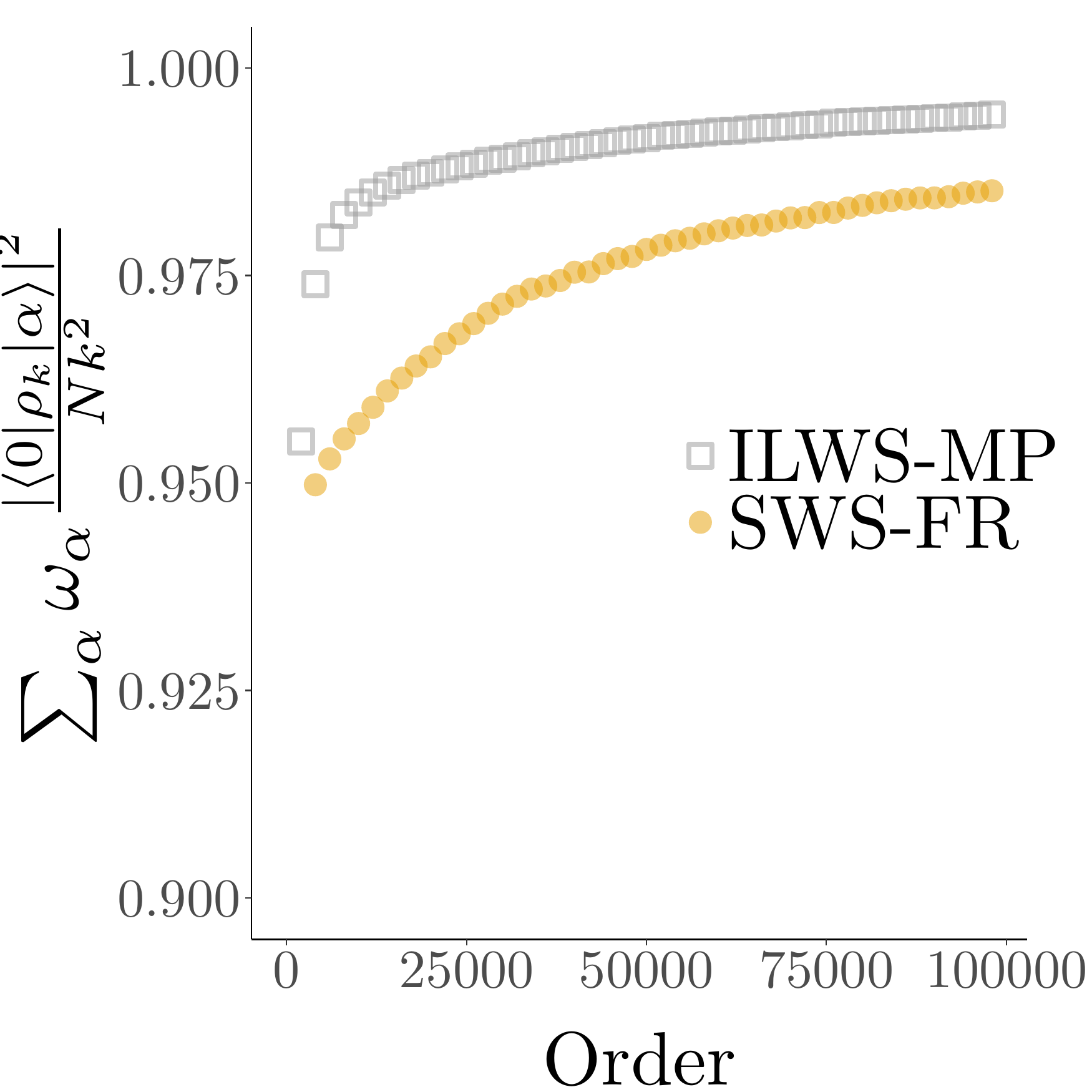}
  }
  \caption{Comparison of the saturation of the $f$ sumrule with the number of
  states included in the summation between the improved momentum preserving
  leapwise scanning (ILWS-MP) and momentum preserving stepwise scanning with
  forced recombinations.
  Starting from the ground state for $(a)$-$(c)$ and the representative thermal
  state at $T=1$ for $(d)$-$(f)$, we generate 10,000 states for $(a)$-$(c)$ and 
  100,000 states for $(d)$-$(f)$  for a target
  momentum of $k=\pi$, and $N = 128 = L$. We plot the sum rule saturation after
  every 200 states for $c=1$ in $(a)$ and $(d)$, for $c=4$ in $(b)$ and $(e)$,
  and for $c=16$ in $(c)$ and $(f)$. Convergence at zero temperature is
  virtually identical, whereas the improved momentum preserving scanning
  outperforms SWS-FR at finite temperature.}
  \label{fig:kf,rho,N=128,D_partial,ABACUS}
\end{figure}

For the ground state dynamical structure factor calculation, illustrated in Fig.
  \ref{subfig:sumrule saturation T=0, c=1, kf rho, ABACUS, D_partial},
  \subref{subfig:sumrule saturation T=0, c=4, kf rho, ABACUS, D_partial},
  \subref{subfig:sumrule saturation T=0, c=16, kf rho, ABACUS, D_partial},
we see that the performance of SWS-FR and our improved momentum preserving
leapwise scanning routine are virtually identical.
Similarly to the discussion comparing momentum preserving stepwise and leapwise
scanning, there is not a lot of freedom on how to generate the states when
starting from the ground state.
The differences which do exist are caused by states being generated in a
different order due to a different way of building the same tree.

At finite temperature, we observe an increase in performance for the improved
momentum preserving leapwise scanning compared to SWS-FR for all values of the
interaction strength as illustrated in Fig.
\ref{fig:kf,rho,N=128,D_partial,ABACUS}.
At finite temperature, the seed states we consider no longer consist of a
contiguous block of quantum numbers with empty positions between neighbouring
quantum numbers being introduced.
In this case, creating a new particle-hole pair leaves a hole that may not
neighbour another quantum number that has not moved yet.
For the improved momentum preserving leapwise scanning algorithm, states with
the same number of particle-hole pairs can be direct descendents of this state
by allowing the quantum number to move by more than one position.
For example, we saw that we can have subtrees like 
\begin{equation*}
\begin{split}
 {\circ} \; {\bullet} \; {\circ} \; {\bullet} \; \stackunder{${\bullet}$}{$\downarrow$} \; {\bullet} \; {\circ} \; {\bullet} \; {\circ} \\
 {\underset{p}{\textcolor{ggBlue}{\bullet}}} \; {\underset{h}{\circ}} \; {\circ} \; {\bullet} \; \stackunder{${\bullet}$}{$\downarrow$} \; {\bullet} \; {\circ} \; {\bullet} \; {\circ} \\
 {\underset{p}{\textcolor{ggBlue}{\bullet}}} \; {\textcolor{ggBlue}{\bullet}} \; {\circ} \; {\underset{h}{\circ}} \; {\bullet} \; {\bullet} \; {\circ} \; {\bullet} \; {\circ} \\
\end{split}
\end{equation*}
For the stewpwise scanning algorithm with forced recombinations such jumps are
not allowed, forcing it to first create an additional particle-hole pair and
move it to annihilate the initial hole leading to the following subtree:
\begin{equation*}
\begin{split}
 {\circ} \; {\bullet} \; {\circ} \; {\bullet} \; \stackunder{${\bullet}$}{$\downarrow$} \; {\bullet} \; {\circ} \; {\bullet} \; {\circ} \\
 {\underset{p}{\textcolor{ggBlue}{\bullet}}} \; {\underset{h}{\circ}} \; {\circ} \; {\bullet} \; \stackunder{${\bullet}$}{$\downarrow$} \; {\bullet} \; {\circ} \; {\bullet} \; {\circ} \\
 {\underset{p}{\textcolor{ggBlue}{\bullet}}} \; {\underset{h}{\circ}} \; {\underset{p}{\textcolor{ggBlue}{\bullet}}} \; {\underset{h}{\circ}} \; \stackunder{${\bullet}$}{$\downarrow$} \; {\bullet} \; {\circ} \; {\bullet} \; {\circ} \\
 {\underset{p}{\textcolor{ggBlue}{\bullet}}} \; {\textcolor{ggBlue}{\bullet}} \; {\circ} \; {\underset{h}{\circ}} \; {\bullet} \; {\bullet} \; {\circ} \; {\bullet} \; {\circ} \\
\end{split}
\end{equation*}
The difference may look innocent in this simple example, but as the distances
between quantum numbers in the initial state grow and the number of such
isolated quantum numbers increases, the number of additional states that have to
be generated in order to reach all states with the same number of particle-hole
pairs grows rapidly.

\begin{figure}
  \subcaptionbox{$c=1$, $T=1$ \label{subfig:sumrule saturation T=1, c=1, kf rho, ABACUS, D_partial histogram}} {
    \centering
    \includegraphics[width=0.3\textwidth]{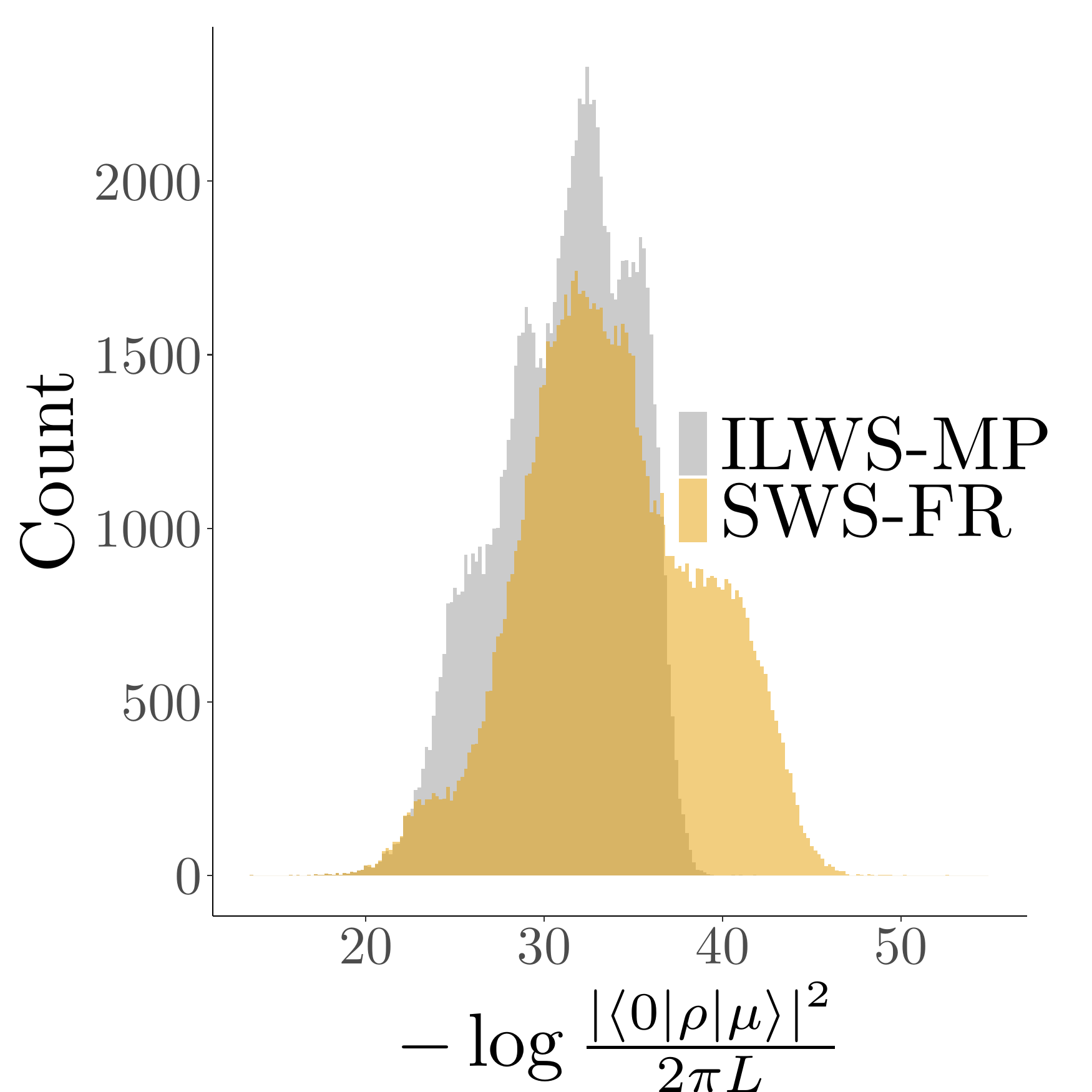}
  }
  \subcaptionbox{$c=4$, $T=1$ \label{subfig:sumrule saturation T=1, c=4, kf rho, ABACUS, D_partial histogram}} {
    \centering
    \includegraphics[width=0.3\textwidth]{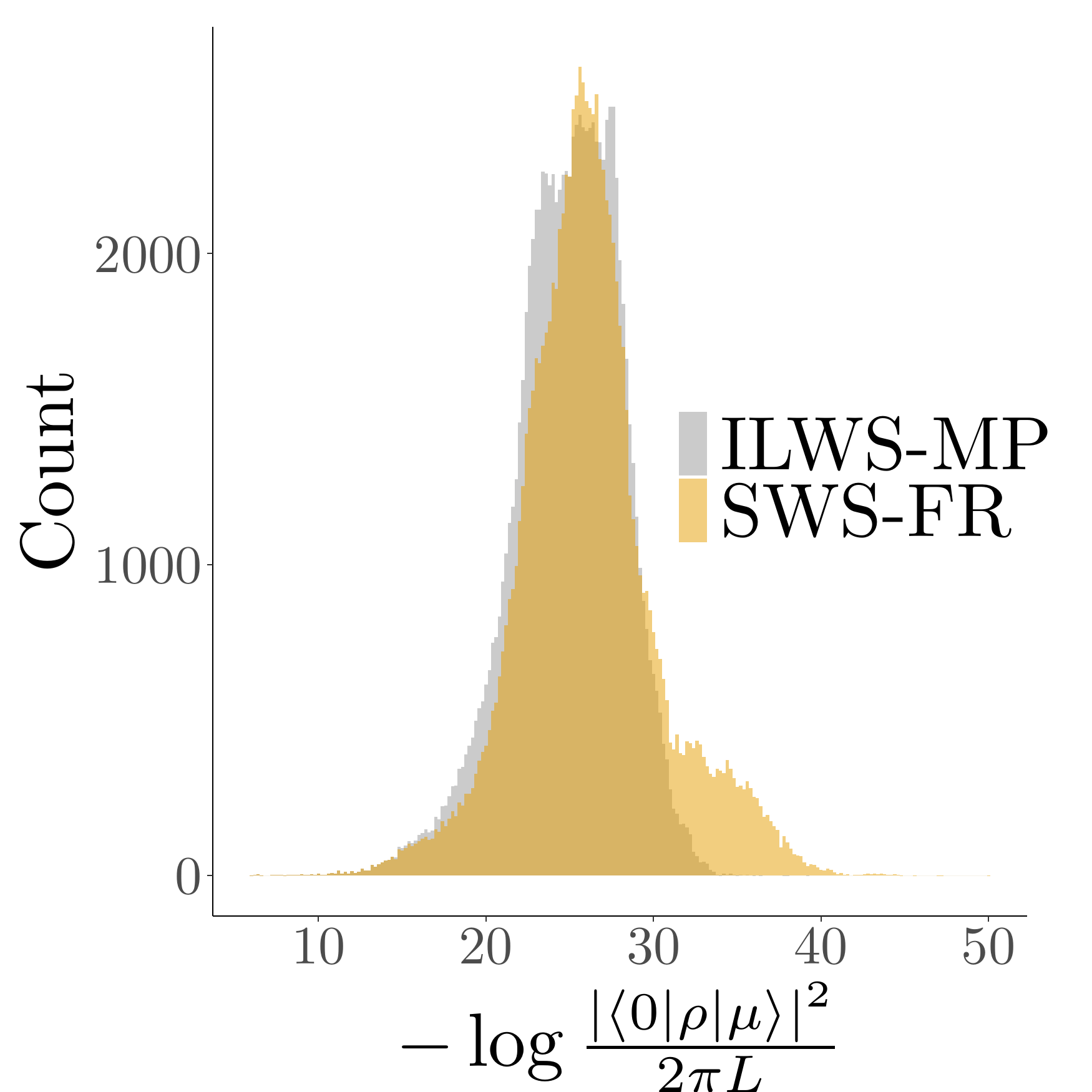}
  }
  \subcaptionbox{$c=16$, $T=1$ \label{subfig:sumrule saturation T=1, c=16, kf rho, ABACUS, D_partial histogram}} {
    \centering
    \includegraphics[width=0.3\textwidth]{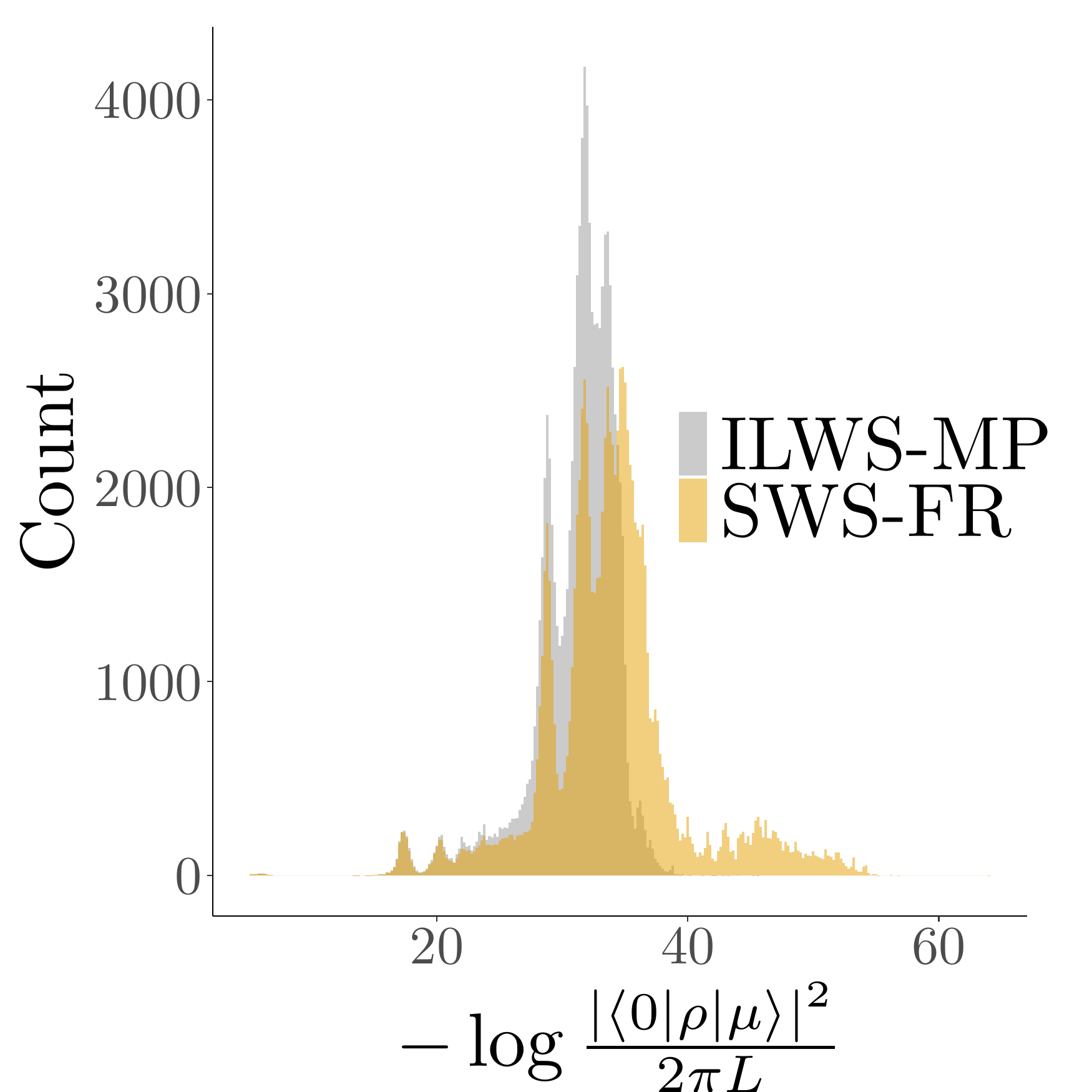}
  }
  \caption{Comparison of the histograms of the $f$ sumrule weights between the improved
  momentum preserving leapwise scanning (ILWS-MP) and SWS-FR. Starting from the
  representative thermal state at $T=1$, we generate 10,000 states for a target
  momentum of $k=\pi$, and $N = 128 = L$. We plot the resulting histogram for
  $c=1$ in $(a)$, for $c=4$ in $(b)$, and for $c=16$ in $(c)$. We see that
  the improved momentum preserving leapwise scanning and
  SWS-FR find the same states with very large weights (those on the left), but
  SWS-FR generates more less important states (the states on the right).}
  \label{fig:kf,rho,N=128,D_partial,ABACUS,distributions}
\end{figure}

The fact that SWS-FR generates such less important intermediate states can be seen from Fig.
\ref{fig:kf,rho,N=128,D_partial,ABACUS,distributions}.
where we consider the histograms of the first 100,000 states generated by each
algorithm. In these histograms, the $x$ axis measures the importance of an
eigenstate for the saturation of the f-sum rule, where the importance decreases
as $x$ increases. The rightmost gray part of the histograms represent the
intermediate states that are generated too soon due to the sub-optimal topology
of the tree that SWS-FR is building. Since this problem is related to the
topology of the tree, it occurs independently of the interaction strength
considered.

\subsection{Generating a basis for the interaction quench}
\label{subsec:generating a basis for the interaction quench}

\begin{figure}
  \subcaptionbox{$c=1$, $T=0$ \label{subfig:sumrule bases saturation T=0, c=1, kf rho, ABACUS, D_partial}} {
    \centering
    \includegraphics[width=0.3\textwidth]{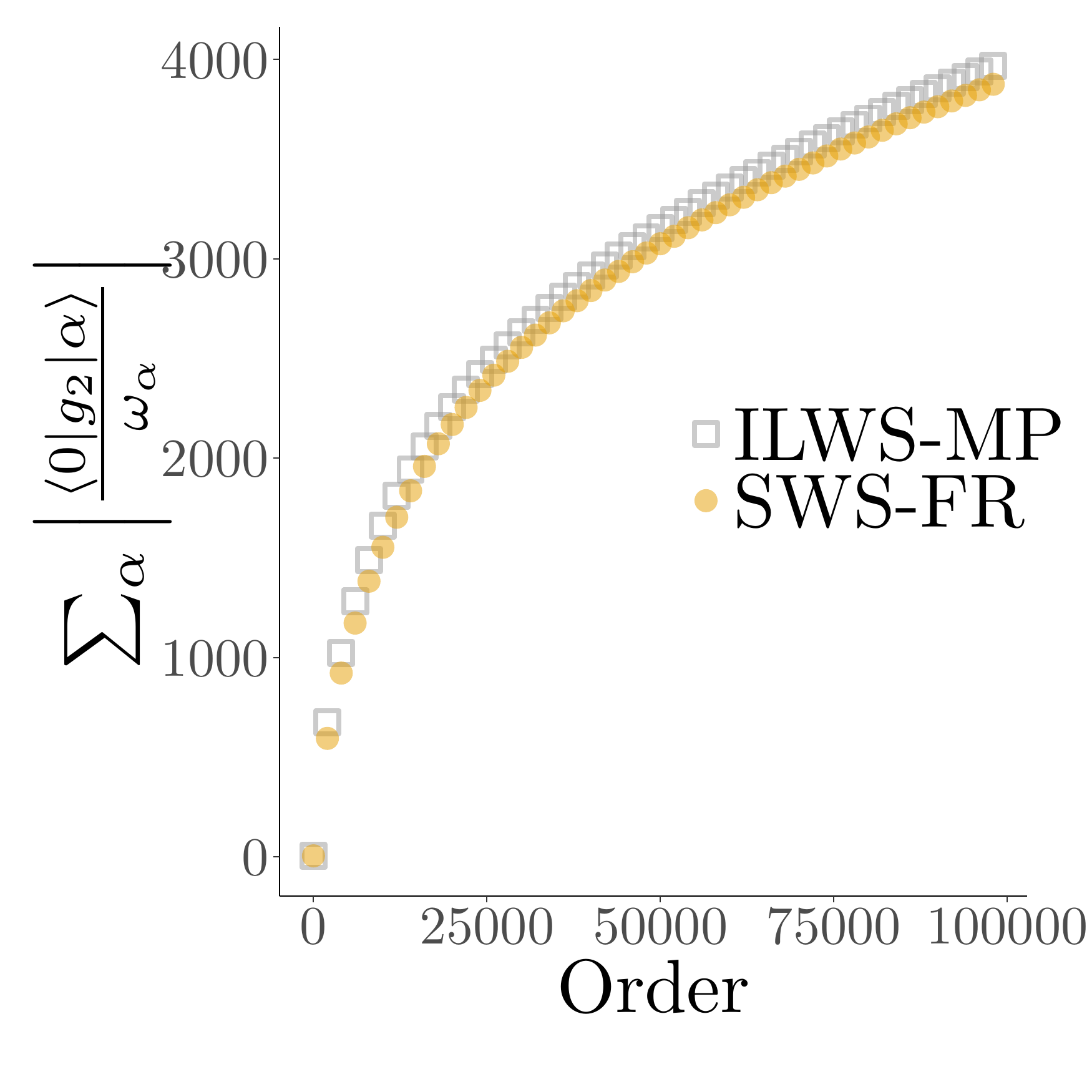}
  }
  \subcaptionbox{$c=4$, $T=0$ \label{subfig:sumrule bases saturation T=0, c=4, kf rho, ABACUS, D_partial}} {
    \centering
    \includegraphics[width=0.3\textwidth]{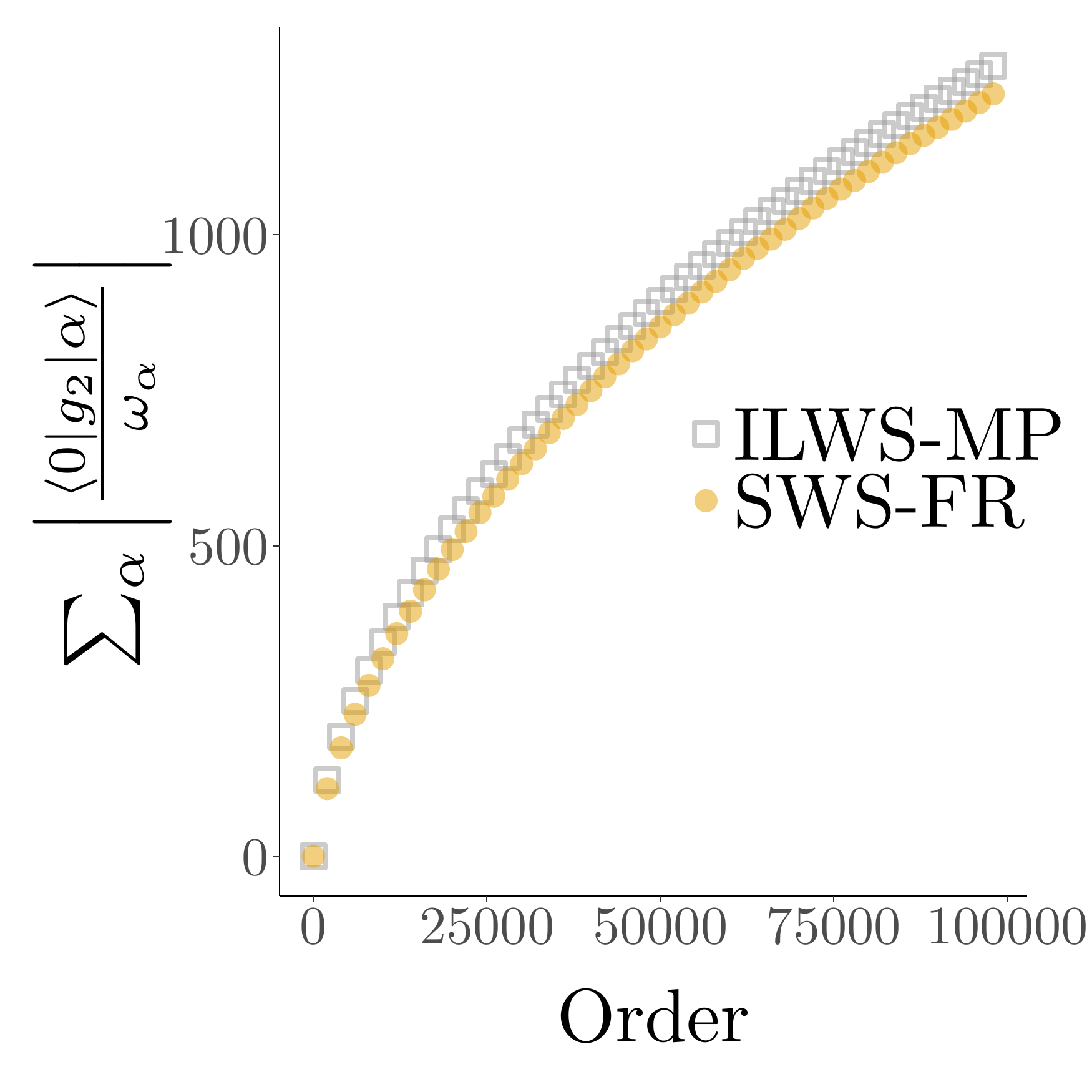}
  }
  \subcaptionbox{$c=16$, $T=0$ \label{subfig:sumrule bases saturation T=0, c=16, kf rho, ABACUS, D_partial}} {
    \centering
    \includegraphics[width=0.3\textwidth]{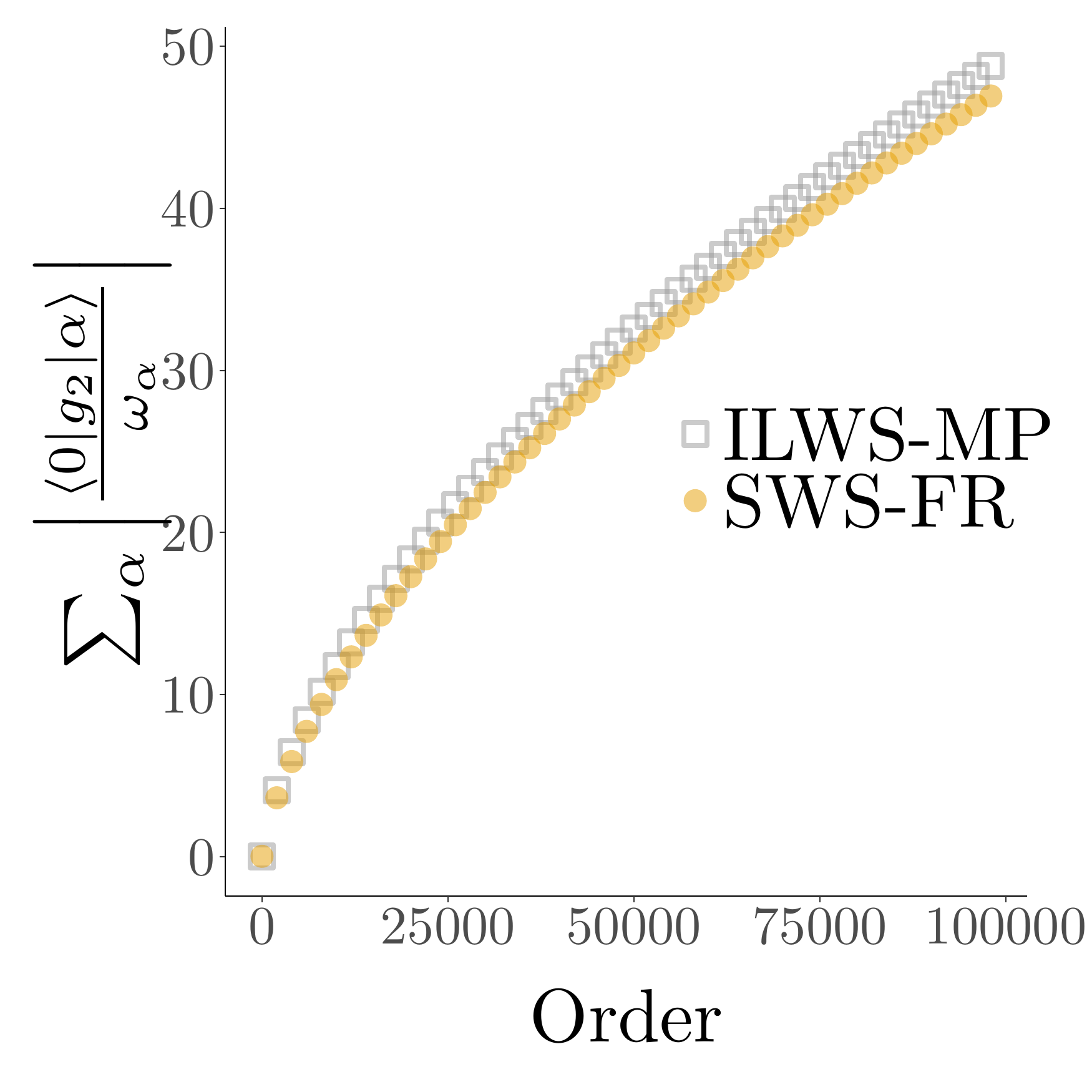}
  }
  \\
  \subcaptionbox{$c=1$, $T=1$ \label{subfig:sumrule bases saturation T=1, c=1, kf rho, ABACUS, D_partial}} {
    \centering
    \includegraphics[width=0.3\textwidth]{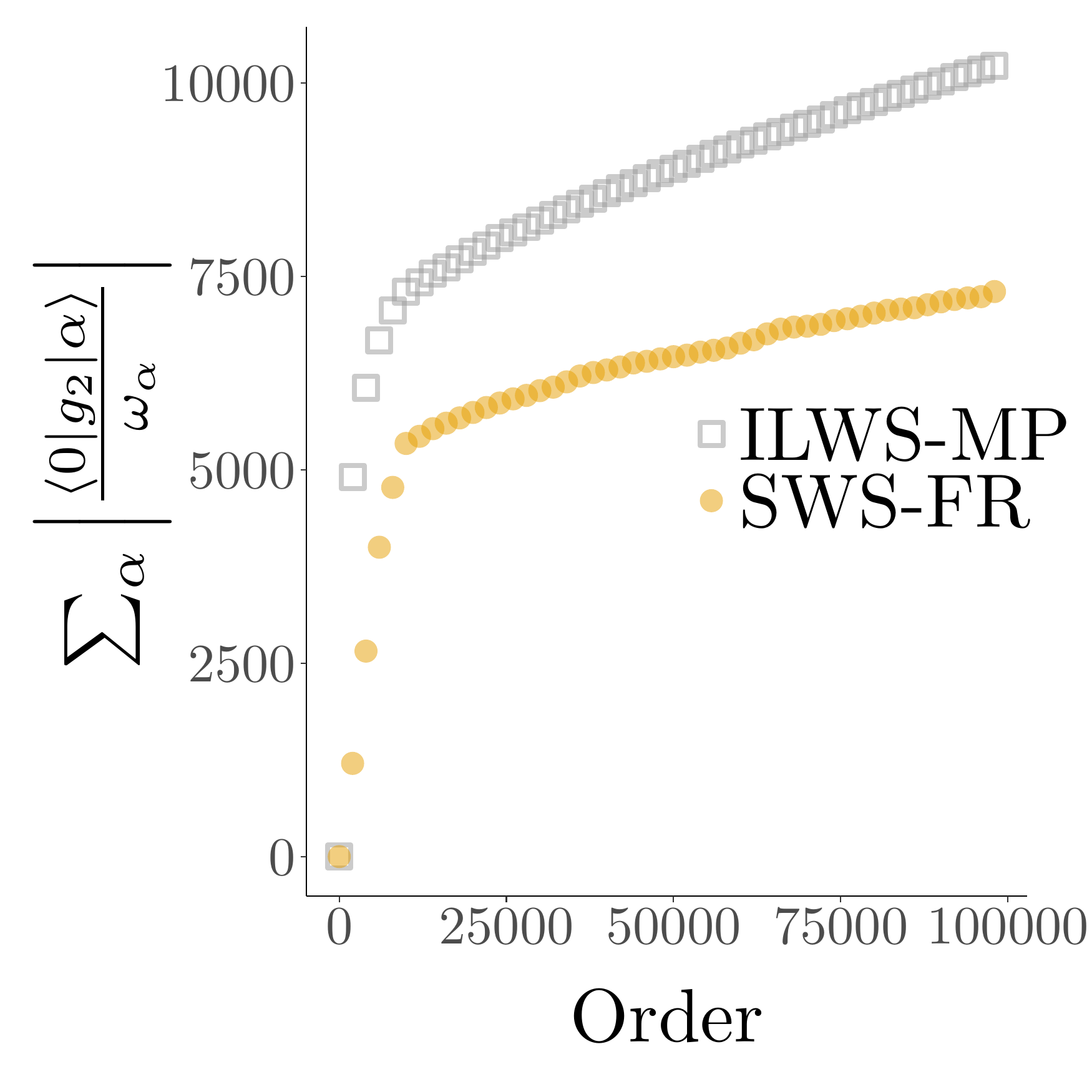}
  }
  \subcaptionbox{$c=4$, $T=1$ \label{subfig:sumrule bases saturation T=1, c=4, kf rho, ABACUS, D_partial}} {
    \centering
    \includegraphics[width=0.3\textwidth]{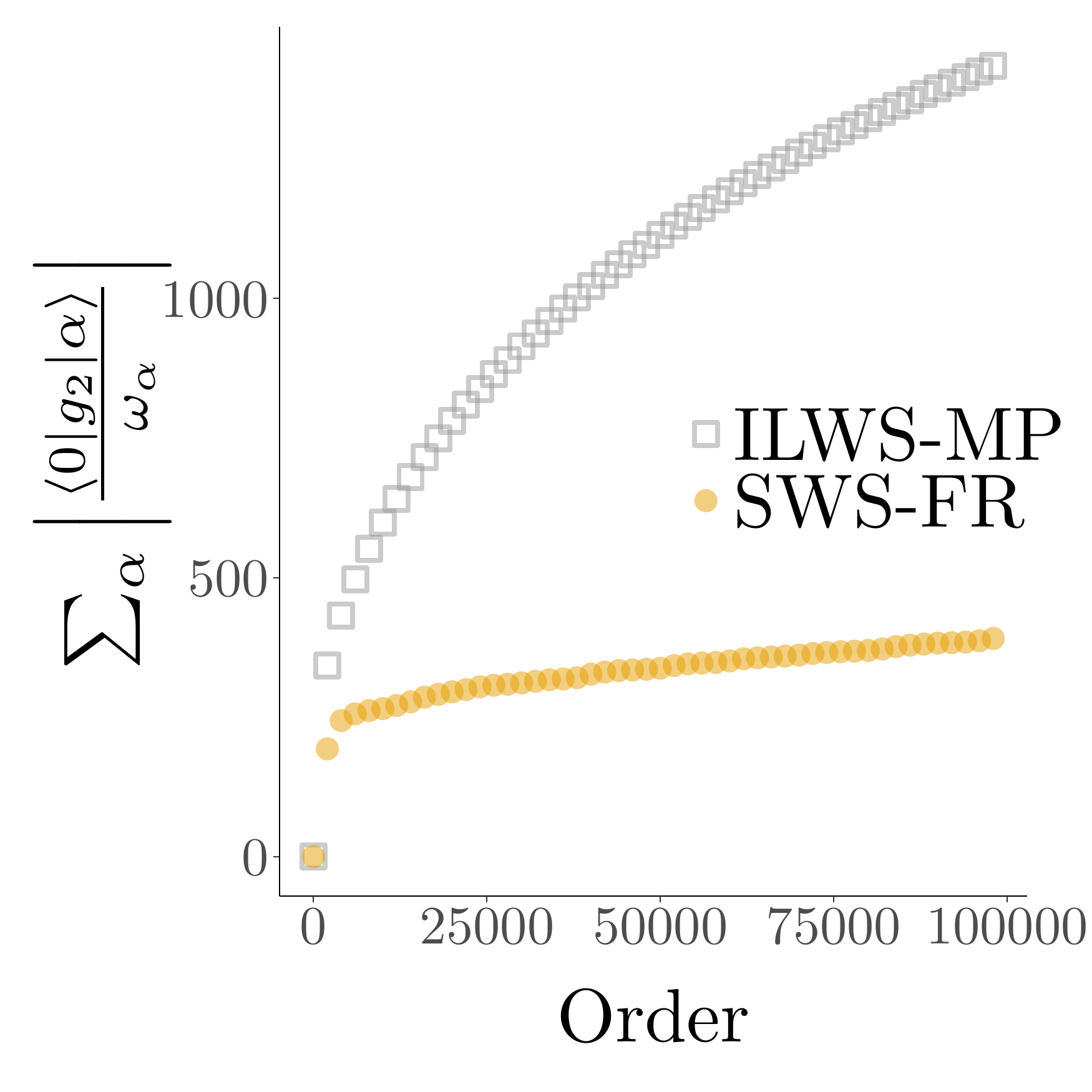}
  }
  \subcaptionbox{$c=16$, $T=1$ \label{subfig:sumrule bases saturation T=1, c=16, kf rho, ABACUS, D_partial}} {
    \centering
    \includegraphics[width=0.3\textwidth]{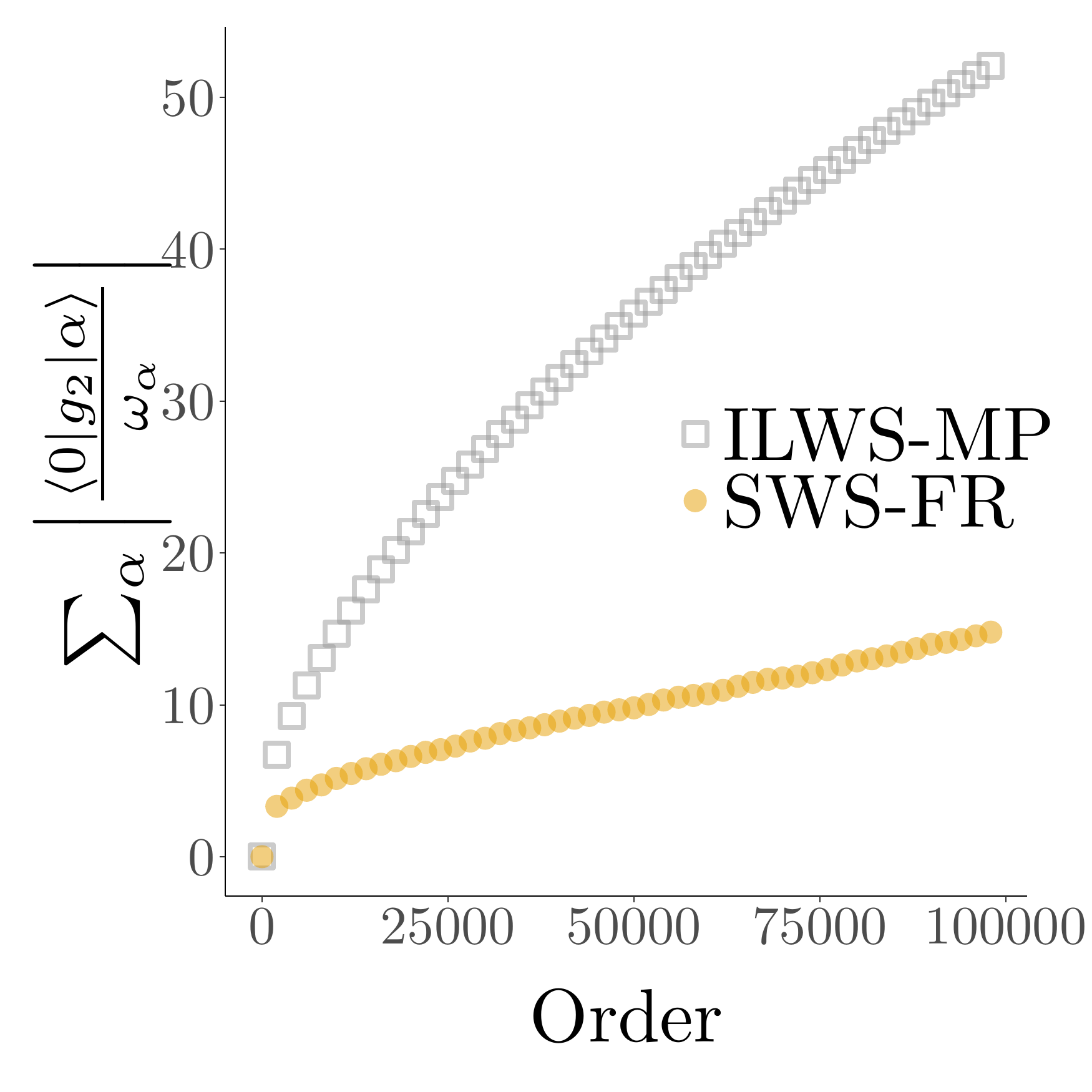}
  }
  \caption{
  Comparison of the sum of eigenstate weights with the number of states included
  in the summation between the improved momentum preserving leapwise scanning (ILWS-MP)
  and momentum preserving stepwise scanning with forced recombinations (SWS-FR).
  Starting from the ground state for $(a)-(c)$ and the representative thermal
  state at $T=1$ for $(d)-(f)$, we generate 10,000 states and 100,000 states respectively for a target momentum
  of $k=\pi$, and $N = 128 = L$. We plot the sum rule saturation after every 200
  states for $c=1$ in $(a)$ and $(d)$, for $c=4$ in $(b)$ and $(e)$, and for
  $c=16$ in $(c)$ and $(f)$. The differences between SWS-FR and the improved
  momentum preserving leapwise scanning for the sums at zero temperature are
  small whereas at finite temperature the latter outperforms SWS-FR by a large
  margin.}
  \label{fig:kf,rho,N=128,D_partial,ABACUS,bases}
\end{figure}

Another way to benchmark our algorithm is by considering the problem of a quench
in the interaction strength from $c_i$ to $c_f$
\cite{robinson_computing_2021-1}.
In this case, we want to find an approximate expansion of some initial
eigenstate $|\Psi_0 \rangle$ of $H(c_i)$ in terms of eigenstates of $H(c_f)$.
Truncated spectrum methods can be used to obtain the expansion coefficients
$b_\alpha$ in
\begin{equation}
  |\Psi_0 (t) \rangle = \sum_\alpha b_\alpha e^{-iE_\alpha t} |\alpha \rangle.
\end{equation}
In order to obtain an expansion that captures the time evolution following the
quench accurately, we need to choose our basis states $|\alpha \rangle$ wisely.
In \cite{robinson_computing_2021-1} we showed that a good estimate of the
importance of an eigenstate $|\alpha \rangle$ is given by
\begin{equation}
  w(|\alpha \rangle) = \left| \frac{\langle \Psi_0 | g_2 | \alpha \rangle }{E_{\Psi_0} - E_\alpha} \right|.
\end{equation}
The task of our scanning algorithms is therefore to generate the eigenstates
with the largest weights.

In order to compare our algorithm to the momentum preserving stepwise scanning
algorithm with recombinations, we compare the sums of the weights generated
having generated $i$ eigenstates as shown in Fig.
\ref{fig:kf,rho,N=128,D_partial,ABACUS,bases}.
We see that at zero temperature, the results from the improved momentum
preserving leapwise scanning are only marginally better than those of SWS-FR,
which can again be explained by the fact that at zero temperature the topology
of trees generated by both algorithms is identical. At finite temperature
however, we see a much more pronounced difference than we saw for the finite
temperature calculation of the dynamical structure factor.

\begin{figure}
  \subcaptionbox{$c=1$, $T=1$ \label{subfig:sumrule bases saturation T=1, c=1, kf rho, ABACUS, D_partial histogram}} {
    \centering
    \includegraphics[width=0.3\textwidth]{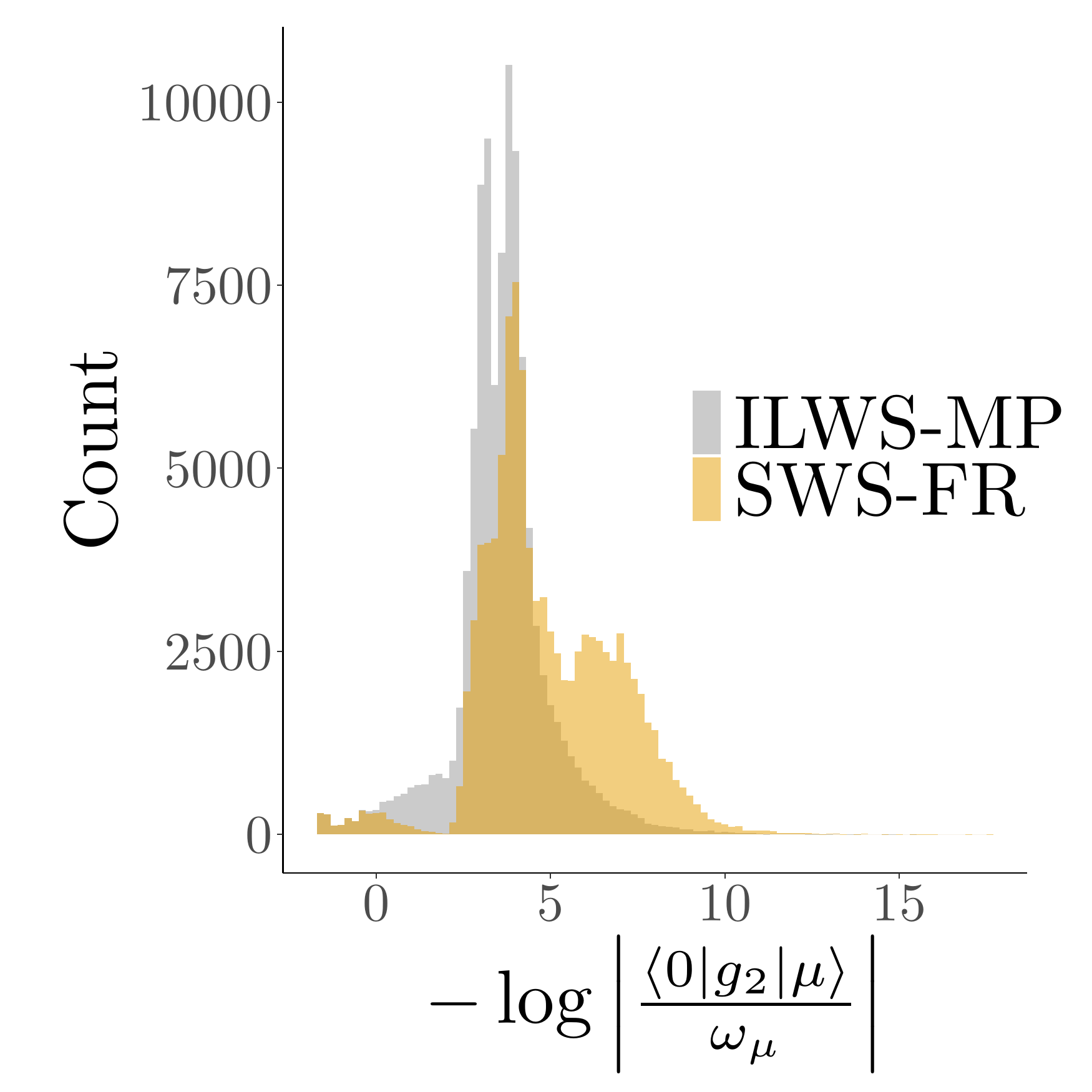}
  }
  \subcaptionbox{$c=4$, $T=1$ \label{subfig:sumrule bases saturation T=1, c=4, kf rho, ABACUS, D_partial histogram}} {
    \centering
    \includegraphics[width=0.3\textwidth]{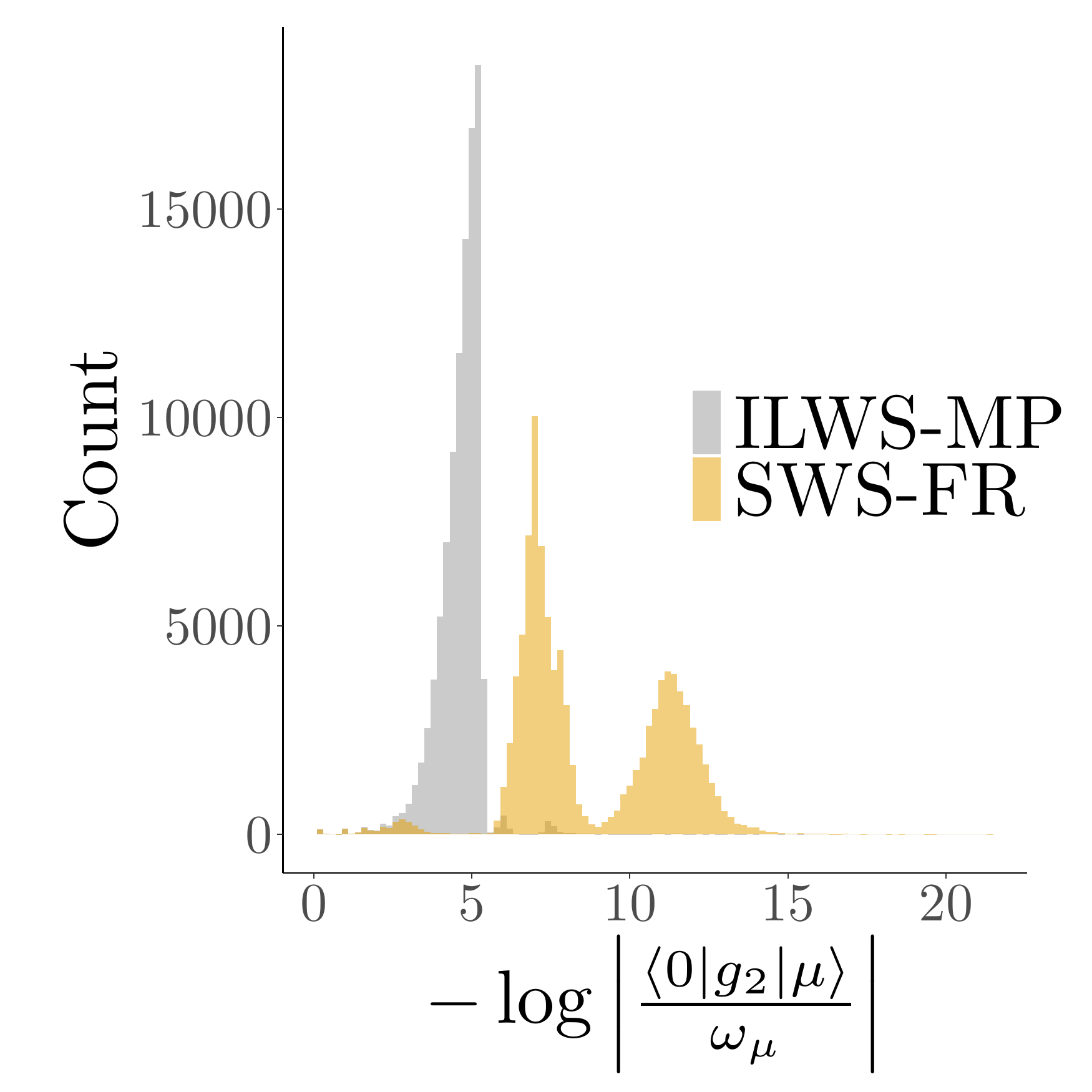}
  }
  \subcaptionbox{$c=16$, $T=1$ \label{subfig:sumrule bases saturation T=1, c=16, kf rho, ABACUS, D_partial histogram}} {
    \centering
    \includegraphics[width=0.3\textwidth]{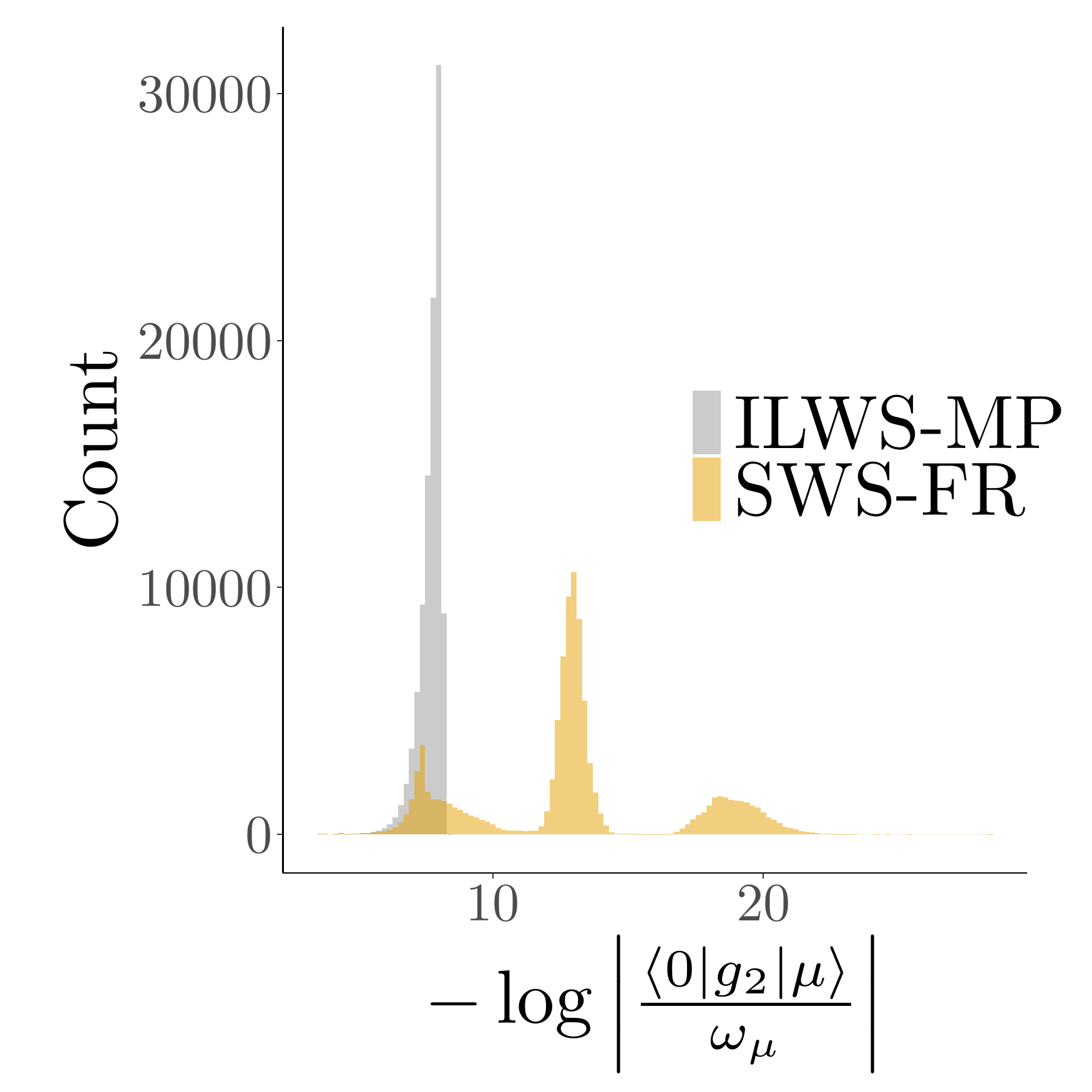}
  }
  \caption{Comparison of the histograms of the first 100,000 states generated by
  SWS-FR and the improved momentum preserving leapwise scanning (ILWS-MP)
  for the basis generation problem at $k=0$ and $N=128=L$ at $T=1$. We see that
  there is some overlap in which states are generated by both algorithms for the
  large weight states (on the left) but we also see that SWS-FR generates far
  more low weight states. At larger interaction strengths, we see that these
  lower weight states generated by SWS-FR are grouped in two distinct bumps.}
  \label{fig:T=1,k=0,IC,N=128,D_partial,ABACUS,bases}
\end{figure}

\begin{figure}
  \subcaptionbox{$c=1$, $T=0$ \label{subfig:sumrule bases saturation T=0, c=1, kf rho, ABACUS, D_partial ph breakdown}} {
    \centering
    \includegraphics[width=0.3\textwidth]{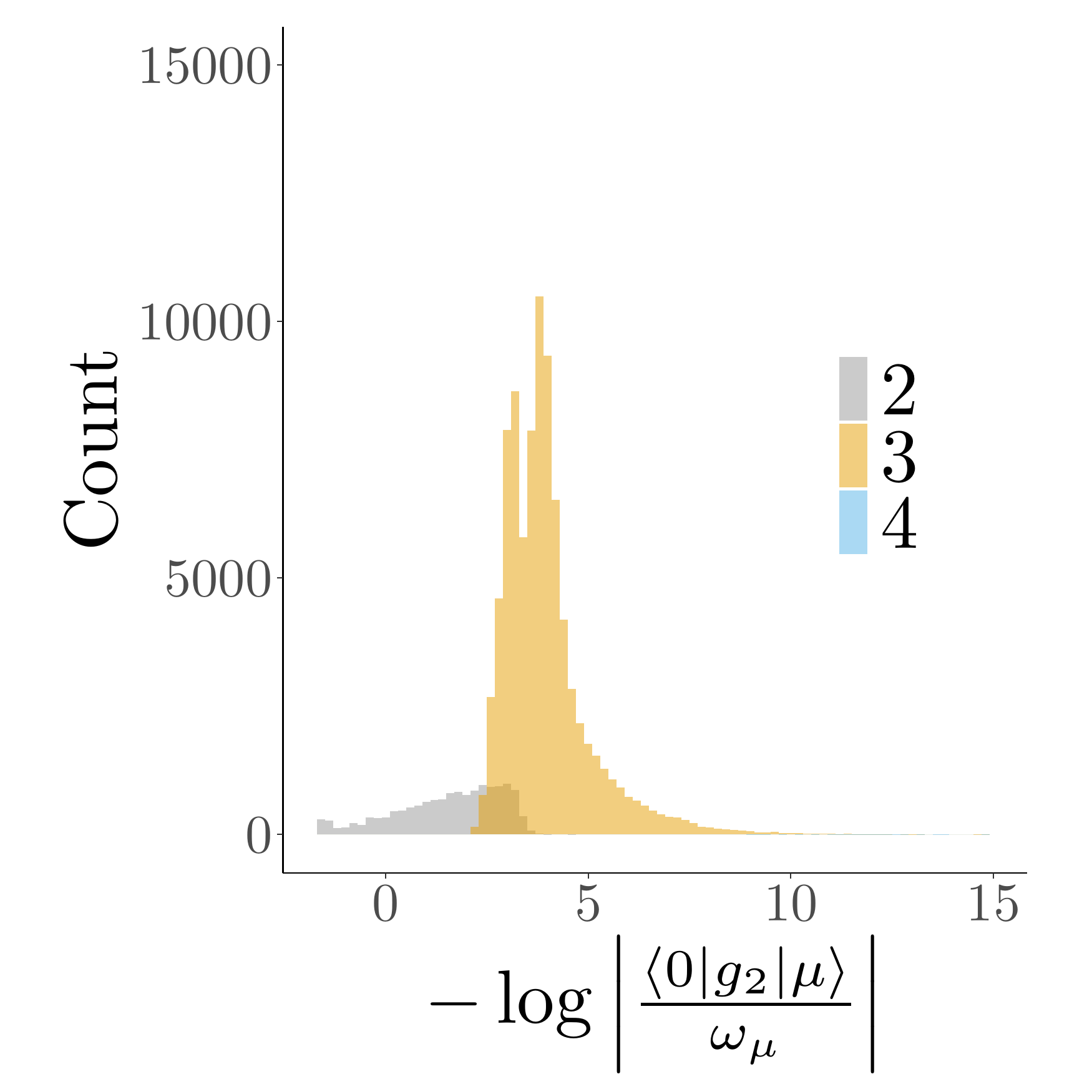}
  }
  \subcaptionbox{$c=4$, $T=0$ \label{subfig:sumrule bases saturation T=0, c=4, kf rho, ABACUS, D_partial ph breakdown}} {
    \centering
    \includegraphics[width=0.3\textwidth]{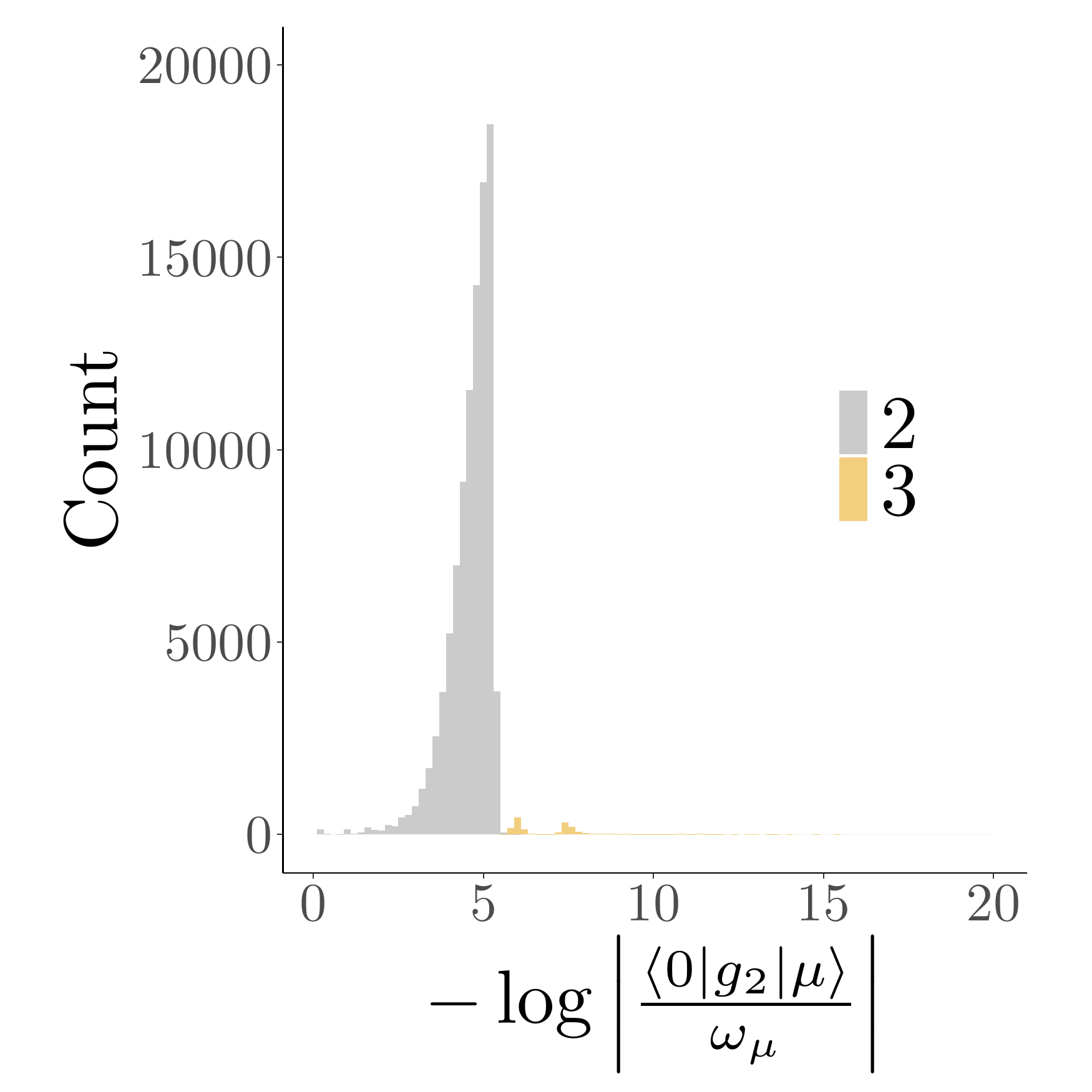}
  }
  \subcaptionbox{$c=16$, $T=0$ \label{subfig:sumrule bases saturation T=0, c=16, kf rho, ABACUS, D_partial ph breakdown}} {
    \centering
    \includegraphics[width=0.3\textwidth]{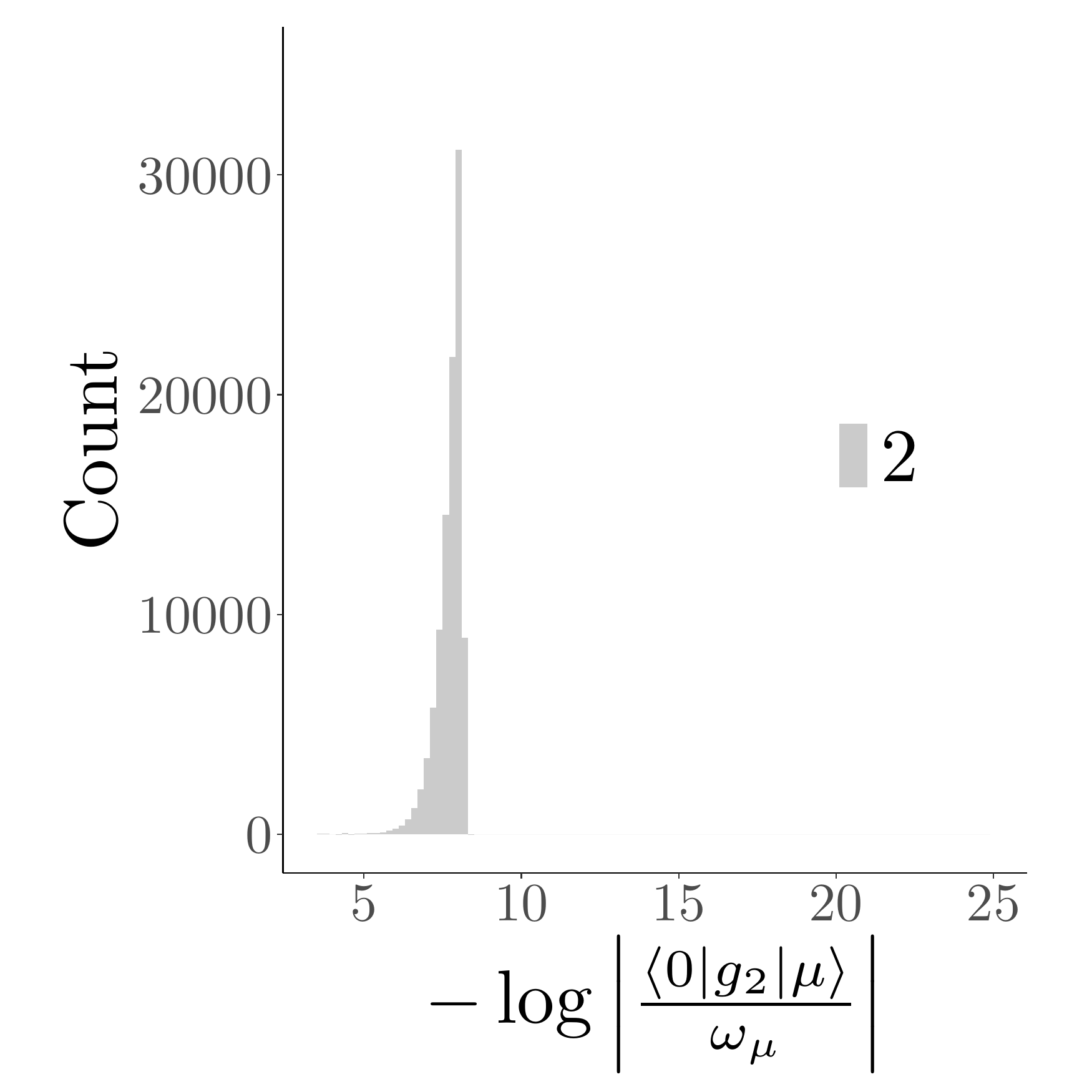}
  }
  \\
  \subcaptionbox{$c=1$, $T=1$ \label{subfig:sumrule bases saturation T=1, c=1, kf rho, ABACUS, D_partial ph breakdown}} {
    \centering
    \includegraphics[width=0.3\textwidth]{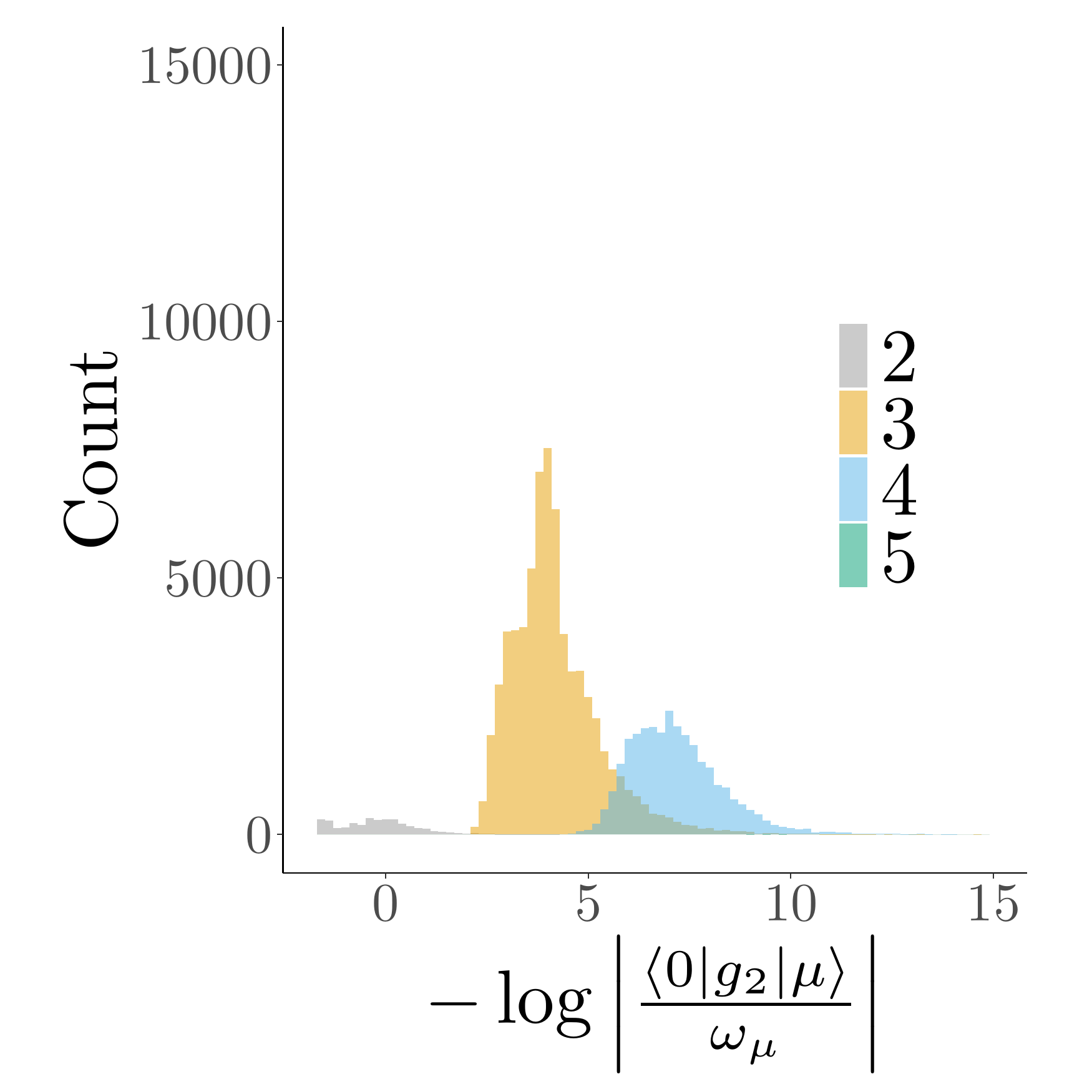}
  }
  \subcaptionbox{$c=4$, $T=1$ \label{subfig:sumrule bases saturation T=1, c=4, kf rho, ABACUS, D_partial ph breakdown}} {
    \centering
    \includegraphics[width=0.3\textwidth]{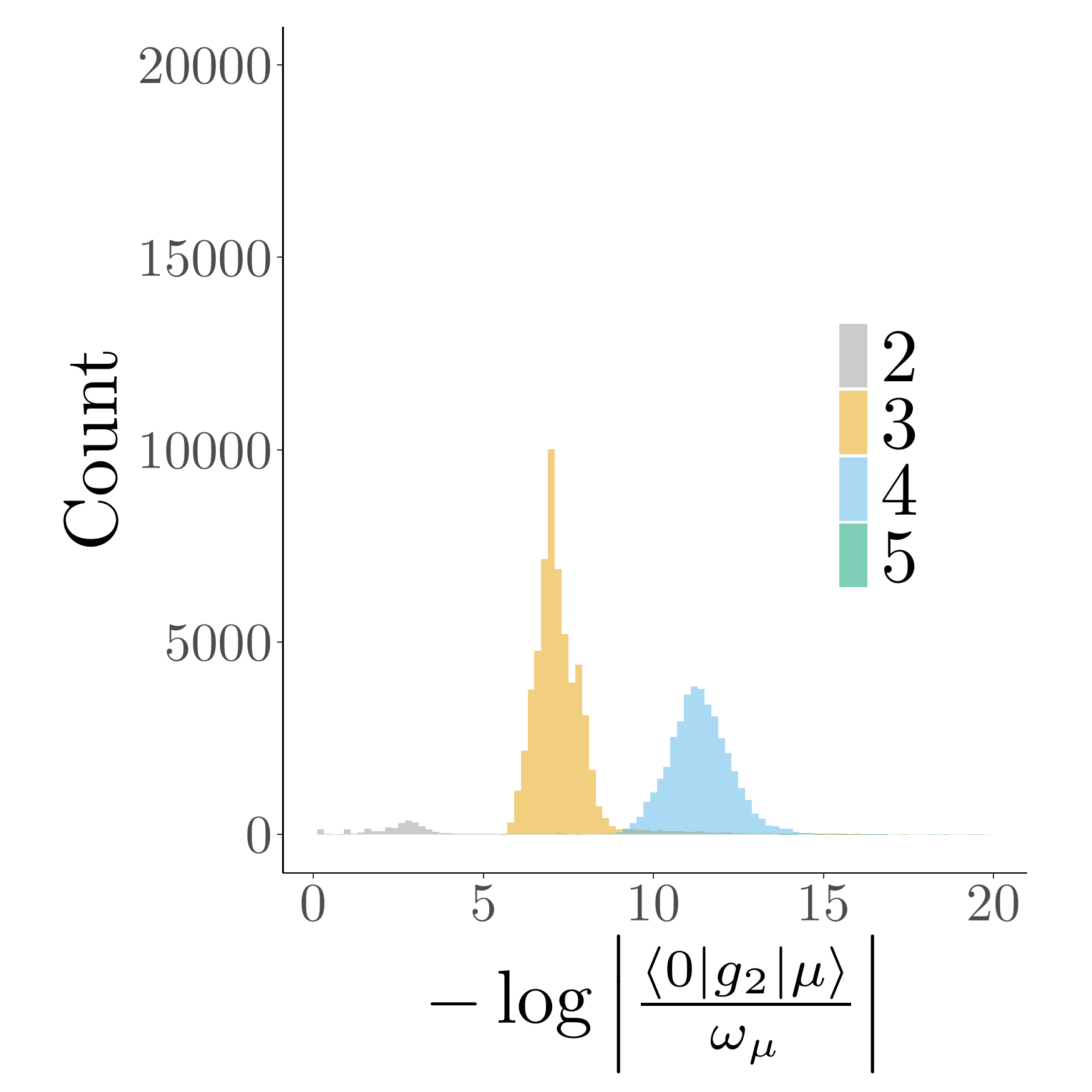}
  }
  \subcaptionbox{$c=16$, $T=1$ \label{subfig:sumrule bases saturation T=1, c=16, kf rho, ABACUS, D_partial ph breakdown}} {
    \centering
    \includegraphics[width=0.3\textwidth]{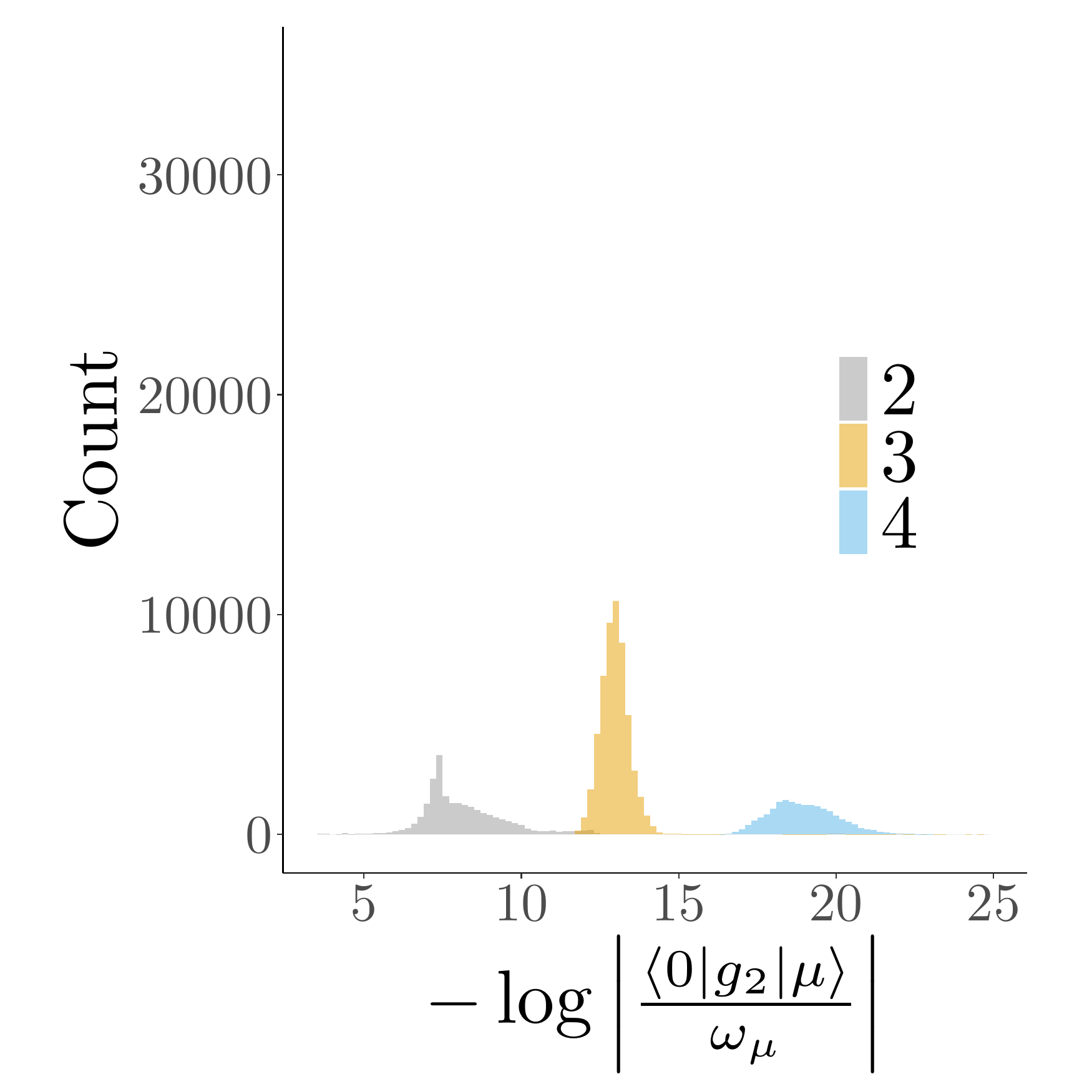}
  }
  \caption{Breakdown of the histograms shown in Fig.
  \ref{fig:T=1,k=0,IC,N=128,D_partial,ABACUS,bases} based on the number of
  particle-hole pairs for improved momentum preserving leapwise scanning
  (ILWS-MP) $(a)$-$(c)$ and SWS-FR $(d)$-$(f)$. We see that the bumps of
  unimportant states generated by SWS-FR are states from the three and four
  particle-hole sectors whereas improved momentum preserving leapwise scanning
  is capable of sticking to lower particle-hole sectors.}
  \label{fig:T=1,k=0,IC,N=128,D_partial,ABACUS,bases,breakdown}
\end{figure}

So why is the difference between SWS-FR and the improved momentum preserving
leapwise scanning for this basis generation problem so much more pronounced
compared to the finite temperature dynamical structure factor calculation? Note
that the $f$ sum rule for the dynamical structure factor is dominated by the one
particle-hole sector, which both algorithms are perfectly capable of generating
in full, as it
contains at most $N$ states at fixed momentum. For the basis generation problem
considered here however, the states contributing most strongly are those in the
two particle-hole sector (states in the one particle-hole sector cannot have the
same momentum as the seed state). The process of generating a given state in the
two particle-hole sector is however very different between the improved momentum
preserving leapwise scanning and SWS-FR. As we mentioned, SWS-FR is forced to
generate intermediate states, of which there are increasingly many as
temperature increases. The improved momentum preserving leapwise scanning
 on the other hand allows for jumps of quantum numbers by more
than one position allowing it to generate these states without intermediate
states resulting in a more efficient calculation.
The choice of topology of the tree combined with the versatile way of building
the tree is thus what allows the algorithm to focus on the contributions that
are most important at a given point in the calculation,
as illustrated by the histograms of contributions in Fig. 
\ref{fig:T=1,k=0,IC,N=128,D_partial,ABACUS,bases}

In order to substantiate our claim that our algorithm outperforms SWS-FR due to
the different topology of the tree, consider
Fig. \ref{fig:T=1,k=0,IC,N=128,D_partial,ABACUS,bases,breakdown}.
Here we see a breakdown of the histograms of weights generated by either
the improved momentum preserving leapwise scanning $(a)$ through $(c)$ or SWS-FR
$(d)$ through $(f)$ based on the number of particle-hole sectors the states whose
weights are displayed are from previously shown in Fig.
\ref{fig:T=1,k=0,IC,N=128,D_partial,ABACUS,bases}
based on the number of particle-hole pairs of a given contribution.
We see that SWS-FR spends its time generating a lot of unimportant states from
the three and four particle-hole sectors in order to generate the important
contributions from the two particle-hole sector.
Since for the improved momentum preserving leapwise scanning any state can be
generated without intermediate states with more particle-hole pairs, it is able
to avoid such problems.
We therefore conclude that the improved momentum preserving leapwise scanning
algorithm is better suited for dealing with problems at finite temperature than
algorithms that are not able to strictly preserved the number of particle-hole
pairs like stepwise scanning or SWS-FR.
We have seen that this is especially true if the calculation under consideration
is not one dominated by the one particle-hole sector.

\section{Conclusions}
\label{sec:conclusions}

Despite the powerful analytical tools that have been developed for integrable
systems, performing a numerical evaluation over eigenstates is still often an
inevitable step required to compute correlation functions.
It is therefore crucial that we have good Hilbert space exploration algorithms
that allow us to accurately and efficiently approximate such summations.
Here we have reviewed the basic principles that such an algorithm has to satisfy
and developed a number of concrete examples that satisfy these criteria.
Starting from the most basic algorithm satisfying these basic principles, we
considered its shortcomings one by one and proposed solutions resulting in
incremental changes that led us to the final algorithm.
Finally, we compared this algorithm to the state of the art for the dynamical
structure factor at zero and finite temperature as well as the problem of
generating an efficient basis for a quench in the interaction strength.

Overall, our algorithms can be viewed as an algorithm for building a
single-rooted tree where every node of the tree represents an eigenstate and
the algorithm specifies the topology of the tree.
Since the tree is infinite and we only have finite computational resources it is
also important in what order the nodes of the tree are generated as this
determines what the tree looks like after some finite time.
A key characteristic for the importance of a state is the number of
particle-hole pairs it has with respect to the seed state.
In order to avoid having to consider less relevant states with more
particle-hole pairs we chose our final algorithm to have a topology
where any node with $n$ particle-hole pairs can be generated without
having to generate nodes with more particle-hole pairs.
This topology, in combination with a clever way of choosing the order in which
to generate new states is what led to the most efficient algorithm.

Comparing our final algorithm to the current version of the state of the art
library \textsc{abacus} showed that we outperform the latter when considering
the dynamical structure factor at finite temperature as well as the problem of
generating an efficient basis for a quench in the interaction strength at finite
temperature.
In the latter case the difference is particularly pronounced, emphasizing the
importance of the choice of topology.
After all, the reason that our algorithm outperforms \textsc{abacus} is
primarily due to the fact that if \textsc{abacus} wants to generate certain
states with $n$ particle-hole pairs it has to go through states with more
particle-hole pairs.
The examples considered show that the improved momentum preserving leapwise
scanning algorithm offers significant advantages for finite temperature
calculations.
One of the main advantages is due to the topology of the tree being generated,
which allows any state from the $n$ particle-hole sector to be generated without
generating states from higher particle-hole sectors.
This is especially important when considering calculations not dominated by the
one particle-hole sector as the number of intermediate states required to
construct a state from the $n$ particle-hole sector grows dramatically with
increasing $n$ at finite temperature.
Improved momentum preserving leapwise scanning is therefore more suited to
dealing with problems involving for example the $g_2$ and $g_3$ operators, the
latter being relevant to modelling three-body losses.
Because of these advantages, future versions of \textsc{abacus} will incorporate this
approach to be better equiped to deal with finite temperature calculations.

In this work we considered calculations where the number of particles of the
states we explore is equal to the number of particles of the reference state.
However, for some problems this is not the case (e.g. when calculating the
Green's function).
In these cases additional difficulties arise for finite temperature states since
in this case the number of particle-hole pairs is no longer well-defined.
Future research is required to come up with strategies to efficiently deal with
these problems.
Another interesting direction for future research direction would be the extension
of the ideas in this paper to spin chains.
Despite their differences, such as the existence of string solutions, the spin
chain and the Bose gas also share key characteristics.
For example, the quantum numbers of the spin chain can be viewed as a multi-level
version of those of the Bose gas, each lattice representing the quantum numbers
of a given string sector which can be visualized as:
\begin{equation*}
\begin{split}
  &\vdots  \\
  \textrm{2-strings:} \hspace{0.5cm} \cdots \; {\circ} \; {\circ} \; {\circ} \; {\circ} \; &{\bullet} \; {\circ} \; {\bullet} \; {\circ} \; {\circ} \; \cdots \\
  \textrm{1-strings:} \hspace{0.5cm} \cdots \; {\circ} \; {\circ} \; {\bullet} \; {\bullet} \; &{\bullet} \; {\bullet} \; {\circ} \; {\bullet} \; {\circ} \; \cdots
\end{split}
\end{equation*}
An algorithm like the one we developed can then be applied to the quantum
numbers of every string sector.
However, additional complications due to the differences are bound to arise.
For example, in the spin chain more care has to be taken to avoid overcounting
since a given eigenstate can be represented by multiple quantum number
configurations.
Furthermore, additional constraints such as the magnetization, as well as the
periodicity of the momentum will require careful consideration.

\section{Acknowledgements}

We are grateful to Neil Robinson for his valuable insights and his suggestions
during the preparation of this manuscript.

\section{Funding information}

This work received funding from the European Research Council under ERC Advanced
grant No 743032 DYNAMINT.

\bibliography{References}
\end{document}